\documentclass[12pt]{article}
\usepackage{a4wide}
\usepackage{latexsym}
\usepackage{amsmath}
\usepackage{amssymb}
\usepackage{amsfonts}
\usepackage{amscd}
\usepackage{cite}
\usepackage{placeins}
\usepackage{longtable}

\usepackage{pslatex}
\usepackage{graphicx}
\usepackage{float}
\usepackage[latin1,utf8]{inputenc}
\usepackage[T1]{fontenc}

\usepackage{xcolor}

\allowdisplaybreaks

\newcommand{\bq}{\begin{eqnarray}}
\newcommand{\eq}{\end{eqnarray}}
\newcommand{\eps}{\varepsilon}

\newcommand{\NF}{N_F}
\newcommand{\NL}{N_L}

\begin{document}

\thispagestyle{empty}

\begin{flushright}
  MITP/24-086
\end{flushright}

\vspace{1.5cm}

\begin{center}
  {\Large\bf Electroweak double-box integrals for M{\o}ller scattering\\
  }
  \vspace{1cm}
  {\large Niklas Schwanemann and Stefan Weinzierl \\
  \vspace{1cm}
      {\small \em PRISMA Cluster of Excellence, Institut f{\"u}r Physik,}\\
      {\small \em Johannes Gutenberg-Universit{\"a}t Mainz,}\\
      {\small \em D - 55099 Mainz, Germany}\\
  } 
\end{center}

\vspace{2cm}

\begin{abstract}\noindent
  {
We present for M{\o}ller scattering 
planar and non-planar two-loop double-box integrals where three electroweak gauge bosons are exchanged between the fermion lines,
among which at least one is a photon.
These integrals are relevant for the NNLO electroweak corrections to M{\o}ller scattering.
   }
\end{abstract}

\vspace*{\fill}

\newpage

\section{Introduction}
\label{sect:intro}

M{\o}ller scattering ($e^- e^- \rightarrow e^- e^-$) can be used to measure the weak mixing angle at low energies \cite{Benesch:2014bas}.
In order to match the anticipated experimental precision reliable calculations on the theory side are required. 
As higher precision on the theory side is reached by including higher-order perturbative corrections, 
electroweak two-loop corrections come into 
focus \cite{Dubovyk:2016aqv,Du:2019evk,Erler:2022ckm,Dubovyk:2022frj,Armadillo:2022ugh,Aleksejevs:2015dba,Aleksejevs:2013gxa,Aleksejevs:2012xua,Delto:2023kqv}.
The most complicated Feynman integrals at the two-loop level relevant to this process are the planar and the non-planar
double-box integrals. 
These integrals are the topic of this article.

We further note that the scattering amplitudes for Bhabha scattering ($e^+ e^- \rightarrow e^+ e^-$) and M{\o}ller scattering
are related by crossing symmetry, and the Feynman integrals
computed in this paper are useful for both processes.
Bhabha scattering is typically used to monitor the luminosity 
at an $e^+ e^-$-collider \cite{FCC:2018evy}.
The integrals are also relevant to electroweak two-loop corrections to the Drell-Yan process and to quark-pair production in electron-positron annihilation.

In this paper we present planar and non-planar two-loop double-box integrals where three electroweak gauge bosons are exchanged between the fermion lines,
with the additional requirement that at least one exchanged gauge-boson is a photon.

Throughout this paper we treat electrons (and neutrinos) as massless.
The non-zero kinematic variables are the two Mandelstam variables $s$ and $t$ (the third Mandelstam variable is given by $u=-s-t$)
and the mass $m$ of the heavy gauge boson (either a $Z$-boson or a $W$-boson).
Diagrams with two massive gauge bosons will necessarily have either two $Z$-bosons or two $W$-bosons.
The mixed case is excluded by the assumed initial and final states.
For each diagram we distinguish the mass configuration: We consider the cases with zero, one or two internal massive gauge bosons.
For one and two internal massive gauge bosons we have for each topology two inequivalent possibilities to assign the masses, leading
to five different mass configurations for each topology.
The complexity of the calculation increases with the number of internal massive gauge bosons:
The simplest case are the Feynman integrals with three internal photons.
These have been computed long time ago \cite{Smirnov:1999gc,Tausk:1999vh}.
This case is included here only for completeness.
On the other side, the non-planar integrals with two internal massive gauge bosons involve elliptic curves and require state-of-the-art techniques.
This paper is an example, how techniques developed for elliptic Feynman integrals \cite{Sabry:1962,Broadhurst:1993mw,Laporta:2004rb,MullerStach:2011ru,Adams:2013nia,Bloch:2013tra,Remiddi:2013joa,Adams:2014vja,Bloch:2014qca,Adams:2015gva,Adams:2015ydq,Bloch:2016izu,Adams:2017ejb,Bogner:2017vim,Adams:2018yfj,Honemann:2018mrb,Sogaard:2014jla,Tancredi:2015pta,Primo:2016ebd,Remiddi:2016gno,Adams:2016xah,Bonciani:2016qxi,vonManteuffel:2017hms,Adams:2017tga,Ablinger:2017bjx,Primo:2017ipr,Passarino:2017EPJC,Remiddi:2017har,Bourjaily:2017bsb,Hidding:2017jkk,Broedel:2017kkb,Broedel:2017siw,Broedel:2018iwv,Adams:2018bsn,Adams:2018kez,Broedel:2018qkq,Ablinger:2018zwz,Broedel:2019hyg,Blumlein:2019svg,Broedel:2019tlz,Bogner:2019lfa,Kniehl:2019vwr,Broedel:2019kmn,Abreu:2019fgk,Duhr:2019rrs,Weinzierl:2020fyx,Walden:2020odh,Bezuglov:2020ywm,Kristensson:2021ani,Bourjaily:2021vyj,Badger:2021owl,Broedel:2021zij,Frellesvig:2021hkr,Duhr:2021fhk,Wilhelm:2022wow,Morales:2022csr,Pogel:2022yat,Giroux:2022wav,Bourjaily:2022bwx,Jiang:2023jmk,Frellesvig:2023iwr,Gorges:2023zgv,Ahmed:2024tsg,Duhr:2024bzt,Forner:2024ojj,Duhr:2024uid}
are applied to phenomenologically relevant processes.

In total we present ten families of Feynman integrals, which differ by the topology (planar or non-planar) and the mass configuration.
We label the planar families $A$, $B$, $C$, $D$ and $E$, where $A$ denotes the most complicated one and $E$ the simplest one.
The non-planar families are labelled in a similar way by $\tilde{A}$, $\tilde{B}$, $\tilde{C}$, $\tilde{D}$ and $\tilde{E}$.
The planar double-box diagrams and the non-planar double-box diagrams are shown in fig.~\ref{fig:doubleboxes_topos}.

We do not include in this paper the exchange of three massive gauge bosons between the fermion lines.
There are two reasons for this: 
On the one hand we expect the contributions from these integrals to M{\o}ller scattering to be suppressed by a (negative) power of the heavy gauge
boson mass.
On the other hand (and this is the main reason), this case goes beyond the current state-of-the-art: it is known that the non-planar double-box integral
with three internal massive gauge bosons is related to a curve of genus two \cite{Marzucca:2023gto}.
We will report on these integrals in a future publication.

We note that for the calculation of scattering amplitudes at the two-loop order one will need additional master integrals.
Worth mentioning is for example a further double-box graph, where two $W$-bosons are exchanged between the fermion lines and a photon or a $Z$-boson
is exchanged between the $W$-bosons.
We expect that the photon-exchange graph can be calculated along the lines of this article.
The $Z$-boson exchange graph will be more complicated.
The purpose of this article is to show that some of the most complicated integrals can be computed systematically,
and to lay the ground for the computation of the calculation of the double-box graphs with the exchange of three massive gauge bosons between the fermion lines.

To compute the Feynman integrals we follow to a large extent the setup and the notation 
used in \cite{Bottcher:2023wsr}, where the (simpler) box integrals with self-energy insertion 
have been computed for M{\o}ller scattering.
In particular, the original kinematic variables are defined in the same way.
The actual calculation proceeds along the following steps:
Using integration-by-parts identities \cite{Tkachov:1981wb,Chetyrkin:1981qh} we first derive a differential equation \cite{Kotikov:1990kg,Kotikov:1991pm,Remiddi:1997ny,Gehrmann:1999as}
for a pre-canonical basis of master integrals. This differential equation is in general not in an $\eps$-factorised form.
This first step only uses standard techniques.
The essential (second) step of our work is to construct a new basis, 
such that the differential equation is transformed to an $\eps$-factorised form \cite{Henn:2013pwa}.
This is the new result of our work.
As some Feynman integrals are related to elliptic integrals, we profit from the techniques developed in \cite{Adams:2018bsn,Adams:2018kez,Muller:2022gec}.
In a third step we determine for all integrals the boundary values.
This again is standard, and so is the final fourth step,
where we solve the $\eps$-factorised differential equation order-by-order in $\eps$.
We express the results for the simpler topologies in terms of multiple polylogarithms, the results for the more complicated topologies in terms of iterated
integrals.
For all integrals we provide numerical evaluation routines.
As an additional side-result we can look at potentially numerically large contributions, which arise from large logarithms.
The large logarithms are easily obtained from our full result.

This paper is organised as follows:
In the next section we introduce our notation.
In section~\ref{sect:master_integrals} we discuss the master integrals, which put the differential equation into an $\eps$-factorised form.
In section~\ref{sect:integration} we discuss the integration of the $\eps$-factorised differential equation.
Numerical results for a benchmark point are provided in section~\ref{sect:numerical_results}.
In section~\ref{sect:logarithms} we discuss large logarithms.
Finally, section~\ref{sect:conclusions} contains our conclusions.
The article includes three appendices: 
In appendix~\ref{sect:diagrams} we show for all master integrals the corresponding diagrams.
In appendix~\ref{sect:master_integrals_list} we list all master integrals, which put the differential equation into an $\eps$-factorised form.
In appendix~\ref{sect:supplement} we describe the content of the supplementary electronic file
attached to the arxiv version of this article.


\section{Notation and setup}
\label{sect:notation}


\subsection{Kinematics}

\begin{figure}
\begin{center}
\includegraphics[scale=0.5]{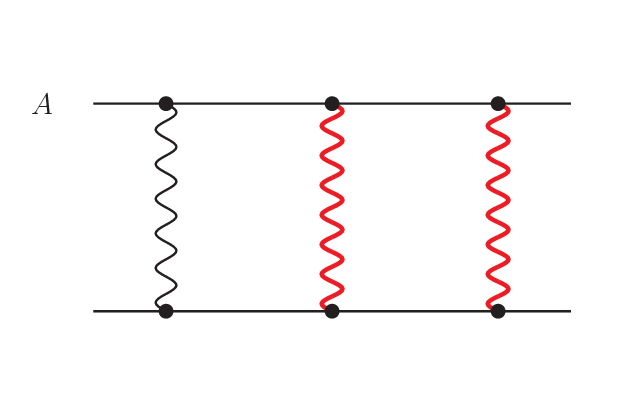}
\hspace*{5mm}
\includegraphics[scale=0.5]{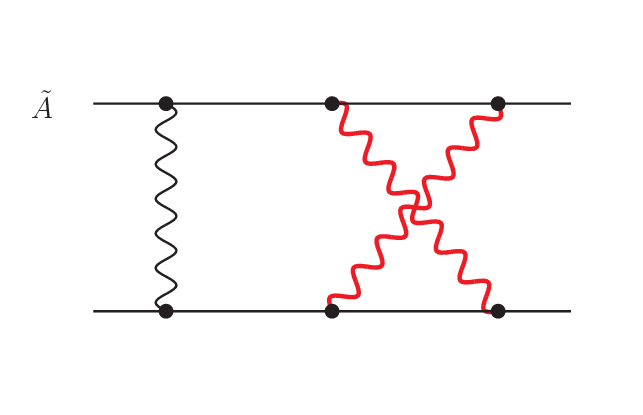}

\includegraphics[scale=0.5]{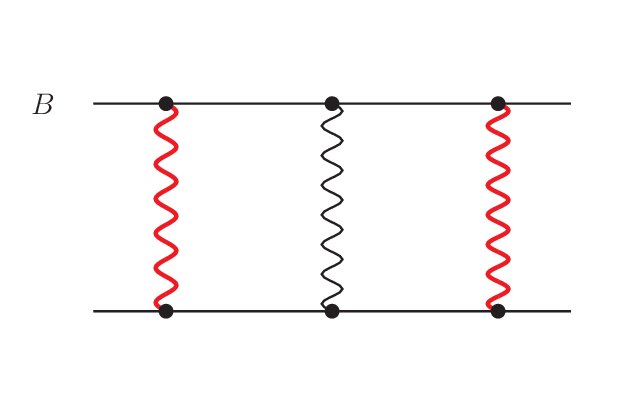}
\hspace*{5mm}
\includegraphics[scale=0.5]{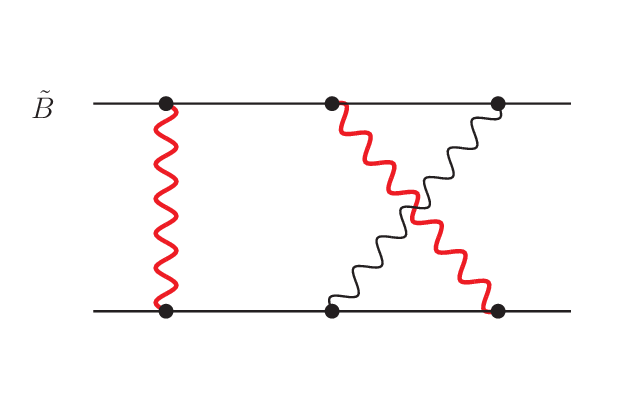}

\includegraphics[scale=0.5]{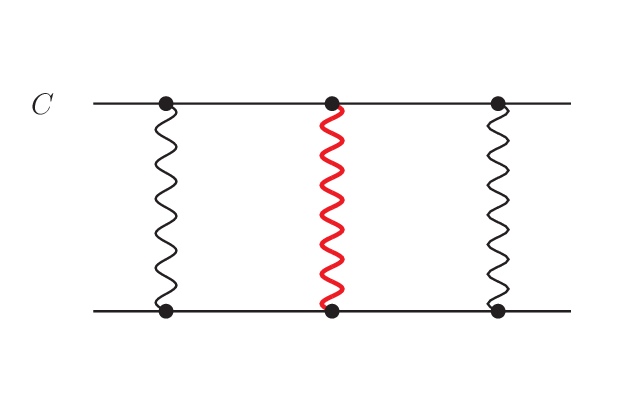}
\hspace*{5mm}
\includegraphics[scale=0.5]{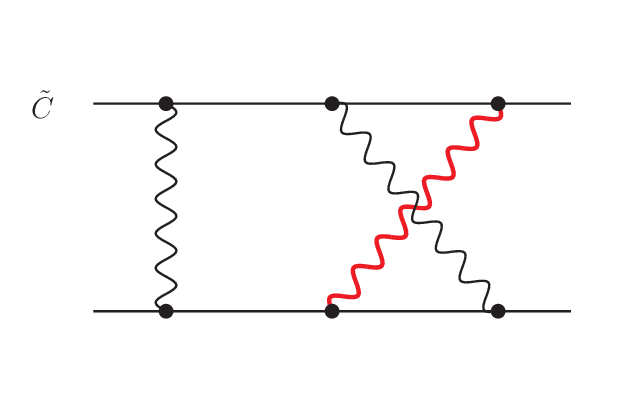}

\includegraphics[scale=0.5]{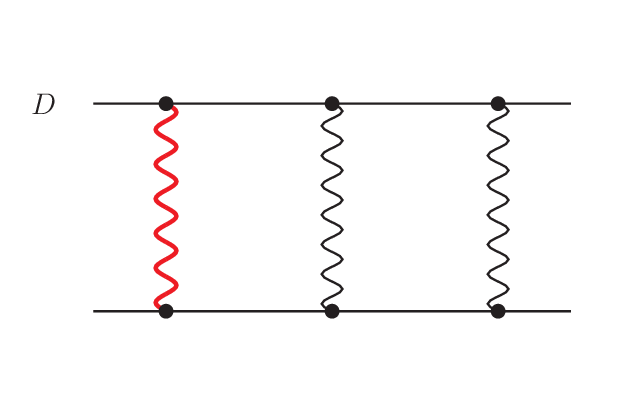}
\hspace*{5mm}
\includegraphics[scale=0.5]{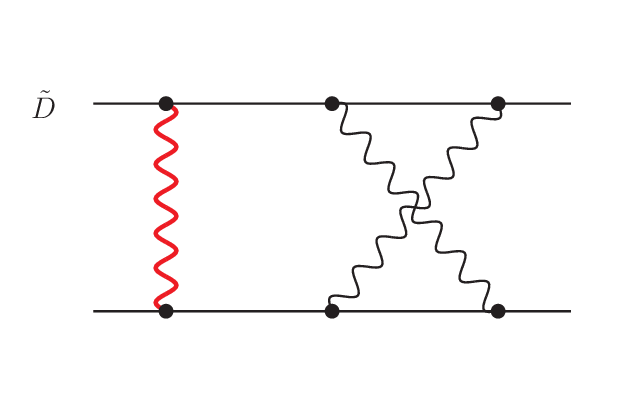}

\includegraphics[scale=0.5]{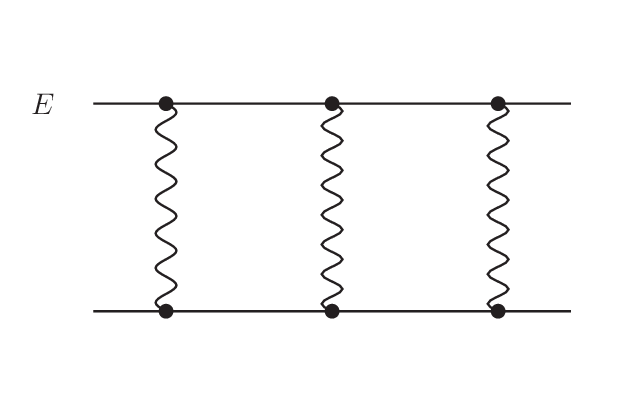}
\hspace*{5mm}
\includegraphics[scale=0.5]{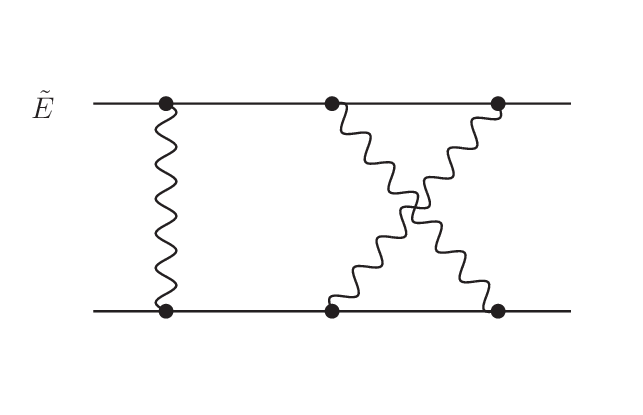}
\end{center}
\caption{
The planar double-box diagrams (left) and the non-planar double-box diagrams (right).
The solid lines are electrons, the black wavy lines are photons and the red wavy lines are heavy gauge bosons.
}
\label{fig:doubleboxes_topos}
\end{figure}

We consider two-loop electroweak corrections to M{\o}ller scattering
\bq
 e^-\left(-p_1\right) \; + \; e^-\left(-p_2\right) & \rightarrow & e^-\left(p_3\right) \; + \; e^-\left(p_4\right).
\eq
It will be convenient to take all momenta as outgoing, hence 
the momenta of the two incoming particles have a minus sign.
Momentum conservation reads
\bq
 p_1+ p_2 + p_3 + p_4 & = & 0.
\eq
We assume the electrons to be massless. 
This implies that the external momenta satisfy
\bq
 p_1^2 \; = \; p_2^2 \; = \; p_3^2 \; = \; p_4^2 \; = \; 0.
\eq
The Mandelstam variables are defined by
\bq
\label{def_Mandelstam}
 s \; = \; \left(p_1+p_2\right)^2,
 \;\;\;\;\;\;
 t \; = \; \left(p_2+p_3\right)^2,
 \;\;\;\;\;\;
 u \; = \; \left(p_1+p_3\right)^2.
\eq
Among the most complicated loop diagrams contributing to the electroweak NNLO corrections
are the planar and non-planar double-box diagrams, shown in fig.~\ref{fig:doubleboxes_topos}.
The wavy lines are either photons or heavy gauge bosons (i.e. either $Z$-bosons or $W$-bosons).
The case where three heavy gauge bosons are exchanged is expected to be numerically suppressed by 
$|t|/m^2$ \cite{Erler:2022ckm}, where $m$ denotes the mass of a heavy gauge boson.
In addition it is from a computational point of view significantly more involved.
For these reasons we focus here on the cases, where there is a least one photon exchanged.
In this case there will be at most two massive gauge bosons. 
If there are two massive gauge bosons, they will have the same mass $m$.

We are interested in the case where
\bq
\label{eq_kinematic_region}
 -t \;\; \lesssim \;\; s \;\; \ll \;\; m^2.
\eq


\subsection{The families of Feynman integrals}

\begin{figure}
\begin{center}
\includegraphics[scale=0.8]{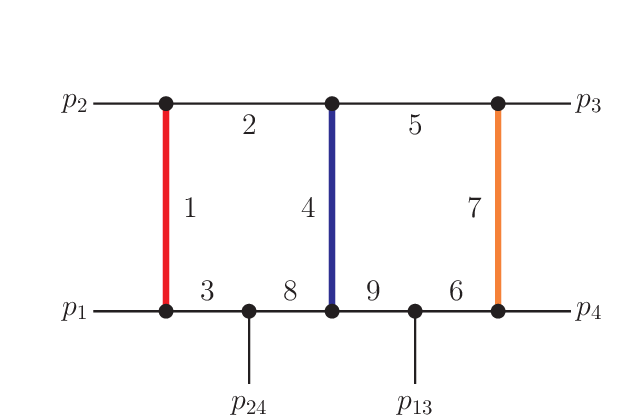}
\end{center}
\caption{
The auxiliary graph for the planar topologies. 
The coloured legs carry the masses $m^X_1$, $m^X_4$ and $m^X_7$, respectively, 
which can either be the mass of a heavy gauge boson or zero. All other lines are massless.
}
\label{fig:planar_doublebox_aux_topology}
\end{figure}

\begin{figure}
\begin{center}
\includegraphics[scale=0.8]{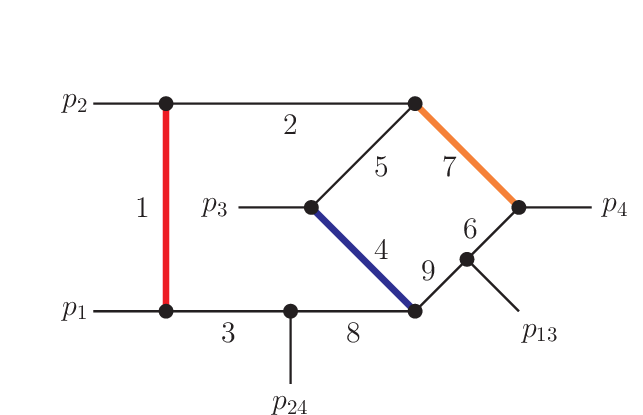}
\end{center}
\caption{
The auxiliary graph for the non-planar topologies. 
The coloured legs carry the masses $m^X_1$, $m^X_4$ and $m^X_7$, respectively, 
which can either be the mass of a heavy gauge boson or zero. All other lines are massless.
}
\label{fig:nonplanar_doublebox_aux_topology}
\end{figure}

We consider the integrals
\bq
\label{def_integral}
 I^X_{\nu_1 \nu_2 \nu_3 \nu_4 \nu_5 \nu_6 \nu_7 \nu_8 \nu_9}
 & = &
 e^{2 \gamma_E \eps}
 \left(\mu^2\right)^{\nu-D}
 \int \frac{d^Dk_1}{i \pi^{\frac{D}{2}}} \frac{d^Dk_2}{i \pi^{\frac{D}{2}}} 
 \prod\limits_{j=1}^9 \frac{1}{ \left(P^X_j\right)^{\nu_j} },
 \nonumber \\
 & & X \in \{A,B,C,D,E,\tilde{A},\tilde{B},\tilde{C},\tilde{D},\tilde{E}\},
\eq
where $D=4-2\eps$ denotes the number of space-time dimensions,
$\gamma_E$ denotes the Euler-Mascheroni constant, 
$\mu$ is an arbitrary scale introduced to render the Feynman integral dimensionless,
and the quantity $\nu$ is defined by
\bq
 \nu & = &
 \sum\limits_{j=1}^9 \nu_j.
\eq
We will further use the notation $p_{ij}=p_i+p_j$, $p_{ijk}=p_i+p_j+p_k$.
The labels $\{A,B,C,D,E\}$ denote planar topologies, the labels $\{\tilde{A},\tilde{B},\tilde{C},\tilde{D},\tilde{E}\}$ denote non-planar topologies.
The inverse propagators for the planar topologies are given by
\begin{align}
 P^X_1 & = -\left(k_1-p_1\right)^2 +\left(m^X_1\right)^2,
 &
 P^X_2 & = -\left(k_1-p_{12}\right)^2,
 &
 P^X_3 & = -k_1^2,
 \nonumber \\
 P^X_4 & = -\left(k_1+k_2\right)^2 + \left(m^X_4\right)^2,
 &
 P^X_5 & = -\left(k_2+p_{12}\right)^2,
 &
 P^X_6 & = -k_2^2,
 \nonumber \\
 P^X_7 & = -\left(k_2+p_{123}\right)^2 + \left(m^X_7\right)^2,
 & 
 P^X_8 & = -\left(k_1-p_{13}\right)^2,
 &
 P^X_9 & = -\left(k_2+p_{13}\right)^2.
\end{align}
The auxiliary graph for the planar topologies is shown in fig.~\ref{fig:planar_doublebox_aux_topology}.
The planar topologies $\{ A, B, C, D, \allowbreak E \}$ differ by the assignment of the masses $(m_1^X,m_4^X,m_7^X)$.
The cases which we consider in this paper are summarised in table~\ref{tab:planar_topologies}.
\begin{table}
\begin{center}
\begin{tabular}{|c|l|c|l|}\hline
  Topology & Masses & \# MIs & Square roots\\\hline
 A & $(m^A_1,m^A_4,m^A_7) = (0,m,m)$ & 35 & $r_1$, $r_2$, $r_3$, $r_4$\\
 B & $(m^B_1,m^B_4,m^B_7) = (m,0,m)$ & 26 & $r_1$, $r_4$ \\
 C & $(m^C_1,m^C_4,m^C_7) = (0,m,0)$ & 19 & $-$\\
 D & $(m^D_1,m^D_4,m^D_7) = (m,0,0)$ & 21 & $-$\\
 E & $(m^E_1,m^E_4,m^E_7) = (0,0,0)$ &  8 & $-$\\\hline
\end{tabular}
\caption{
The planar topologies and the corresponding mass configurations, number of master integrals and occurring square roots.}
\label{tab:planar_topologies}
\end{center}
\end{table}
The first graph polynomial \cite{Bogner:2010kv} for the auxiliary graph shown in fig.~(\ref{fig:planar_doublebox_aux_topology}) reads
\bq
\label{def_U_planar}
\lefteqn{
 {\mathcal U}^X\left(a_1,a_2,a_3,a_4,a_5,a_6,a_7,a_8,a_9\right)
 = } & & \nonumber \\ 
 & &
 \left(a_1 + a_2 + a_3 + a_8\right)\left(a_5 + a_6 + a_7 + a_9\right) + a_4 \left(a_1+a_2+a_3+a_5+a_6+a_7+a_8+a_9\right).
 \nonumber \\
 & & X \; \in \; \{A,B,C,D,E\}.
\eq
The inverse propagators for the non-planar topologies $\{\tilde{A},\tilde{B},\tilde{C},\tilde{D},\tilde{E}\}$ are given by
\begin{align}
 P^X_1 & = -\left(k_1-p_1\right)^2 +\left(m^X_1\right)^2,
 &
 P^X_2 & = -\left(k_1-p_{12}\right)^2,
 &
 P^X_3 & = -k_1^2,
 \nonumber \\
 P^X_4 & = -\left(k_1+k_2\right)^2 + \left(m^X_4\right)^2,
 &
 P^X_5 & = -\left(k_{12}+p_3\right)^2,
 &
 P^X_6 & = -k_2^2,
 \nonumber \\
 P^X_7 & = -\left(k_2+p_{123}\right)^2 + \left(m^X_7\right)^2,
 & 
 P^X_8 & = -\left(k_1-p_{13}\right)^2,
 &
 P^X_9 & = -\left(k_2+p_{13}\right)^2.
\end{align}
Note that the momentum assignment differs between the planar topologies and the non-planar topologies
only for $P^X_5$.
The auxiliary graph for the non-planar topologies is shown in fig.~\ref{fig:nonplanar_doublebox_aux_topology}.
The non-planar topologies $\{\tilde{A},\tilde{B},\tilde{C},\tilde{D},\tilde{E}\}$ 
differ again by the assignment of the masses $(m_1^X,m_4^X,m_7^X)$.
The cases which we consider in this paper are summarised in table~\ref{tab:nonplanar_topologies}.
\begin{table}
\begin{center}
\begin{tabular}{|c|l|c|l|l|}\hline
  Topology & Masses & \# MIs & Square roots & Elliptic curves \\\hline
 $\tilde{A}$ & $(m^{\tilde{A}}_1,m^{\tilde{A}}_4,m^{\tilde{A}}_7) = (0,m,m)$ & 48 & $r_1$, $r_2$, $r_3$, $r_6$, & $E^{(b)}, E^{(c)}$ \\
 $\tilde{B}$ & $(m^{\tilde{B}}_1,m^{\tilde{B}}_4,m^{\tilde{B}}_7) = (m,0,m)$ & 64 & $r_1$, $r_3$, $r_4$, $r_7$, $r_8$ & $E^{(a)}$ \\
 $\tilde{C}$ & $(m^{\tilde{C}}_1,m^{\tilde{C}}_4,m^{\tilde{C}}_7) = (0,m,0)$ & 41 & $r_5$ & $-$ \\
 $\tilde{D}$ & $(m^{\tilde{D}}_1,m^{\tilde{D}}_4,m^{\tilde{D}}_7) = (m,0,0)$ & 23 & $-$ & $-$ \\
 $\tilde{E}$ & $(m^{\tilde{E}}_1,m^{\tilde{E}}_4,m^{\tilde{E}}_7) = (0,0,0)$ & 12 & $-$ & $-$ \\\hline
\end{tabular}
\caption{
The non-planar topologies and the corresponding mass configurations, number of master integrals and occurring square roots.}
\label{tab:nonplanar_topologies}
\end{center}
\end{table}
The first graph polynomial for the auxiliary graph shown in fig.~(\ref{fig:nonplanar_doublebox_aux_topology}) reads
\bq
\label{def_U_nonplanar}
\lefteqn{
 {\mathcal U}^X\left(a_1,a_2,a_3,a_4,a_5,a_6,a_7,a_8,a_9\right)
 = } & & \nonumber \\ 
 & &
 \left(a_1 + a_2 + a_3 + a_6 + a_7 + a_8 + a_9\right) \left(a_4 + a_5\right)
 + \left(a_1 + a_2 + a_3\right) \left(a_6 + a_7 + a_9\right) 
 \nonumber \\
 & & 
 + \left(a_6 + a_7 + a_9\right) a_8,
 \;\;\;\;\;\;\;\;\;\;\;\;
 X \; \in \; \{\tilde{A},\tilde{B},\tilde{C},\tilde{D},\tilde{E}\}.
\eq
For each integral $I^X_{\nu_1 \nu_2 \nu_3 \nu_4 \nu_5 \nu_6 \nu_7 \nu_8 \nu_9}$ 
we define its sector id (or topology id) by
\bq
\label{def_sector_id}
 \mathrm{id}
 & = & \sum\limits_{j=1}^9 2^{j-1} \Theta\left(\nu_j\right).
\eq
Here, $\Theta(x)$ denotes the Heaviside step function, defined by $\Theta(x)=1$ for $x>0$ and $\Theta(x)=0$ otherwise.


\subsection{The method of calculation}

Integration-by-parts identities allow us to express any integral from a family of Feynman integrals 
as a linear combination of master integrals \cite{Tkachov:1981wb,Chetyrkin:1981qh}. 
We use the program {\tt Kira} \cite{Maierhoefer:2017hyi,Klappert:2020nbg}
for the integration-by-parts reduction.
The set of master integrals is finite \cite{Smirnov:2010hn}.
Therefore, we only need to compute the master integrals.
We use the method of differential equations \cite{Kotikov:1990kg,Kotikov:1991pm,Remiddi:1997ny,Gehrmann:1999as}
to compute the latter:
By differentiating under the integral sign and by using integration-by-parts identities
we obtain a system of differential equations for the master integrals $I(x,\eps)$
with respect to the kinematic variables $x$
\bq
\label{diff_eq_pre_canonical}
 d I\left(x,\eps\right) & = & \tilde{A}\left(x,\eps\right) \; I\left(x,\eps\right).
\eq
Here, the symbol $d$ is the differential with respect to all kinematic variables $x$,
the variable $\eps$ denotes the dimensional regularisation parameter, and $\tilde{A}$ is a square matrix with dimensions
equal to the number of master integrals. 
The entries of the matrix $\tilde{A}$ are differential one-forms, rational in $x$ and $\eps$. 
The essential step in computing Feynman integrals is to find a basis of master integrals
$J(x,\eps)$
\bq
\label{def_fibre_transformation}
 J\left(x,\eps\right) & = & U\left(x,\eps\right) I\left(x,\eps\right),
\eq 
such that
system of differential equations is in an $\eps$-factorised form \cite{Henn:2013pwa}
\bq
\label{eps_factorised_deq}
 d J\left(x,\eps\right) & = & \eps A\left(x\right) J\left(x,\eps\right).
\eq
The essential point is that the entries of $A$ are now differential one-forms,
depending on the kinematic variables, but independent of the dimensional regularisation parameter $\eps$. 
A differential equation in $\eps$-factorised form can be solved systematically order-by-order in $\eps$
in terms of iterated integrals \cite{Chen}.

We introduce an operator ${\bf i}^+$, 
which raises the power of the propagator $i$ by one and multiplies by $\nu_i$, e.g.
\bq
 {\bf 1}^+ I^X_{\nu_1 \nu_2 \nu_3 \nu_4 \nu_5 \nu_6 \nu_7 \nu_8 \nu_9} & = &
 \nu_1 \cdot I^X_{(\nu_1+1) \nu_2 \nu_3 \nu_4 \nu_5 \nu_6 \nu_7 \nu_8 \nu_9}.
\eq
The notation with an extra prefactor $\nu_j$ follows ref.~\cite{Weinzierl:2022eaz}.
In addition we define the operator ${\bf D}^-$,
which lowers the dimension of space-time by two units through
\bq
 {\bf D}^- I^X_{\nu_1 \nu_2 \nu_3 \nu_4 \nu_5 \nu_6 \nu_7 \nu_8 \nu_9}\left( D \right)
 & = &
 I^X_{\nu_1 \nu_2 \nu_3 \nu_4 \nu_5 \nu_6 \nu_7 \nu_8 \nu_9}\left( D-2 \right).
\eq
The dimensional shift relations read \cite{Tarasov:1996br,Tarasov:1997kx}
\bq
\label{dim_shift_eq}
 {\bf D}^- I^X_{\nu_1 \nu_2 \nu_3 \nu_4 \nu_5 \nu_6 \nu_7 \nu_8 \nu_9}\left(D\right)
 =
 {\mathcal U}^X\left( {\bf 1}^+, {\bf 2}^+, {\bf 3}^+, {\bf 4}^+, {\bf 5}^+ , {\bf 6}^+, {\bf 7}^+ , {\bf 8}^+ , {\bf 9}^+ \right)
 I^X_{\nu_1 \nu_2 \nu_3 \nu_4 \nu_5 \nu_6 \nu_7 \nu_8 \nu_9}\left(D\right),
\eq
where ${\mathcal U}^X$ denotes the first graph polynomial, defined in eq.~(\ref{def_U_planar}) and eq.~(\ref{def_U_nonplanar}) 
for the planar and non-planar topologies, respectively.


\subsection{Square roots}

In defining master integrals of uniform transcendental weight
we will encounter eight square roots. 
These are given by
\bq
\label{def_square_roots}
 r_1 & = & \sqrt{-t\left(4m^2-t\right)},
 \nonumber \\
 r_2 & = & \sqrt{s\left(4m^2+s\right)},
 \nonumber \\
 r_3 & = & \sqrt{-t\left[4m^2s^2-t\left(m^2-s\right)^2\right]},
 \nonumber \\
 r_4 & = &
 \sqrt{- s t \left[4m^2\left(m^2+s\right)-st\right]},
 \nonumber \\
 r_5 & = &
 \sqrt{- t m^2 \left[-4s\left(s+t\right)-tm^2\right]},
 \nonumber \\
 r_6 & = &
 \sqrt{- t \left(s+t\right) \left[4m^2\left(m^2-s-t\right)+t\left(s+t\right)\right]},
 \nonumber \\
 r_7 & = &
 \sqrt{- t \left[4m^2\left(s+t\right)^2-t\left(m^2+s+t\right)^2 \right]},
 \nonumber \\
 r_8 & = &
 \sqrt{s \left[-4 tm^4+s\left(s+t\right)^2\right]}.
\eq
We have chosen the arguments of the eight square roots such that in the region of interest 
($s>0$, $t<0$, $u<0$, $m^2 \gg s,(-t)$)
the arguments of all eight roots are positive.
In this region we chose the sign of the square roots such that all eight roots are positive.


\subsection{Elliptic curves}

The non-planar topologies $\tilde{A}$ and $\tilde{B}$ involve elliptic curves.
In more detail, there are two elliptic curves related to topology $\tilde{A}$ and one elliptic curve
related to topology $\tilde{B}$.
In this section we discuss these elliptic curves.
The defining equations for the elliptic curves are obtained from the maximal cut 
in the loop-by-loop Baikov representation \cite{Frellesvig:2017aai}.

In this section we denote by $(u,v)$ coordinates in a plane\footnote{The variable $u$ in this section is not related to the Mandelstam variable $u$ defined in eq.~(\ref{def_Mandelstam}).}.
We encounter elliptic curves defined by a quartic polynomial $P_4(u)$:
\bq
 E & : & v^2 - \frac{1}{s^2} P_4\left(u\right) \; = \; 0.
\eq
The factor of $1/s^2$ normalises the coefficient of $u^4$ in $P_4(u)$
and is chosen such that we have
\bq
 E & : & v^2 - \left(u-u_1\right)\left(u-u_2\right)\left(u-u_3\right)\left(u-u_4\right) \; = \; 0,
\eq
where $u_1, \dots, u_4$ are the roots of the polynomial $P_4(u)$.
We set
\begin{align}
U_1 = (u_3 - u_2)(u_4 - u_1), \quad U_2 = (u_2 - u_1)(u_4 - u_3), \quad U_3 = (u_3 - u_1) (u_4 - u_2),
\label{eq:elliptic_curves_Us}
\end{align}
and we define the modulus $k$ and the complementary modulus $\bar{k}$ of the elliptic curve by
\begin{align}
k^2 = \frac{U_1}{U_3}, \quad \bar{k}^2 = 1-k^2 = \frac{U_2}{U_3}.
\end{align}
We define the periods $\psi_1$ and $\psi_2$ of the elliptic curve as 
\begin{alignat}{3}
\label{def_periods}
\psi_1 &= \frac{4\mu^2K(k)}{U_3^\frac{1}{2}}, \quad &\psi_2 &= \frac{4i\mu^2 K(\bar{k})}{U_3^\frac{1}{2}}.
\end{alignat}
For topology $\tilde{B}$ we encounter in sector $123$ the elliptic curve $E^{(a)}$ defined by the quartic polynomial
\bq
 P_4^{(a)} 
 & = &
 u^2 \left[ s\left(u+m^2\right)+m^2 \left(2s+t\right)\right]^2
 + m^2 \left(s+t\right) 
  \left[ 
    2 s \left(2m^2-t\right)\left(u+3m^2\right) u
 \right. \nonumber \\
 & & \left.
    + 2 m^2 t \left(2m^2-2s-t\right) u
    + m^2 \left( \left(4m^4-4m^2t+t^2\right)\left(s+t\right)-8m^2st\right)
  \right].
\eq
For topology $\tilde{A}$ we encounter two elliptic curves $E^{(b)}$ and $E^{(c)}$: The elliptic curve $E^{(b)}$ occurs in sector $123$ and
is defined by the quartic polynomial
\bq
 P_4^{(b)} 
 =
 s^2 u^2 \left(u+m^2\right)^2 + t \left(s+t\right) m^2 \left[ m^2 \left(2u+2m^2-t\right)\left(2u+2m^2-s-t\right) - 2s\left(u+m^2\right)^2 \right].
\eq
The elliptic curve $E^{(c)}$ occurs in sectors $126$ and $127$ and 
is defined by the quartic polynomial
\bq
 P_4^{(c)} 
 & = & 
 s^2
 u
 \left(u-s\right)
 \left[ 4 m^2 \left(u+m^2-s\right)+u\left(u-s\right)\right].
\eq
The elliptic curve $E^{(c)}$ does not vary with $t$.

The roots of the polynomials $P_4^{(X)}$ (with $X \in \{a,b,c\}$) are in general algebraic functions of the kinematic variables.
For concreteness, we list how we label them. Together with eq.~(\ref{def_periods}) this defines the periods.
The elliptic curves $E^{(a)}$ and $E^{(b)}$ are defined by irreducible polynomials of degree $4$.
The roots of $P_4^{(a)}$ are 
\begin{align}
 u_1^{(a)}
 & =
 \frac{-2 m^2 s \left(3s+t\right) + \sqrt{D^{(a)}_1} - \sqrt{D^{(a)+}_{2}}}{4s^2},
 &
 u_2^{(a)}
 & = 
 \frac{-2 m^2 s \left(3s+t\right) - \sqrt{D^{(a)}_1} - \sqrt{D^{(a)-}_{2}}}{4s^2},
 \nonumber \\
 u_3^{(a)}
 & = 
 \frac{-2 m^2 s \left(3s+t\right) - \sqrt{D^{(a)}_1} + \sqrt{D^{(a)-}_{2}}}{4s^2},
 &
 u_4^{(a)}
 & = 
 \frac{-2 m^2 s \left(3s+t\right) + \sqrt{D^{(a)}_1} + \sqrt{D^{(a)+}_{2}} }{4s^2},
\end{align}
with
\bq
 D^{(a)}_1
 & = &
 \frac{1}{3} B^{(a)} -\frac{2}{3} s^2 R^{(a)} + \frac{8 m^4 s^2}{3 R^{(a)}} 
 \left[ 16\,{s}^{2}{t}^{2} \left( s+t \right) ^{2}+ \left( s-t \right) ^{4}{m}^{4}-8\,st \left( {s}^{2}-{t}^{2} \right)  \left( 2\,s+t \right) {m}^{2} \right],
 \nonumber \\
 D^{(a)\pm}_{2}
 & = &
 B^{(a)} - D^{(a)}_1 \pm \frac{64 m^4 s^5 t \left(s+t\right)}{D^{(a)}_1},
 \\
 R^{(a)}
 & = &
 \left( R^{(a)}_1 + 12 \sqrt{R^{(a)}_2} \right)^{\frac{1}{3}},
 \nonumber \\
 B^{(a)}
 & = &
 4\,{m}^{2}{s}^{2} \left( 4\,st \left( s+t \right) +{m}^{2} \left( s-t \right) ^{2} \right),
 \nonumber \\
 R^{(a)}_1
 & = &
 8\,{m}^{6} \left( 64\, \left( s+t \right) ^{3}{s}^{3}{t}^{3}+24\, \left( 5\,{s}^{2}+2\,st+2\,{t}^{2} \right)  \left( s+t \right) ^{2}{m}^{2}{s}^{2}{t}^{2}
 \right. \nonumber \\
 & & \left.
 -12\, \left( s+t \right)  \left( 2\,s+t \right)  \left( s-t \right) ^{3}{m}^{4}st+ \left( s-t \right) ^{6}{m}^{6} \right),
 \nonumber \\
 R^{(a)}_2
 & = &
 768\,{m}^{14}{s}^{5}{t}^{4} \left( s+t \right) ^{3} \left( 16\, \left( s+t \right) ^{2}{s}^{2}t- \left( s-t \right) ^{3}{m}^{4}- \left( s+t \right) 
 \left( {s}^{2}-20\,st-8\,{t}^{2} \right) {m}^{2}s \right).
 \nonumber
\eq
The roots of $P_4^{(b)}$ are 
\begin{align}
 u_1^{(b)}
 & = 
 \frac{-2 m^2 s^2 + \sqrt{D^{(b)}_1} - \sqrt{D^{(b)+}_{2}}}{4s^2},
 &
 u_2^{(b)}
 & = 
 \frac{-2 m^2 s^2 - \sqrt{D^{(b)}_1} - \sqrt{D^{(b)-}_{2}}}{4s^2},
 \nonumber \\
 u_3^{(b)}
 & = 
 \frac{-2 m^2 s^2 - \sqrt{D^{(b)}_1} + \sqrt{D^{(b)-}_{2}}}{4s^2},
 &
 u_4^{(b)}
 & =
 \frac{-2 m^2 s^2 + \sqrt{D^{(b)}_1} + \sqrt{D^{(b)+}_{2}}}{4s^2},
\end{align}
with
\bq
 D^{(b)}_1
 & = &
 \frac{1}{3} B^{(b)} -\frac{2}{3} s^2 R^{(b)} + \frac{8 m^4 s^2}{3 R^{(b)}} 
  \left[ 16\,{s}^{2}{t}^{2} \left( s+t \right) ^{2}+ \left( s+2\,t \right) ^{4}{m}^{4}
 \right. \nonumber \\
 & & \left.
 -8\,ts \left( s+2\,t \right)  \left( 2\,s+t \right)  \left( s+t \right) {m}^{2} \right],
 \nonumber \\
 D^{(b)\pm}_{2}
 & = &
 B^{(b)} - D^{(b)}_1 \pm \frac{64\,{m}^{4}{s}^{4}t \left( s+t \right)  \left( s+t-{m}^{2} \right)}{D^{(b)}_1},
 \\
 R^{(b)}
 & = &
 \left( R^{(b)}_1 + 12 \sqrt{R^{(b)}_2} \right)^{\frac{1}{3}},
 \nonumber \\
 B^{(b)}
 & = &
 4\,{m}^{2}{s}^{2} \left( 4\,st \left( s+t \right) + \left( {s}^{2}-8\,st-8\,{t}^{2} \right) {m}^{2} \right),
 \nonumber \\
 R^{(b)}_1
 & = &
 8\,{m}^{6} \left( 64\, \left( s+t \right) ^{3}{s}^{3}{t}^{3}
 -12\,ts \left( s+t \right)  \left( 2\,s+t \right)  \left( s+2\,t \right) ^{3}{m}^{4}
 + \left( s+2\,t \right) ^{6}{m}^{6}
 \right. \nonumber \\
 & & \left. 
 +24\,{t}^{2}{s}^{2} \left( 5\,{s}^{2}+8\,st+5\,{t}^{2} \right)  \left( s+t \right) ^{2}{m}^{2} \right),
 \nonumber \\
 R^{(b)}_2
 & = &
 768\,{m}^{14}{s}^{3}{t}^{4} \left( s+t \right) ^{5} \left( 16\, \left( s+t \right) ^{2}{s}^{2}t+ \left( s+2\,t \right) ^{3}{m}^{4}- \left( s+t \right) 
 \left( {s}^{2}+22\,st+13\,{t}^{2} \right) {m}^{2}s \right).
 \nonumber
\eq
The roots of the elliptic curve $E^{(c)}$ are simpler:
 \begin{align}
 u_1^{(c)}
 & = 
 \frac{1}{2} \left[ s -4 m^2 + \sqrt{s^2+8m^2s} \right],
 &
 u_2^{(c)}
 & =
 0,
 &
 u_3^{(c)}
 & =
 s,
 &
 u_4^{(c)}
 & = 
 \frac{1}{2} \left[ s -4 m^2 - \sqrt{s^2+8m^2s} \right].
\end{align}
In the differential equation for the Feynman integrals it is advisable to avoid algebraic extensions as far as possible.
In particular, we don't want to use the explicit expressions for the roots of the polynomial $P_4^{(X)}$.
This can be done as follows:
For each elliptic curve we may  express the derivatives of $\psi^{(X)}_i$ (for $X \in \{a,b,c\}$ and $i \in \{1,2\}$) with respect to the kinematic variables
as a linear combination of $\psi^{(X)}_i$
and $\frac{\partial}{\partial m^2} \psi^{(X)}_i$ with coefficients, which are rational functions.
The method is explained in section 5.1 of ref.~\cite{Adams:2018kez}.
The relevant formulae for the elliptic curve $E^{(a)}$ are
\bq
\label{derivatives_E_a}
 \frac{\partial}{\partial s} \psi^{(a)}_i
 & = &
 {\frac { \left( 9\,t{m}^{2}s+{m}^{2}{s}^{2}+4\,t{s}^{2}+2\,{m}^{2}{t}^{2}+4\,s{t}^{2} \right) }
 {s \left( s+t \right)  \left( 3\,{m}^{2}s-3\,t{m}^{2}+4\,st+4\,{s}^{2} \right) }}
 \psi^{(a)}_i
 \nonumber \\
 & &
 +{\frac {{m}^{2} \left( 4\,s{t}^{2}+4\,t{s}^{2}+{m}^{2}{s}^{2}+12\,t{m}^{2}s+5\,{m}^{2}{t}^{2} \right) }
  {s \left( s+t \right)  \left( 3\,{m}^{2}s-3\,t{m}^{2}+4\,st+4\,{s}^{2} \right) }}
 \frac{\partial}{\partial m^2} \psi^{(a)}_i,
 \nonumber \\
 \frac{\partial}{\partial t} \psi^{(a)}_i
 & = &
 -{\frac { \left( 4\,{m}^{2}{s}^{2}+9\,t{m}^{2}s-{m}^{2}{t}^{2}+4\,{s}^{3}+12\,t{s}^{2}+8\,s{t}^{2} \right)}
  {t \left( s+t \right)  \left( 3\,{m}^{2}s-3\,t{m}^{2}+4\,st+4\,{s}^{2} \right) }}
 \psi^{(a)}_i
 \nonumber \\
 & &
 -2\,{\frac {{m}^{2} \left( 2\,{m}^{2}{s}^{2}+6\,t{m}^{2}s+{m}^{2}{t}^{2}+2\,{s}^{3}+6\,t{s}^{2}+4\,s{t}^{2} \right)}
     {t \left( s+t \right)  \left( 3\,{m}^{2}s-3
\,t{m}^{2}+4\,st+4\,{s}^{2} \right) }}
 \frac{\partial}{\partial m^2} \psi^{(a)}_i,
 \nonumber \\
 \left( \frac{\partial}{\partial m^2} \right)^2 \psi^{(a)}_i
 & = &
 {\frac {N_1^{(a)}}{ \left( 3\,{m}^{2}s-3\,t{m}^{2}+4\,st+4\,{s}^{2} \right) {m}^{4} Q^{(a)}}}
 \psi^{(a)}_i
 \nonumber \\
 & &
 + {\frac {N_2^{(a)}}{{m}^{2} \left( 3\,{m}^{2}s-3\,t{m}^{2}+4\,st+4\,{s}^{2} \right) Q^{(a)}}}
 \frac{\partial}{\partial m^2} \psi^{(a)}_i,
\eq
with
\bq
 N_1^{(a)}
 & = &
 48\,{s}^{5}t{m}^{2}+28\,{m}^{4}{s}^{4}t+152\,{m}^{2}{t}^{2}{s}^{4}-28\,{m}^{4}{s}^{2}{t}^{3}
 +6\,{m}^{4}{t}^{2}{s}^{3}+144\,{m}^{2}{t}^{3}{s}^{3}+2\,{m}^{4}s{t}^{4}
 \nonumber \\
 & &
 +44\,{m}^{2}{s}^{2}{t}^{4}+12\,{m}^{6}s{t}^{3}-18\,{m}^{6}{t}^{2}{s}^{2}+12\,{m}^{6}t{s}^{3}
 -3\,{m}^{6}{s}^{4}-3\,{m}^{6}{t}^{4}-8\,{m}^{4}{s}^{5}-4\,{m}^{2}{s}^{6}
 \nonumber \\
 & &
 +16\,{s}^{6}t
 +48\,{s}^{5}{t}^{2}+48\,{s}^{4}{t}^{3}+16\,{s}^{3}{t}^{4},
 \nonumber \\
 N_2^{(a)}
 & = &
 36\,{m}^{6}t{s}^{3}-54\,{m}^{6}{t}^{2}{s}^{2}+48\,{m}^{2}{s}^{2}{t}^{4}+128\,{s}^{6}t
 +384\,{s}^{5}{t}^{2}+384\,{s}^{4}{t}^{3}+128\,{s}^{3}{t}^{4}+36\,{m}^{6}s{t}^{3}
 \nonumber \\
 & &
 +152\,{m}^{4}{s}^{4}t+54\,{m}^{4}{t}^{2}{s}^{3}-152\,{m}^{4}{s}^{2}{t}^{3}-32\,{m}^{4}s{t}^{4}
 +264\,{s}^{5}t{m}^{2}+612\,{m}^{2}{t}^{2}{s}^{4}
 \nonumber \\
 & &
 +384\,{m}^{2}{t}^{3}{s}^{3}-9\,{m}^{6}{s}^{4}
 -9\,{m}^{6}{t}^{4}-22\,{m}^{4}{s}^{5}-12\,{m}^{2}{s}^{6},
 \nonumber \\
 Q^{(a)}
 & = &
 {m}^{4}{s}^{3}-3\,{m}^{4}{s}^{2}t+3\,{m}^{4}s{t}^{2}-{m}^{4}{t}^{3}-16\,{s}^{2}{t}^{3}-32\,{s}^{3}{t}^{2}
 -16\,{s}^{4}t+{m}^{2}{s}^{4}-19\,{m}^{2}{s}^{3}t
 \nonumber \\
 & &
 -28\,{m}^{2}{t}^{2}{s}^{2}-8\,{m}^{2}s{t}^{3}.
\eq
For the elliptic curve $E^{(b)}$ we have
\bq
\label{derivatives_E_b}
 \frac{\partial}{\partial s} \psi^{(b)}_i
 & = &
 {\frac {{m}^{2} \left( s-4\,t \right)}{ \left( s+t \right)  \left( -4\,st-4\,{s}^{2}+3\,{m}^{2}s+6\,t{m}^{2} \right) }}
 \psi^{(b)}_i
 \nonumber \\
 & &
 +{\frac {{m}^{2} \left( 4\,s{t}^{2}+4\,t{s}^{2}+{m}^{2}{s}^{2}-10\,t{m}^{2}s-6\,{m}^{2}{t}^{2} \right) }
  {s \left( s+t \right)  \left( -4\,st-4\,{s}^{2}+3\,{m}^{2}s+6\,t{m}^{2} \right) }}
 \frac{\partial}{\partial m^2} \psi^{(b)}_i,
 \nonumber \\
 \frac{\partial}{\partial t} \psi^{(b)}_i
 & = &
 -{\frac { \left( 4\,{m}^{2}{s}^{2}+5\,t{m}^{2}s+6\,{m}^{2}{t}^{2}-4\,{s}^{3}-8\,t{s}^{2}-4\,s{t}^{2} \right) }
  {t \left( s+t \right)  \left( -4\,st-4\,{s}^{2}+3\,{m}^{2}s+6\,t{m}^{2} \right) }}
 \psi^{(b)}_i
 \nonumber \\
 & &
 -{\frac {{m}^{2}s \left( 4\,{m}^{2}s-t{m}^{2}-4\,st-4\,{s}^{2} \right)}
  {t \left( s+t \right)  \left( -4\,st-4\,{s}^{2}+3\,{m}^{2}s+6\,t{m}^{2} \right) }}
 \frac{\partial}{\partial m^2} \psi^{(b)}_i,
 \nonumber \\
 \left( \frac{\partial}{\partial m^2} \right)^2 \psi^{(b)}_i
 & = &
 {\frac {N_1^{(b)}}{{m}^{4} \left( -4\,st-4\,{s}^{2}+3\,{m}^{2}s+6\,t{m}^{2} \right) Q^{(b)}}}
 \psi^{(b)}_i
 \nonumber \\
 & &
 + {\frac {N_2^{(b)}}{{m}^{2} \left( -4\,st-4\,{s}^{2}+3\,{m}^{2}s+6\,t{m}^{2} \right) Q^{(b)}}}
 \frac{\partial}{\partial m^2} \psi^{(b)}_i,
\eq
with
\bq
 N_1^{(b)}
 & = &
 -3\,{m}^{6}{s}^{4}-4\,{m}^{2}{s}^{6}+8\,{m}^{4}{s}^{5}-72\,{m}^{6}{t}^{2}{s}^{2}-24\,{m}^{6}t{s}^{3}
 -16\,{m}^{2}{s}^{2}{t}^{4}+76\,{m}^{4}s{t}^{4}-96\,{m}^{6}s{t}^{3}
 \nonumber \\
 & &
 +48\,{s}^{5}{t}^{2}+48\,{s}^{4}{t}^{3}+16\,{s}^{3}{t}^{4}+16\,{s}^{6}t-48\,{m}^{6}{t}^{4}
 +68\,{m}^{4}{s}^{4}t-148\,{m}^{2}{t}^{2}{s}^{4}+186\,{m}^{4}{t}^{2}{s}^{3}
 \nonumber \\
 & &
 -72\,{s}^{5}t{m}^{2}
 +202\,{m}^{4}{s}^{2}{t}^{3}-96\,{m}^{2}{t}^{3}{s}^{3},
 \nonumber \\
 N_2^{(b)}
 & = &
 -216\,{m}^{6}{t}^{2}{s}^{2}-72\,{m}^{6}t{s}^{3}+128\,{s}^{3}{t}^{4}+384\,{s}^{4}{t}^{3}
 +384\,{s}^{5}{t}^{2}+128\,{s}^{6}t-252\,{m}^{2}{s}^{2}{t}^{4}
 \nonumber \\
 & &
 -816\,{m}^{2}{t}^{3}{s}^{3}-888\,{m}^{2}{t}^{2}{s}^{4}-336\,{s}^{5}t{m}^{2}+284\,{m}^{4}s{t}^{4}
 +818\,{m}^{4}{s}^{2}{t}^{3}+774\,{m}^{4}{t}^{2}{s}^{3}
 \nonumber \\
 & &
 +262\,{m}^{4}{s}^{4}t-288\,{m}^{6}s{t}^{3}
 -12\,{m}^{2}{s}^{6}+22\,{m}^{4}{s}^{5}-144\,{m}^{6}{t}^{4}-9\,{m}^{6}{s}^{4},
 \nonumber \\
 Q^{(b)}
 & = &
 16\,{s}^{2}{t}^{3}+32\,{s}^{3}{t}^{2}+16\,{s}^{4}t+{m}^{4}{s}^{3}+6\,{m}^{4}{s}^{2}t+12\,{m}^{4}s{t}^{2}
 +8\,{m}^{4}{t}^{3}-{m}^{2}{s}^{4}
 \nonumber \\
 & &
 -23\,{m}^{2}{s}^{3}t-35\,{m}^{2}{t}^{2}{s}^{2}-13\,{m}^{2}s{t}^{3}.
\eq
Finally, for the elliptic curve $E^{(c)}$ we have
\bq
\label{derivatives_E_c}
 \frac{\partial}{\partial s} \psi^{(c)}_i
 & = &
 -{\frac {1}{s}}
 \psi^{(c)}_i
-{\frac {{m}^{2}}{s}}
 \frac{\partial}{\partial m^2} \psi^{(c)}_i,
 \\
 \frac{\partial}{\partial t} \psi^{(c)}_i
 & = &
 0,
 \nonumber \\
 \left( \frac{\partial}{\partial m^2} \right)^2 \psi^{(c)}_i
 & = &
 -2\,{\frac { \left( -s+4\,{m}^{2} \right) }{{m}^{2} \left( 8\,{m}^{2}+s \right)  \left( {m}^{2}-s \right) }}
 \psi^{(c)}_i
 -{\frac { \left( -14\,{m}^{2}s+24\,{m}^{4}-{s}^{2} \right) }{{m}^{2} \left( 8\,{m}^{2}+s \right)  \left( {m}^{2}-s \right) }}
 \frac{\partial}{\partial m^2} \psi^{(c)}_i.
 \nonumber
\eq
The Wronskian $W^{(X)}_{m^2}$ (for $X \in \{a,b,c\}$) is defined by
\bq
 W^{(X)}_{m^2}
 & = &
 \mu^2
 \left(
 \psi^{(X)}_1 \frac{\partial}{\partial m^2} \psi^{(X)}_2
 -
 \psi^{(X)}_2 \frac{\partial}{\partial m^2} \psi^{(X)}_1
 \right).
\eq
The Wronskian is also a rational function in the kinematic variables.
Explicitly we have
\begin{align*}
W^{(a)}_{m^2} &= - \frac{2 \pi i \mu^6 s^2 \big[3m^2(s-t) + 4s(s+t)\big]}{m^4\Big[m^4(s-t)^3 - 16s^2 t (s+t)^2 + m^2 s (s+t) \big(s^2 - 20s t - 8t^2\big)\Big]}.
\end{align*}
\begin{align*}
W^{(b)}_{m^2} = \frac{2\pi i \mu^6 s^2 \big[4s(s+t) - 3m^2(s+2t)\big]}{m^4 \big[16s^2 t (s+t)^2 + m^4 (s+2t)^3 - m^2 s (s+t) (s^2+22st + 13t^2)\big]}.
\end{align*}
\begin{align*}
W^{(c)}_{m^2} &= -\frac{12 \pi i \mu^6}{m^2(m^2-s)(8m^2+s)}.
\end{align*}
All three elliptic curves degenerate for generic values of $s$ and $t$ in the limit $m^2\rightarrow \infty$ to a genus zero curve.
More concretely, the curve $E^{(a)}$ degenerates for $s-t \neq 0$ in the limit $m^2\rightarrow \infty$ to a genus zero curve,
the curve $E^{(b)}$ degenerates for $s+2t \neq 0$ in the limit $m^2\rightarrow \infty$ to a genus zero curve.
There are no restrictions for curve $E^{(c)}$.

The expansion of $\psi^{(X)}_1$ for $X \in \{a,b,c\}$ around $m^2 = \infty$ reads
\bq
 \psi^{(a)}_1
 & = &
 2 \pi i \mu^2 \left[ 
                - \frac{s}{\left(s-t\right)} \frac{1}{m^2}
                - 2 \frac{s^2t\left(s+t\right)\left(2s+t\right)}{\left(s-t\right)^4} \frac{1}{m^4}
                - 6 \frac{s^3 t^2\left(s+t\right)^2\left(6s^2+8st+t^2\right)}{\left(s-t\right)^7} \frac{1}{m^6}
 \right]
 \nonumber \\
 & &
 + {\mathcal O}\left(m^{-8}\right),
 \nonumber \\
 \psi^{(b)}_1
 & = &
 2 \pi i \mu^2 \left[
                  \frac{s}{\left(s+2t\right)} \frac{1}{m^2}
                + 2 \frac{s^2t\left(s+t\right)\left(2s+t\right)}{\left(s+2t\right)^4} \frac{1}{m^4}
                + 6 \frac{s^3 t^2\left(s+t\right)^2\left(6s^2+4st-t^2\right)}{\left(s+2t\right)^7} \frac{1}{m^6}
 \right]
 \nonumber \\
 & &
 + {\mathcal O}\left(m^{-8}\right),
 \nonumber \\
 \psi^{(c)}_1
 & = &
 2 \pi i \mu^2 \left[
                - \frac{1}{2 m^2}
                - \frac{s}{8 m^4}
                - \frac{5 s^2}{64 m^6}
 \right]
 + {\mathcal O}\left(m^{-8}\right).
\eq
For these expansions we have to choose the signs of several square roots. 
We adopted the convention, that all square roots are continuous 
under Feynman's $i\delta$-prescription.
For the kinematic region of interest the periods are continuous under the substitution
 $s \rightarrow s + i \delta$ with $\delta \ge 0$.


\section{Master integrals}
\label{sect:master_integrals}

The essential step in the computation of the Feynman integrals is to find a basis of master integrals $J(x,\eps)$ (see eq.~(\ref{def_fibre_transformation}))
\bq
\label{def_fibre_transformation_recall}
 J\left(x,\eps\right) & = & U\left(x,\eps\right) I\left(x,\eps\right),
\eq 
such that
system of differential equations is in an $\eps$-factorised form (see eq.~(\ref{eps_factorised_deq}))
\bq
\label{eps_factorised_deq_recall}
 d J\left(x,\eps\right) & = & \eps A\left(x\right) J\left(x,\eps\right).
\eq
Currently it is an open question, if such a basis exists for any family of Feynman integrals.
Given a pre-canonical basis $I(x,\eps)$ with differential equation as in eq.~(\ref{diff_eq_pre_canonical})
\bq
\label{diff_eq_pre_canonical_recall}
 d I\left(x,\eps\right) & = & \tilde{A}\left(x,\eps\right) \; I\left(x,\eps\right).
\eq
and a second basis $J(x,\eps)$ related to $I(x,\eps)$ as in eq.~(\ref{def_fibre_transformation_recall}), it is easy to check
if the differential equation for $J(x,\eps)$ is $\eps$-factorised: One simply computes
\bq
 U\left(x,\eps\right) \tilde{A}\left(x,\eps\right) U^{-1}\left(x,\eps\right) - U\left(x,\eps\right) d U^{-1}\left(x,\eps\right)
\eq
and checks if $\eps$ factors out.
Thus a heuristic method to guess $J(x,\eps)$ will work, as one can always check easily that a given guess will factor the variable $\eps$ out.
In practice, we make an educated guess based on the information available to us.
This includes analysing the maximal cut in the
loop-by-loop Baikov representation \cite{Frellesvig:2017aai}, 
as well as information from simpler or similar Feynman integrals, where an $\eps$-factorised form is known.
With this heuristic method we were able to construct a basis $J(x,\eps)$ for all topologies.
The list of master integrals is rather lengthy and given in appendix~\ref{sect:master_integrals_list}.
We note that this list of master integrals will be a valuable input to construct a basis of master integrals for more complicated Feynman integrals
(in particular the Feynman integrals with the exchange of three massive gauge bosons).

We remark that the basis of master integrals $J(x,\eps)$ is valid for any crossing we might want to consider.
Thus this basis can be used in particular for M{\o}ller scattering, Bhabha scattering, Drell-Yan and quark-pair production in electron-positron annihilation.
The elliptic master integrals depend on a choice of a period $\psi^{(X)}_1$ of the elliptic curve.
To obtain the $\eps$-factorised differential equation we only need to require that $\psi^{(X)}_1$ satisfies
the differential equations given in eq.~(\ref{derivatives_E_a}), eq.~(\ref{derivatives_E_b}) and eq.~(\ref{derivatives_E_c}) for $X\in\{a,b,c\}$, respectively.
As these linear differential equations are satisfied by $\psi^{(X)}_i$ with $i \in \{1,2\}$ we may replace
$\psi^{(X)}_1$ by
\bq
 \left( \psi^{(X)}_1\right)' & = & c \psi^{(X)}_2 + d \psi^{(X)}_1,
\eq
obtained from a modular transformation
\bq
 \left(\begin{array}{cc}
 a & b \\
 c & d \\
 \end{array} \right)
 & \in &
 \mathrm{SL}_2\left({\mathbb Z}\right).
\eq
This freedom can be used to ensure that the differential one-forms appearing in the differential equation have
at most a simple pole at a chosen boundary point.


\section{Integration of the differential equation}
\label{sect:integration}

In this section we provide details on the integration of the differential equation.
Starting from section~\ref{sect:boundary} we will specialise to the kinematic region for M{\o}ller scattering.
We discuss the alphabets for the various topologies, i.e. the set of differential one-forms appearing 
in the differential equation. 
We express the results for the topologies $B,C,D,E, \tilde{C}, \tilde{D}$ and $\tilde{E}$ 
in terms of multiple polylogarithms. 
This requires the rationalisation of square roots, where the methods of refs.~\cite{Besier:2018jen,Besier:2019kco} are used.
The more complicated topologies $A, \tilde{A}$ and $\tilde{B}$ are expressed in terms of iterated integrals.
We choose a common boundary point for all topologies.
Furthermore we discuss the path of integration for the various topologies.

\subsection{The alphabet}
\label{sect:alphabet}

We may write the matrix $A$ appearing in eq.~(\ref{eps_factorised_deq}) as
\bq
\label{def_matrices_M}
 A & = &
 \sum\limits_{j=1}^{\NL} M_j \omega_j,
\eq
where the $M_j$'s are $(\NF \times \NF)$-matrices with constant entries and the $\omega_j$'s 
differential one-forms.
We may divide the latter into dlog-forms with rational arguments, 
dlog-forms with algebraic (and non-rational) arguments and one-forms related to elliptic curves.
For the dlog-forms with rational arguments we write
\bq
 \omega_j & = & d \ln e_j.
\eq
The rational letters are
\begin{alignat*}{3}
e_1 &= \frac{-s}{\mu^2}, &\quad e_2 &= \frac{-t}{\mu^2}, &\quad e_3 &= \frac{-s-t}{\mu^2},\\
e_4 &= \frac{m^2}{\mu^2}, &\quad e_5 &=  \frac{m^2+s}{\mu^2}, &\quad e_6 &= \frac{m^2-t}{\mu^2},\\
e_7 &= \frac{m^2-s}{\mu^2}, &\quad e_8 &= \frac{m^2(s+t)-st}{\mu^4}, &\quad e_9 &= \frac{4m^2+s}{\mu^2}\\
e_{10} &= \frac{4m^2-t}{\mu^2}, &\quad e_{11} &= \frac{4m^2 s^2 - \left(m^2-s\right)^2 t}{\mu^6}, &\quad e_{12} &= \frac{4m^2\left(m^2+s\right)-st}{\mu^4},\\
e_{13} &= \frac{m^2+s+t}{\mu^2}, &\quad e_{14} &= \frac{m^2-s-t}{\mu^2}, &\quad e_{15} &= \frac{m^2 s-t(s+t)}{\mu^2},\\
e_{16} &= \frac{4s(s+t)+m^2 t}{\mu^2}, &\quad e_{17} &= \frac{m^2 t-s(s+t)}{\mu^2}, &\quad e_{18} &= \frac{4 m^2 (s+t)^2 -t(m^2+s+t)^2}{\mu^2},\\
& &\quad e_{19} &= \frac{4 m^4(-t)+s(s+t)^2}{\mu^2}, &\quad e_{20} &= \frac{4 m^2 (m^2-s-t)+t(s+t)}{\mu^2}. 
\end{alignat*}
For the dlog-forms with algebraic arguments we write
\bq
 \omega_{j} & = & d \ln o_j.
\eq
The algebraic letters involve the square roots $r_1$-$r_8$ and read
\begin{alignat*}{2}
o_{21} &= \frac{2m^2-t-r_1}{2m^2-t+r_1},
&
o_{22} &= \frac{2m^2+s-r_2}{2m^2+s+r_2},
\\
o_{23} &= \frac{2m^2(t-s)+st-r_1 r_2}{2m^2(t-s)+st+r_1 r_2},
&
o_{24} &= \frac{(m^2-s)t - r_3}{(m^2-s)t + r_3},
\\
o_{25} &= \frac{(m^2+s)t - r_3}{(m^2+s)t + r_3},
&
o_{26} &= \frac{st^2 - m^2t(4s+t) - r_1 r_3}{st^2 - m^2t(4s+t) + r_1 r_3},
\\
o_{27} &= \frac{s(2m^2-t)-r_4}{s(2m^2-t)+r_4},
&
o_{28} &= \frac{s\left(2m^2+2s+t\right)-r_4}{s\left(2m^2+2s+t\right)+r_4},
\\
o_{29} &= \frac{t\left[2m^2\left(m^2+2s\right)-st\right] - r_1 r_4}{t\left[2m^2\left(m^2+2s\right)-st\right] + r_1 r_4},
&
o_{30} &= \frac{st \left[s\left(4m^2-t\right)+m^2\left(2m^2+t\right)\right]-r_3 r_4}{st \left[s\left(4m^2-t\right)+m^2\left(2m^2+t\right)\right]+r_3 r_4},
\\
o_{31} &= \frac{m^2 t + 2st - r_5}{m^2 t + 2st + r_5},
&
o_{32} &= \frac{m^2 t - 2t(s+t) - r_5}{m^2 t - 2t(s+t) + r_5},
\\
o_{33} &= \frac{2m^2\left(m^2-s-t\right)+t\left(s+t\right) - i r_6}{2m^2\left(m^2-s-t\right)+t\left(s+t\right) + i r_6}, 
&
o_{34} &= \frac{\left(t-2m^2\right)\left(s+t\right) - i r_6}{\left(t-2m^2\right)\left(s+t\right) + i r_6},
\\
o_{35} &= \frac{t\left(4m^2-t\right)\left(s+t\right)-2m^4t - i r_1 r_6}{t\left(4m^2-t\right)\left(s+t\right)-2m^4t + i r_1 r_6}, \;\;
&
o_{36} &= \frac{-t\left(m^2+s+t\right) - r_7}{-t\left(m^2+s+t\right) + r_7},
\\
o_{37} &= \frac{-t\left(m^2-s-t\right) - r_7}{-t\left(m^2-s-t\right) + r_7},
&
o_{38} &= \frac{-t\left[-s\left(4m^2-t\right)-t\left(3m^2-t\right)\right] - r_1 r_7}{-t\left[-s\left(4m^2-t\right)-t\left(3m^2-t\right)\right] + r_1 r_7},
\\
o_{39} &= \frac{s\left(s+t\right) - r_8}{s\left(s+t\right) + r_8},
&
o_{40} &= \frac{-s\left(2m^2-s-t\right) - r_8}{-s\left(2m^2-s-t\right) + r_8},
\\
o_{41} &= \frac{-s\left(2m^2+s+t\right) - r_8}{-s\left(2m^2+s+t\right) + r_8}.
\end{alignat*}
We denote the set of differential one-forms related to the two elliptic curves of topology $\tilde{A}$
by $H^{\tilde{A}}$, the corresponding set of differential one-forms related to the elliptic curve of topology $\tilde{B}$
by $H^{\tilde{B}}$.
The explicit expressions for the one-forms related to elliptic curves are rather lengthy
and we list them in the supplementary electronic file
attached to the arxiv version of this article, see appendix~\ref{sect:supplement}.

Not every differential one-form occurs in every topology.
We call the set of differential one-forms occurring in the differential equation for a particular topology the alphabet
for this topology.
The alphabets for the various topologies are 
\bq
 \mathcal{A}^A
 & = & \left\{ \omega_1, \omega_2, \omega_3, \omega_4, \omega_5, \omega_6, \omega_7, \omega_9, 
               \omega_{10}, \omega_{11}, \omega_{12}, 
               \omega_{21}, \omega_{22}, \omega_{23}, \omega_{24}, \omega_{25}, \omega_{26}, \omega_{27}, \omega_{28}, 
 \right. \nonumber \\
 & & \left.
               \omega_{29}, \omega_{30} \right\},
 \nonumber \\
 \mathcal{A}^B
 & = & \left\{ \omega_1, \omega_2, \omega_3, \omega_4, \omega_5, \omega_6, \omega_{10}, \omega_{12}, \omega_{21}, \omega_{27}, \omega_{28}, \omega_{29} \right\},
 \nonumber \\
 \mathcal{A}^C
 & = & \left\{ \omega_1, \omega_2, \omega_3, \omega_4, \omega_5, \omega_6, \omega_7, \omega_8 \right\},
 \nonumber \\
 \mathcal{A}^D
 & = & \left\{ \omega_1, \omega_2, \omega_3, \omega_4, \omega_5, \omega_6 \right\},
 \nonumber \\
 \mathcal{A}^E 
 & = & \left\{ \omega_1, \omega_2, \omega_3 \right\},
 \nonumber \\
 \mathcal{A}^{\tilde{A}}
 & = & \left\{ \omega_1, \omega_2, \omega_3, \omega_4, \omega_5, \omega_6, \omega_7, \omega_9, 
               \omega_{10}, \omega_{11}, \omega_{14}, 
               \omega_{20}, \omega_{21}, \omega_{22}, \omega_{23}, \omega_{24}, \omega_{25}, \omega_{26}, 
 \right. \nonumber \\
 & & \left. 
               \omega_{33}, \omega_{34}, \omega_{35} 
       \right\} \cup H^{\tilde{A}},
 \nonumber \\
 \mathcal{A}^{\tilde{B}}
 & = & \left\{ \omega_1, \omega_2, \omega_3, \omega_4, \omega_5, \omega_6, \omega_7, 
               \omega_{10}, \omega_{11}, \omega_{12}, \omega_{13}, \omega_{14}, \omega_{17}, \omega_{18}, \omega_{19},
               \omega_{21}, \omega_{24}, \omega_{25}, \omega_{26}, 
 \right. \nonumber \\
 & & \left.
               \omega_{27}, \omega_{28}, \omega_{29},
               \omega_{30}, \omega_{36}, \omega_{37}, \omega_{38}, \omega_{39}, \omega_{40}, \omega_{41} \right\} \cup H^{\tilde{B}},
 \nonumber \\
 \mathcal{A}^{\tilde{C}}
 & = & \left\{ \omega_1, \omega_2, \omega_3, \omega_4, \omega_5, \omega_6, \omega_7, \omega_8, 
               \omega_{14}, \omega_{16},
               \omega_{31}, \omega_{32} \right\},
 \nonumber \\
 \mathcal{A}^{\tilde{D}}
 & = & \left\{ \omega_1, \omega_2, \omega_3, \omega_4, \omega_5, \omega_6, \omega_{13}, \omega_{15} \right\},
 \nonumber \\
 \mathcal{A}^{\tilde{E}}
 & = & \left\{ \omega_1, \omega_2, \omega_3 \right\}.
\eq

\subsection{Boundary values and integration path}
\label{sect:boundary}

We now specialise to the kinematic region relevant for M{\o}ller scattering.
We choose a boundary point and an integration path for this region which avoids analytic continuation.
Of course, with proper analytic continuation according to Feynman's $i\delta$-prescription, our results will be valid anywhere.
However, working out the required analytic continuation can be tedious and it is usually more efficient to repeat the procedure given below for
any other kinematic region one is interested in.

For the Moller experiment we are interested in the region (see eq.~(\ref{eq_kinematic_region}))
\bq
\label{eq_kinematic_region}
 -t \;\; \lesssim \;\; s \;\; \ll \;\; m^2.
\eq
We choose the boundary point and the integration path such that branch cut crossings are avoided
and such that the master integrals have convergent series expansions.
This will optimise the CPU time required to evaluate the master integrals
in this specific region.
As boundary point we choose
\bq
\label{def_boundary_point}
 t \; = \; 0,
 \;\;\;\;\;\;
 m^2 \; = \; \infty.
\eq
Without loss of generality we set $\mu^2=s$.
The Feynman integrals depend then only on dimensionless kinematic variables, which
we may take initially as
\bq
 x_t \; = \; \frac{-t}{s},
 & &
 x^{-1}_{m^2} \; = \; \frac{s}{m^2}.
\eq
Our chosen boundary point corresponds to
\bq
 \left( x_t, x^{-1}_{m^2} \right)
 & = &
 \left( 0, 0 \right).
\eq
At this boundary point the boundary constants are of uniform weight and spanned as a vector space in the lowest
weights by
\bq
\label{constants_boundary_values}
 \mbox{weight 0} &: & 1,
 \nonumber \\
 \mbox{weight 1} &: & i \pi,
 \nonumber \\
 \mbox{weight 2} &: & \zeta_2,
 \nonumber \\
 \mbox{weight 3} &: & \zeta_3, i \pi \zeta_2,
 \nonumber \\
 \mbox{weight 4} &: & \zeta_4, i \pi \zeta_3.
\eq
The non-zero boundary constants of the simpler master integrals are easily calculated.
The following two constraints provide an efficient way to determine the boundary constants
of the more complicated master integrals:
First of all, some of the master integrals must vanish on the hypersurface $m^2=\infty$ due to power counting
in the variable $m^2$.
Secondly, we also note that topologies $A$, $B$, $D$, $\tilde{B}$ and $\tilde{D}$ do not depend 
on the hypersurface $m^2=\infty$ on the Mandelstam variable $t$. 
This can also be used to determine some boundary constants.
In addition we used for some master integrals (of intermediate complexity)
the PSLQ algorithm \cite{Ferguson:1979,Ferguson:1992,Ferguson:1999,Bailey:1999nv}
to determine the boundary values from high-precision numerical results, assuming that the boundary values
are linear combinations of the constants in eq.~(\ref{constants_boundary_values}) 
with rational coefficients.
As there only very few constants in eq.~(\ref{constants_boundary_values}) numerical evaluations with about 50 digits are sufficient.
The methods mentioned above were sufficient to obtain all boundary values.
The explicit values of the boundary constants for all topologies are given in the supplementary electronic file
attached to the arxiv version of this article.

Topologies $E$ and $\tilde{E}$ depend only on $x_t$ and we simply integrate the differential equation
in the variable $x_t$.
The result is expressed in terms of harmonic polylogarithms \cite{Vermaseren:1998uu,Remiddi:1999ew}.

Topologies $C$, $D$ and $\tilde{D}$ depend on both variables 
$x_t$ and $x^{-1}_{m^2}$.
They do not involve any square roots nor any elliptic curves.
We first integrate in $x_t$ at $x^{-1}_{m^2}=0$, followed by an integration in $x^{-1}_{m^2}$
at $x_t=\mathrm{const}$.
The result is expressed in terms of multiple polylogarithms \cite{Goncharov_no_note,Borwein}.

Topology $\tilde{C}$ involves the square root $r_5$.
The square root $r_5$ is rationalised by
\bq
 m^2 = - \frac{s\left(s+t\right)}{t} \frac{\left(1+\hat{z}\right)^2}{\hat{z}},
 & &
 r_5 = s \left(s+t\right) \frac{\left(1-\hat{z}^2\right)}{\hat{z}}.
\eq
The inverse transformation is given by
\bq
 \hat{z} & = & 
 - \frac{2s\left(s+t\right)+m^2t+r_5}{2 s \left(s+t\right)}.
\eq
The boundary point $( x_t, x^{-1}_{m^2} ) = (0,0)$ corresponds to 
$( x_t, \hat{z} ) = (0,0)$. 
We first integrate in $x_t$ at $\hat{z}=0$, followed by an integration in $\hat{z}$ at $x_t=\mathrm{const}$.
The result is again expressed in terms of multiple polylogarithms.

Topology $B$ involves the square root $r_1$ and $r_4$.
These square roots occurred in ref.~\cite{Bottcher:2023wsr} and we may use the same rationalisation.
The square root $r_1$ is rationalised by
\bq
 t = - \frac{\tilde{y}^2}{1-\tilde{y}} m^2,
 & &
 r_1 = \frac{\tilde{y}\left(2-\tilde{y}\right)}{1-\tilde{y}} m^2.
\eq
The inverse transformation is given by
\bq
 \tilde{y}
 & = &
 \frac{t+r_1}{2 m^2}.
\eq
The square root $r_4$ is rationalised by
\bq
 m^2 =
 \frac{\left(1-2\tilde{z}\right)}{4\tilde{z}^2}
 \frac{\left(2-\tilde{y}\right)^2}{\left(1-\tilde{y}\right)} s,
 & &
 r_4 = 
 \frac{\left(1-\tilde{z}\right)\left(1-2\tilde{z}\right)}{4\tilde{z}^3}
 \frac{\tilde{y}\left(2-\tilde{y}\right)^3}{\left(1-\tilde{y}\right)^2} s^2.
\eq
The inverse transformation is given by
\bq
 \tilde{z}
 & = &
 -\frac{\left(4m^2-t\right)}{4 m^2} \left( \frac{s}{m^2} - \frac{r_4}{m^2 r_1} \right).
\eq
The boundary point $( x_t, x^{-1}_{m^2} ) = (0,0)$ corresponds to 
$( \tilde{z}, \tilde{y} ) = (0,0)$. 
We first integrate in $\tilde{z}$ at $\tilde{y}=0$, followed by an integration in $\tilde{y}$ at $\tilde{z}=\mathrm{const}$.
The result is again expressed in terms of multiple polylogarithms.

Topology $A$ involves the four square roots $r_1$, $r_2$, $r_3$ and $r_4$.
An efficient way to treat this case is to introduce the variables
\bq
 w_t \; = \; \sqrt{x_t},
 & &
 w^{-1}_{m^2} \; = \; \sqrt{x^{-1}_{m^2}}.
\eq
The boundary point $( x_t, x^{-1}_{m^2} ) = (0,0)$ corresponds to 
$( w_t, w^{-1}_{m^2} ) = (0,0)$. 
We first integrate in $w_t$ at $w^{-1}_{m^2}=0$, 
the result of this integration is expressed in terms of multiple polylogarithms.
We then integrate in $w^{-1}_{m^2}$ at $w_t=\mathrm{const}$.
In the result of this integration 
we first isolate all logarithms $\ln(w^{-1}_{m^2})$, the remainder is a regular function at $w^{-1}_{m^2}=0$,
which we evaluate as a Taylor series.

Topologies $\tilde{A}$ and $\tilde{B}$ involve elliptic curves.
From a numerical point of view it is again most efficient to evaluate them in a similar way as topology $A$.
In particular we use the same integration path as for topology $A$.
At $m^2=\infty$ (corresponding to $x^{-1}_{m^2}=w^{-1}_{m^2}=0$) the master integrals of topology $\tilde{B}$
are independent of $x_t$, leaving just the integration in the variable $w^{-1}_{m^2}$.
For topology $\tilde{A}$ we first integrate in $x_t$ at $w^{-1}_{m^2}=0$, followed by an integration in the variable $w^{-1}_{m^2}$
at fixed $x_t$.
The result of the integration in $x_t$ is expressed in terms of harmonic polylogarithms \cite{Remiddi:1999ew},
for the integration in $w^{-1}_{m^2}$ we expand all differential one-forms in the variable $w^{-1}_{m^2}$.
Whereas in the case of topology $A$, the expansion of the differential one-forms can be done on the fly to
any desired order, this approach is computationally more expensive for topologies $\tilde{A}$ and $\tilde{B}$
due to the differential one-forms depending on the elliptic curves.
It is more efficient to pre-compute the expansions once and for all to a fixed order in the variable $w^{-1}_{m^2}$.
In the numerical program we use an expansion 
up to order
$(w^{-1}_{m^2})^{25}$.


\section{Numerical results}
\label{sect:numerical_results}

In this section we give numerical results for a few selected Feynman integrals at a specific kinematic point.
As kinematic point we choose
\begin{align}
\label{eq:num_vals_doublebox}
s = 0.0112 \, \text{GeV}^2, \quad t = -2 \cdot 10^{-3} \, \text{GeV}^2, \quad m^2 = m_Z^2 = 8.32 \cdot 10^3 \, \text{GeV}^2.
\end{align}
The Moller experiment has a beam energy of $E_\text{beam} = 11 \, \text{GeV}$ \cite{Benesch:2014bas},
the corresponding value of the Mandelstam variable $s$ is obtained with the help of
\begin{align*}
 s = 2 m_e E_\text{beam},
\end{align*}
where $m_e$ is the mass of the electron. 
For the value of $t$ (or correspondingly the scattering angle) we choose a generic value.
The boundary point and the integration path of section~\ref{sect:boundary} are adapted to this region
and chosen such that we have convergent series expansions for the iterated integrals.
In the supplementary electronic file attached to the arxiv version of this article
we provide {\tt C++}-programs which provide numerical evaluation routines
for all master integrals of a given topology.
The algorithms for the numerical evaluation of multiple polylogarithms are based on \cite{Vollinga:2004sn},
topologies $A$, $\tilde{A}$ and $\tilde{B}$  use in addition 
the class \verb|user_defined_kernel| from ref.~\cite{Walden:2020odh} for the integration in the variable $w^{-1}_{m^2}$.
For a few selected integrals
the numerical results for the first five terms of the $\eps$-expansions 
evaluated at the point in eq.~\eqref{eq:num_vals_doublebox} are listed in table~\ref{table_numerical_results_selected}.
The numerical results for the first five terms of the $\eps$-expansions of all master integrals 
at the point in eq.~\eqref{eq:num_vals_doublebox}
are given in an electronic file attached to the arxiv version of this article.
In addition, we compared our results to the results of the program \verb|AMFlow| \cite{Liu:2022chg,Liu:2017jxz,Liu:2022mfb} and found perfect agreement.
Our numerical evaluation routines are significantly faster than \verb|AMFlow| 
and stable in the limit $m^2\rightarrow \infty$.
{\small
\begin{longtable}{|l|lllll|}
 \hline 
 & $\eps^0$ & $\eps^1$ & $\eps^2$ & $\eps^3$ & $\eps^4$ \\
 \hline 
$J^{A}_{ 32}$ & $        0$ & $        0$ & $ 1.5379816e-05$ & $-8.9633670e-05$ & $ 0.00018204374$ \\ 
 &  &  & $+  3.5742116e-06 i$ & $+  5.5465536e-05 i$ & $-0.00026141618 i$ \\ 
$J^{B}_{ 24}$ & $        0$ & $        0$ & $        0$ & $        0$ & $ 8.2058873e-06$ \\ 
 &  &  &  &  & $+  1.4933021e-10 i$ \\ 
$J^{C}_{ 17}$ & $        1$ & $-9.6153835e-07$ & $-21.384104$ & $-5.6098211$ & $ 91.727546$ \\ 
 &  & $+  6.2831853 i$ & $+  2.4166048e-06 i$ & $-51.676989 i$ & $-35.246805 i$ \\ 
$J^{D}_{ 19}$ & $     0.25$ & $-2.4038459e-07$ & $-4.5235499$ & $ 4.6078066$ & $ 15.084724$ \\ 
 &  & $+  1.5707963 i$ & $+  2.7186833e-06 i$ & $-7.7514954 i$ & $+  28.951873 i$ \\ 
$J^{E}_{ 7}$ & $        4$ & $ 8.613833$ & $-18.738162$ & $-119.78077$ & $-338.51852$ \\ 
 &  & $+  9.424778 i$ & $+  32.473385 i$ & $+  64.835609 i$ & $+  84.701754 i$ \\ 
$J^{\tilde{A}}_{ 45}$ & $        0$ & $        0$ & $        0$ & $        0$ & $ 0.00010223599$ \\ 
 &  &  &  &  & $-0.00022971381 i$ \\ 
$J^{\tilde{B}}_{ 61}$ & $        0$ & $        0$ & $        0$ & $-1.8714536e-06$ & $-6.9818556e-06$ \\ 
 &  &  &  & $+  2.9941309e-13 i$ & $-1.1758654e-05 i$ \\ 
$J^{\tilde{C}}_{ 38}$ & $        0$ & $        0$ & $ 0.82246838$ & $ 0.60101121$ & $-15.152188$ \\ 
 &  &  &  & $+  5.167717 i$ & $+  3.7764321 i$ \\ 
$J^{\tilde{D}}_{ 21}$ & $     -1.5$ & $ 4.3269202e-07$ & $ 44.413222$ & $ 49.885353$ & $-317.79722$ \\ 
 &  & $-9.424778 i$ & $+  2.7186841e-06 i$ & $+  155.03139 i$ & $+  313.43895 i$ \\ 
$J^{\tilde{E}}_{ 11}$ & $        2$ & $-0.17289727$ & $-17.155723$ & $ 5.7871067$ & $ 242.98467$ \\ 
 &  & $+  3.1415927 i$ & $-4.7942473 i$ & $-51.677128 i$ & $+  65.13519 i$ \\ 
 \hline 
\caption{
Numerical results for a few selected master integrals for the first five terms of the $\eps$-expansion 
for the kinematic point specified by eq.~(\ref{eq:num_vals_doublebox}).
}
\label{table_numerical_results_selected}
\end{longtable}
}


\section{Large logarithms}
\label{sect:logarithms}

Of particular interest are potentially numerically large contributions from the master integrals. 
These arise from large logarithms.
In the kinematic region of interest 
\bq
 -t \;\; \lesssim \;\; s \;\; \ll \;\; m^2
\eq
these are logarithms of the form
\bq
 L & = & \ln\left(\frac{s}{m^2}\right).
\eq
At order $\eps^j$ we can have at most $j$ powers of large logarithms.
The leading logarithms are the ones which occur to power $j$ at order $\eps^j$.
We remark that this counting defines at order $\eps^0$ constants as leading logarithms.
The leading logarithms are easily obtained from our full result.
To this aim we write the matrix $A$
in eq.~(\ref{def_matrices_M}) in the form
\bq
 A & = &
 \sum\limits_{j=1}^{\NL} \tilde{M}_j \tilde{\omega}_j,
\eq
such that
\bq
 \tilde{\omega}_1 & = & d\ln L
\eq
and all other $\tilde{\omega}_j$'s are regular at $m^2=\infty$.
$\tilde{M}_1$ (and all others $\tilde{M}_j$'s) 
are $(\NF \times \NF)$-matrices with constant entries.
Let us denote the boundary values of the master integrals at the boundary point 
of eq.~(\ref{def_boundary_point}) by
$J_{\mathrm{boundary}}$ and the $\eps^0$-part by $J_{\mathrm{boundary}}^{(0)}$.
The leading logarithms are then given by
\bq
 J_{\mathrm{LL}} 
 & = &
 \sum\limits_{j=0}^\infty \frac{1}{j!} \left( \eps L \right)^j \tilde{M}_1^j J_{\mathrm{boundary}}^{(0)}.
\eq
We see that the leading logarithms are determined by the matrix $\tilde{M}_1$ and the $\eps^0$-term of the boundary values.
In the supplementary electronic file
attached to the arxiv version of this article we give for all topologies the matrix $\tilde{M}_1$ and the boundary values.


\section{Conclusions}
\label{sect:conclusions}

In this paper we computed planar and non-planar two-loop double-box integrals 
where three electroweak gauge bosons are exchanged between the fermion lines,
among which at least one is a photon.
These integrals are relevant for the NNLO electroweak corrections to M{\o}ller scattering.
The planar and non-planar double-box integrals are distinguished by their mass configurations.
We considered the cases with zero, one or two internal massive gauge bosons.
The complexity of the calculation increases with the number of internal massive gauge bosons.
While the integrals with three internal photons have been computed long time ago, the non-planar double-box integrals
with two internal massive gauge bosons involve elliptic curves and require state-of-the-art techniques.
We presented for all topologies a basis of master integrals, such that the differential equation is in an $\eps$-factorised form.
We expressed the results for the simpler topologies in terms of multiple polylogarithms, the results for the more complicated topologies in terms of iterated
integrals.
For all integrals we provided numerical evaluation routines.
Of particular interest are potentially numerically large contributions, arising from large logarithms.
We extracted for all master integrals the leading logarithms.

The double-box integrals with the exchange of three massive gauge bosons are expected to give numerically suppressed contributions.
But they are of interest from a theoretical perspective, as the non-planar double-box integral
with three internal massive gauge bosons is related to a curve of genus two \cite{Marzucca:2023gto}.
The computation of these integrals is an interesting project for the future.

\subsection*{Acknowledgments}

This work has been supported by the Cluster of Excellence Precision Physics, Fundamental Interactions, and Structure of
Matter (PRISMA EXC 2118/1) funded by the German Research Foundation (DFG) within
the German Excellence Strategy (Project ID 390831469).


\begin{appendix}

\newpage
\section{Feynman diagrams}
\label{sect:diagrams}

In this appendix we show the diagrams for all master sectors. 
Red lines correspond to particles with mass $m$ and the uncoloured lines indicate massless particles.

\subsection{Planar double-box integrals}
\begin{figure}[H]
\begin{center}
\includegraphics[width=0.35\textwidth]{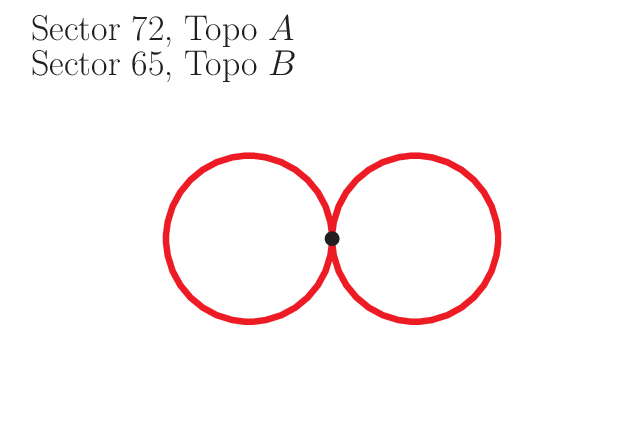}
\includegraphics[width=0.35\textwidth]{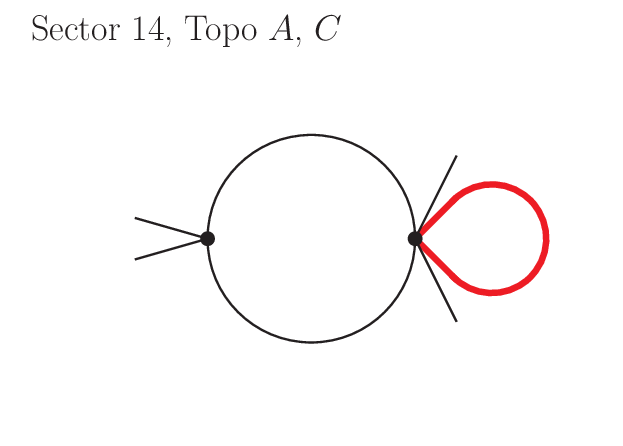}
\includegraphics[width=0.35\textwidth]{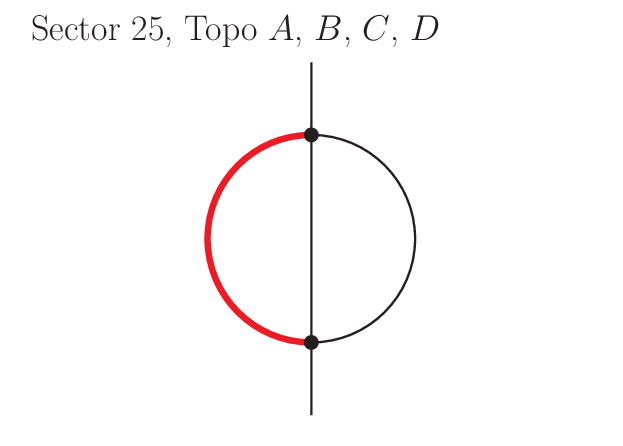}
\includegraphics[width=0.35\textwidth]{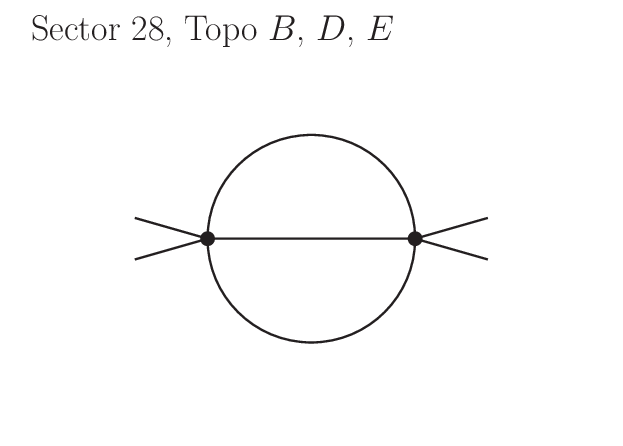}
\includegraphics[width=0.35\textwidth]{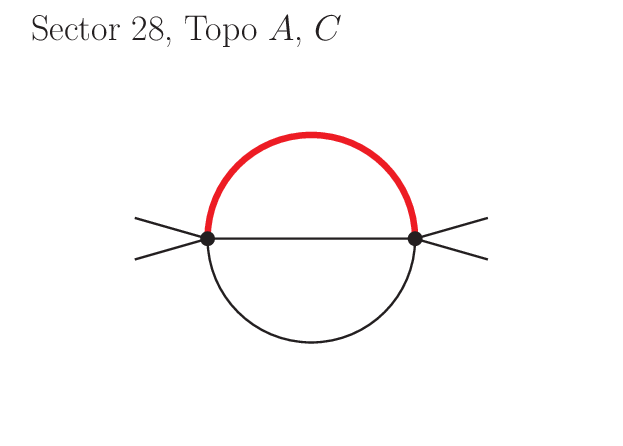}
\includegraphics[width=0.35\textwidth]{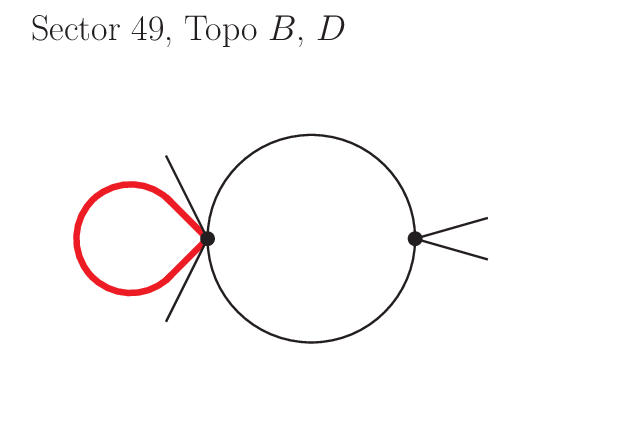}
\includegraphics[width=0.35\textwidth]{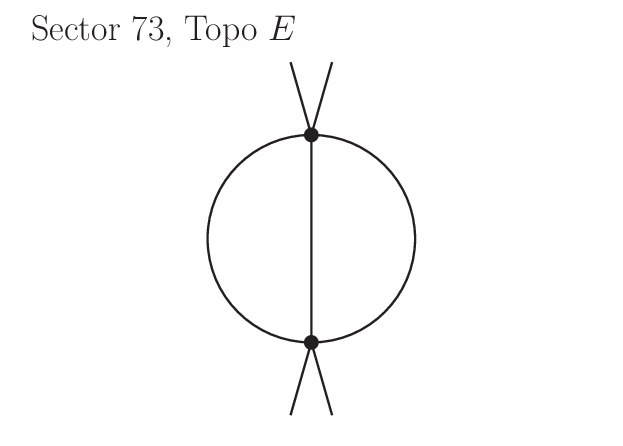}
\includegraphics[width=0.35\textwidth]{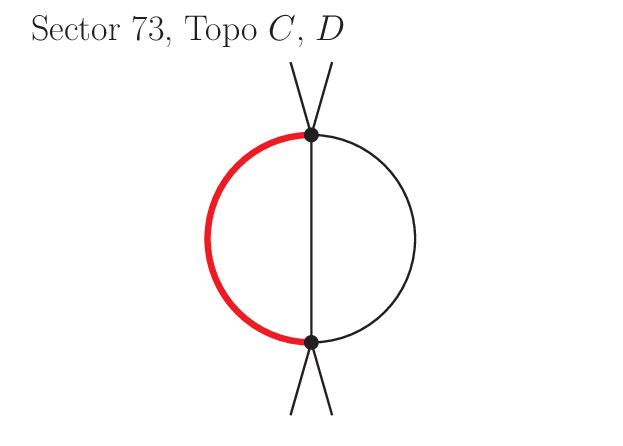}
\end{center}
\caption{\small
Master sectors for planar double-box integrals (part 1).
}
\end{figure}

\begin{figure}[H]
\begin{center}
\includegraphics[width=0.35\textwidth]{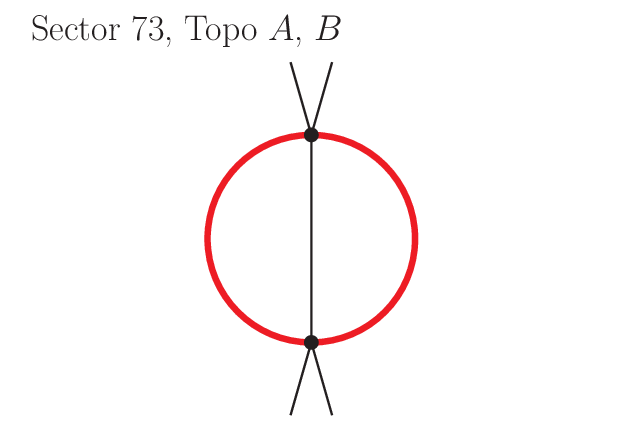}
\includegraphics[width=0.35\textwidth]{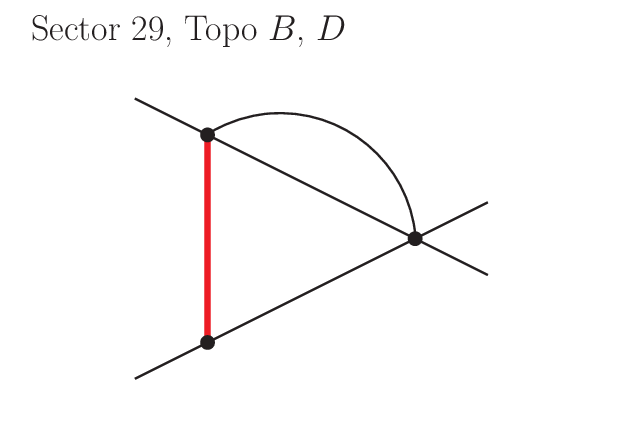}
\includegraphics[width=0.35\textwidth]{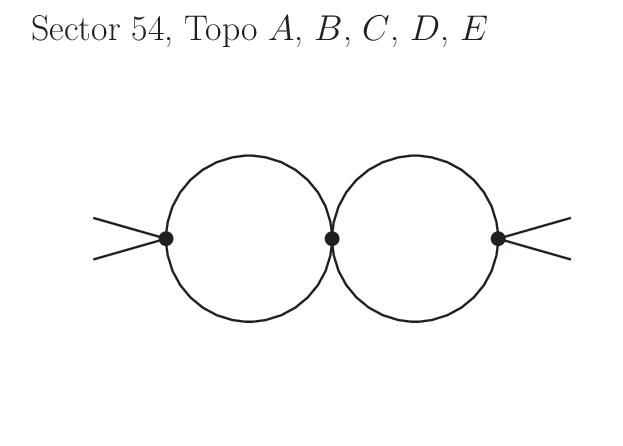}
\includegraphics[width=0.35\textwidth]{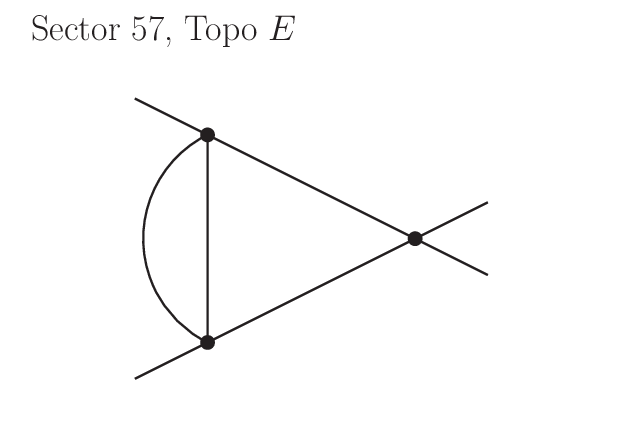}
\includegraphics[width=0.35\textwidth]{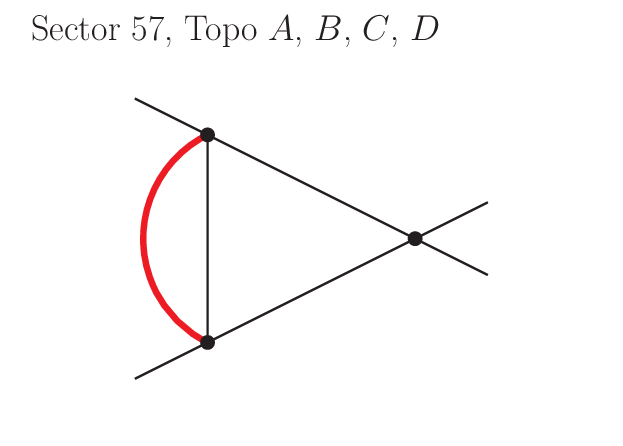}
\includegraphics[width=0.35\textwidth]{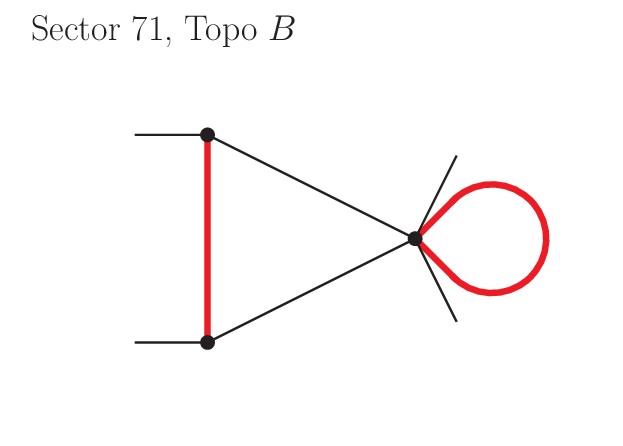}
\includegraphics[width=0.35\textwidth]{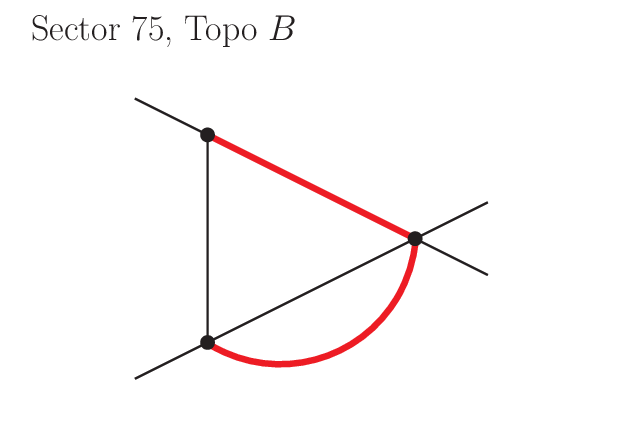}
\includegraphics[width=0.35\textwidth]{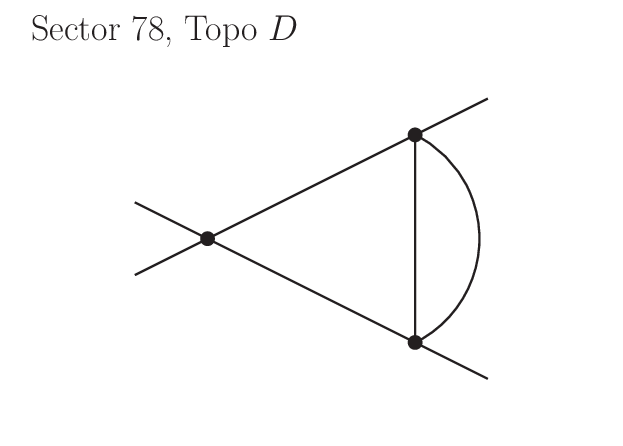}
\includegraphics[width=0.35\textwidth]{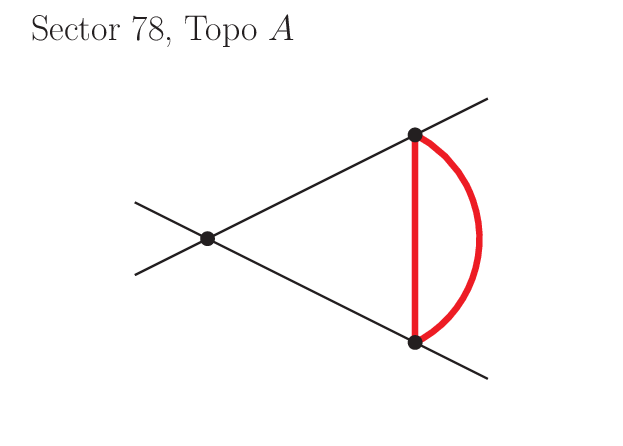}
\includegraphics[width=0.35\textwidth]{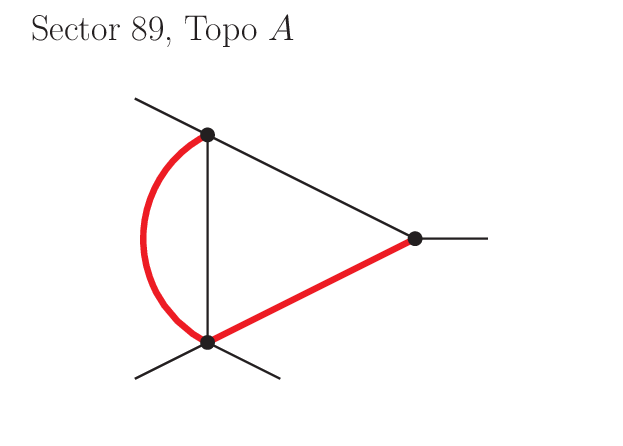}
\end{center}
\caption{\small
Master sectors for planar double-box integrals (part 2).
}
\end{figure}

\begin{figure}[H]
\begin{center}
\includegraphics[width=0.35\textwidth]{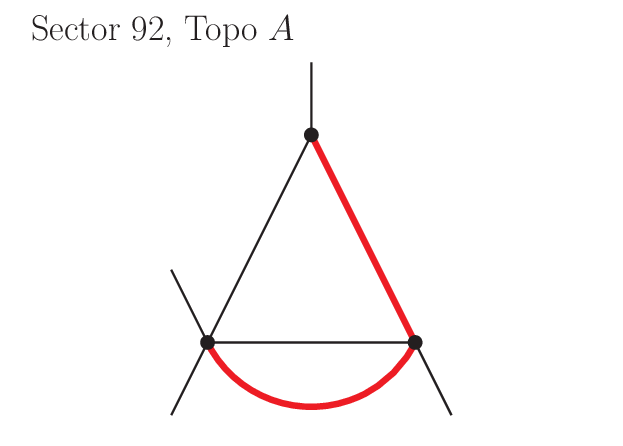}
\includegraphics[width=0.35\textwidth]{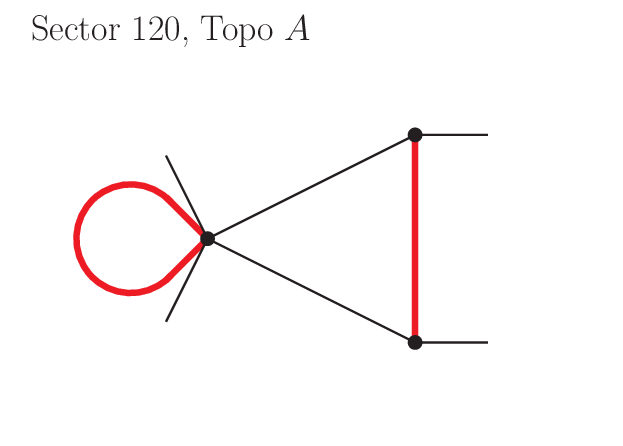}
\includegraphics[width=0.35\textwidth]{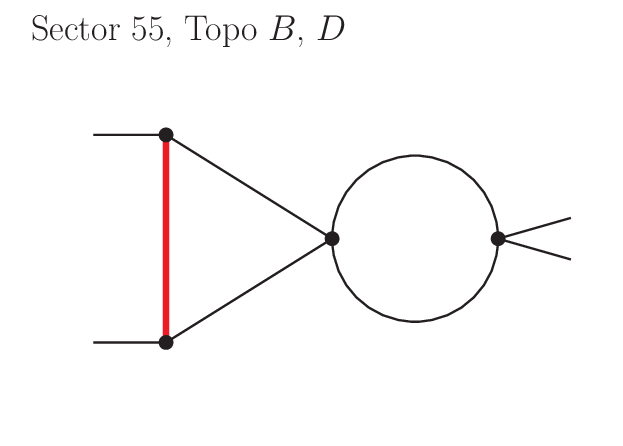}
\includegraphics[width=0.35\textwidth]{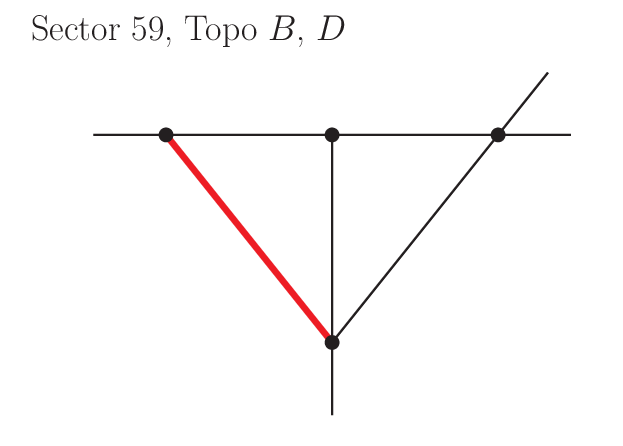}
\includegraphics[width=0.35\textwidth]{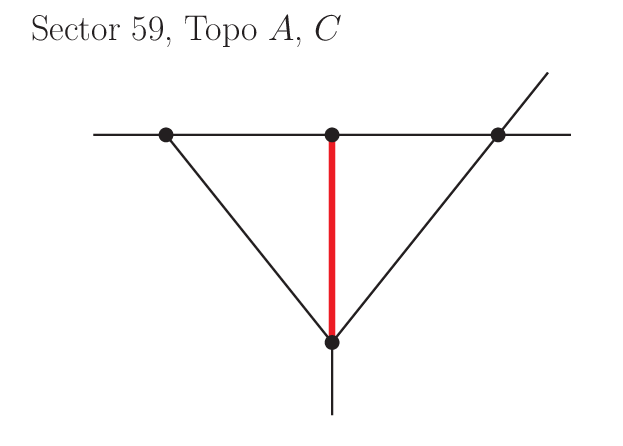}
\includegraphics[width=0.35\textwidth]{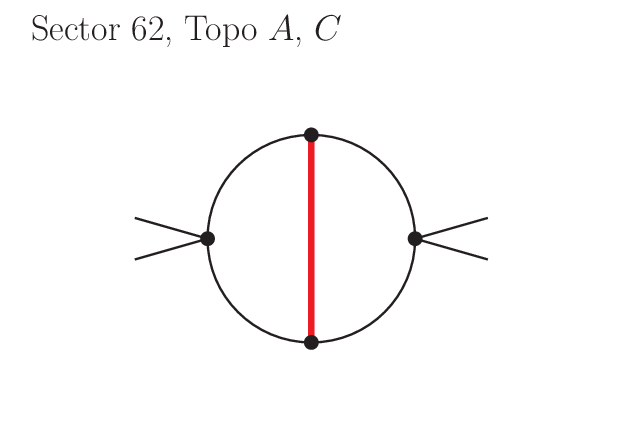}
\includegraphics[width=0.35\textwidth]{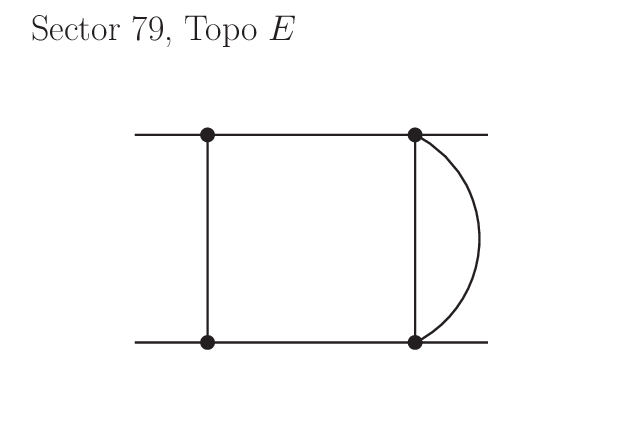}
\includegraphics[width=0.35\textwidth]{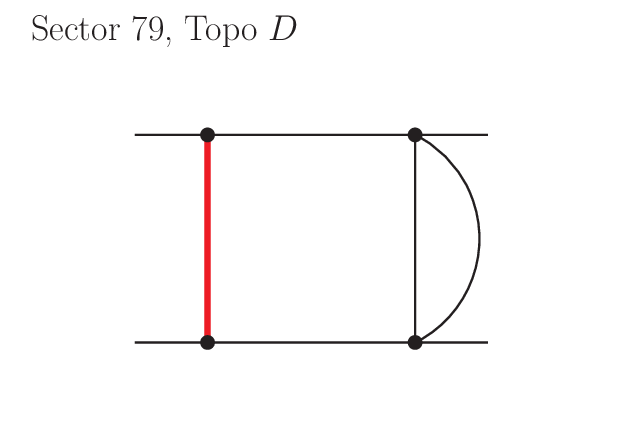}
\includegraphics[width=0.35\textwidth]{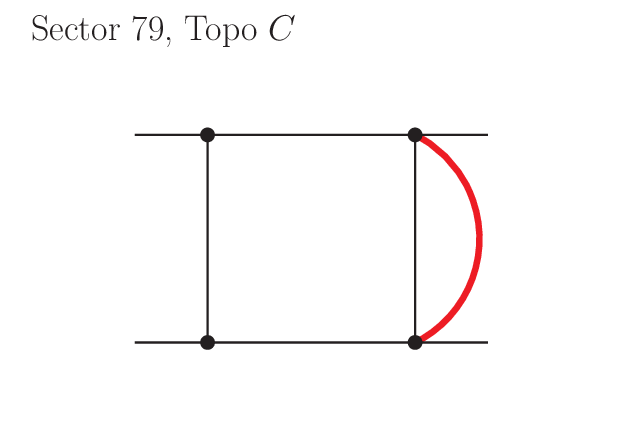}
\includegraphics[width=0.35\textwidth]{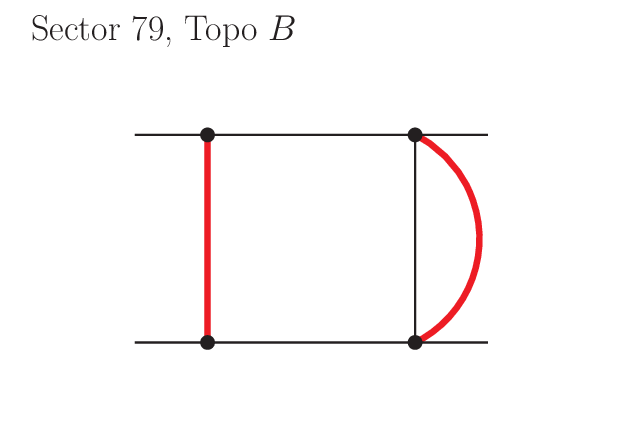}
\end{center}
\caption{\small
Master sectors for planar double-box integrals (part 3).
}
\end{figure}

\begin{figure}[H]
\begin{center}
\includegraphics[width=0.35\textwidth]{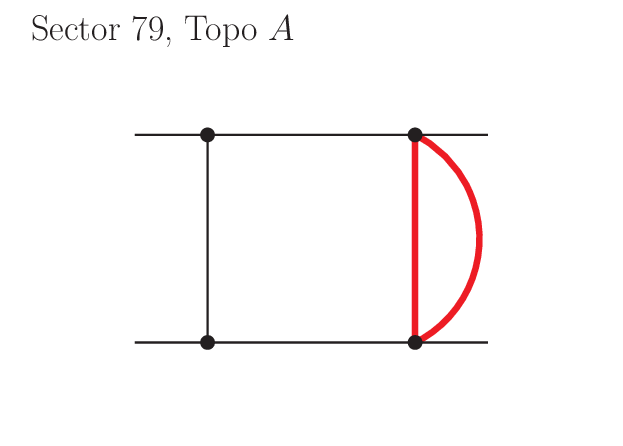}
\includegraphics[width=0.35\textwidth]{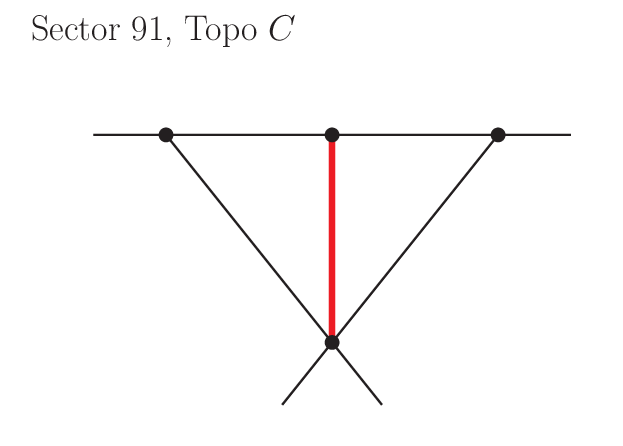}
\includegraphics[width=0.35\textwidth]{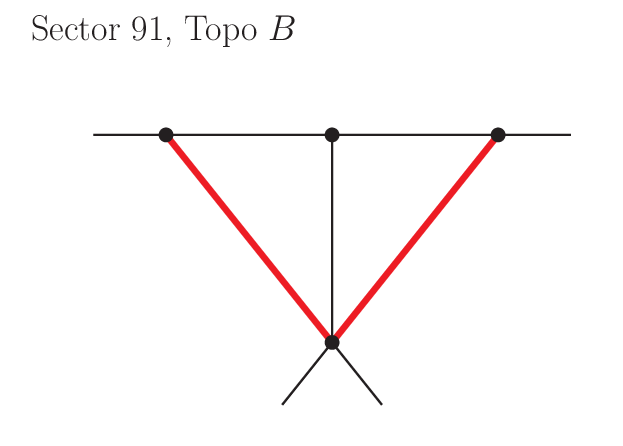}
\includegraphics[width=0.35\textwidth]{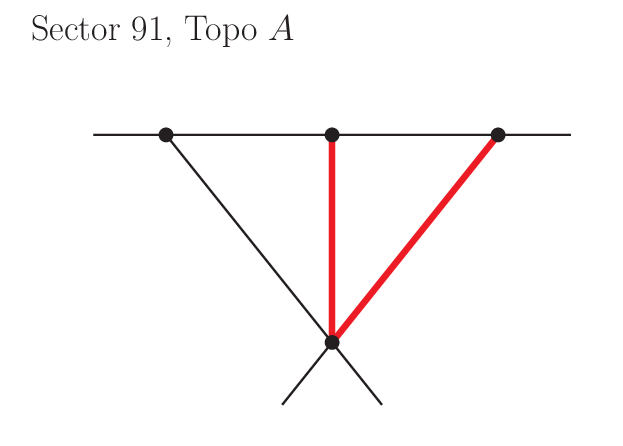}
\includegraphics[width=0.35\textwidth]{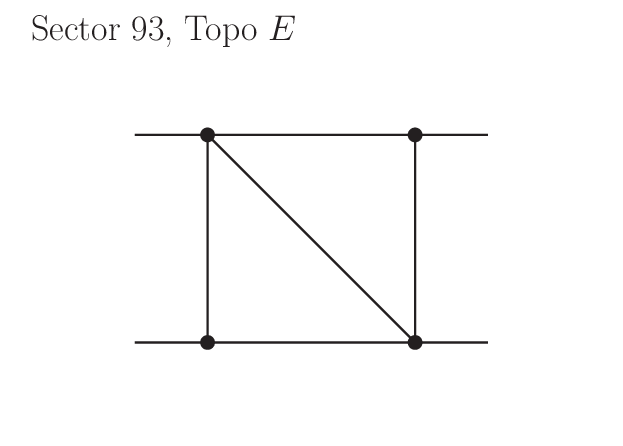}
\includegraphics[width=0.35\textwidth]{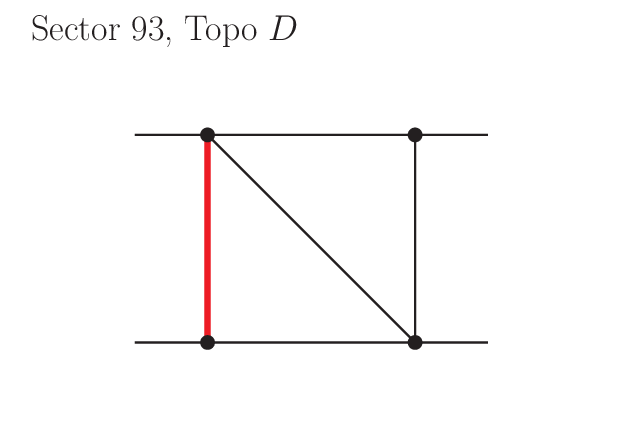}
\includegraphics[width=0.35\textwidth]{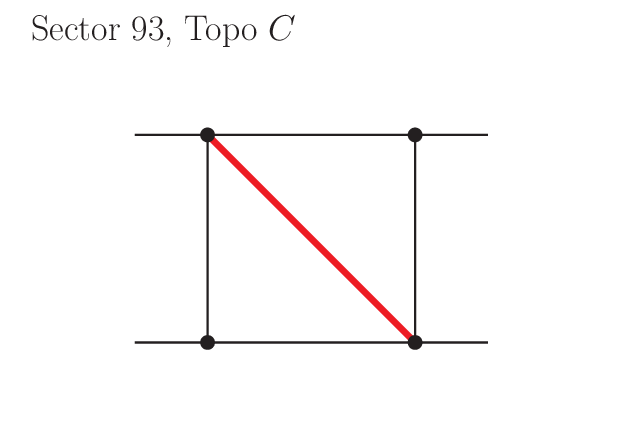}
\includegraphics[width=0.35\textwidth]{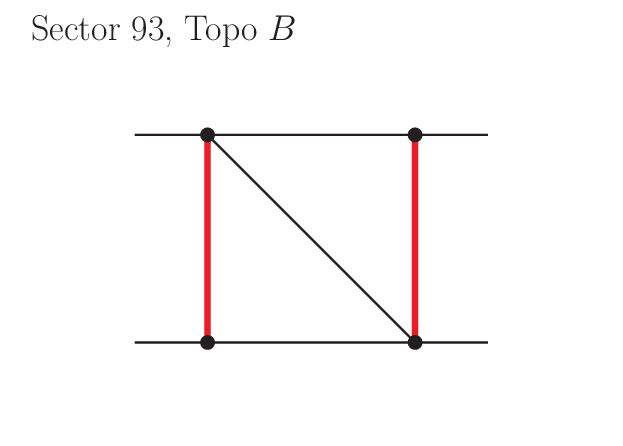}
\includegraphics[width=0.35\textwidth]{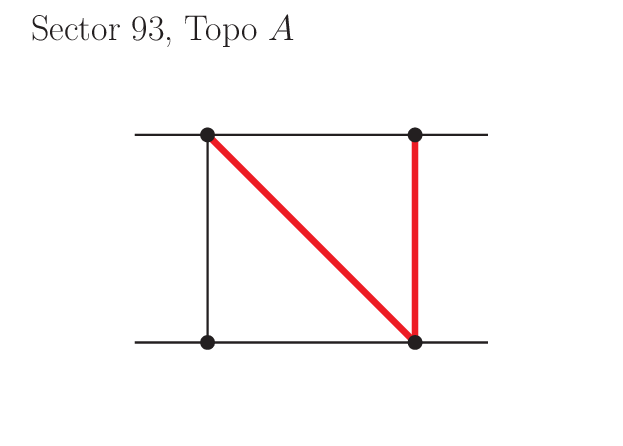}
\includegraphics[width=0.35\textwidth]{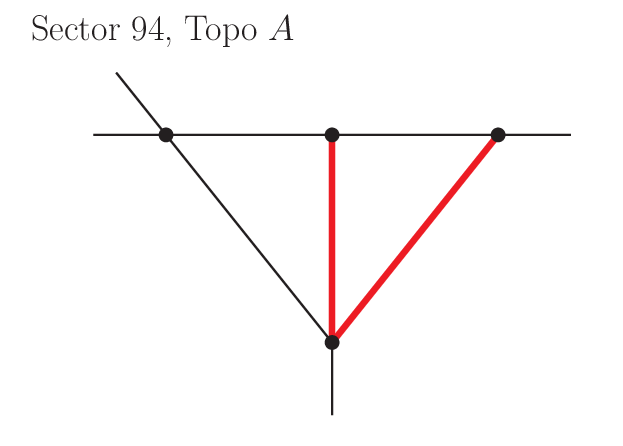}
\end{center}
\caption{\small
Master sectors for planar double-box integrals (part 4).
}
\end{figure}

\begin{figure}[H]
\begin{center}
\includegraphics[width=0.35\textwidth]{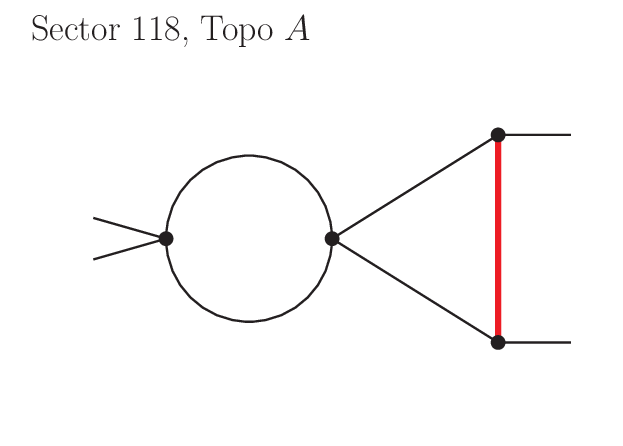}
\includegraphics[width=0.35\textwidth]{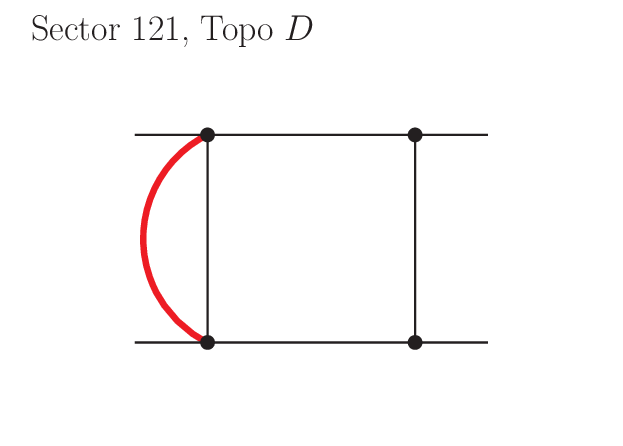}
\includegraphics[width=0.35\textwidth]{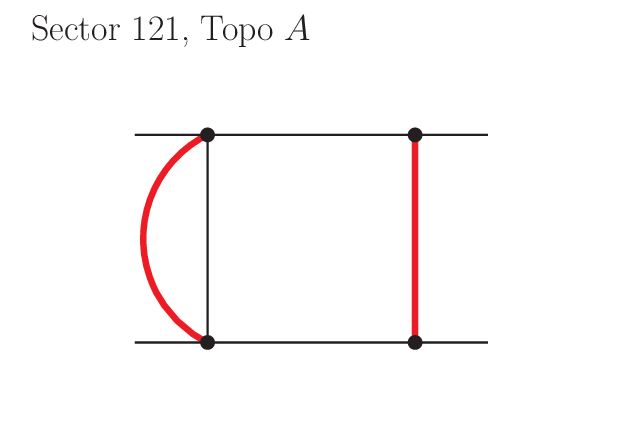}
\includegraphics[width=0.35\textwidth]{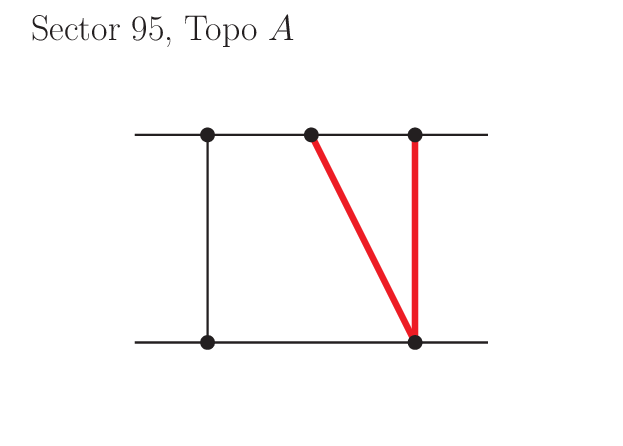}
\includegraphics[width=0.35\textwidth]{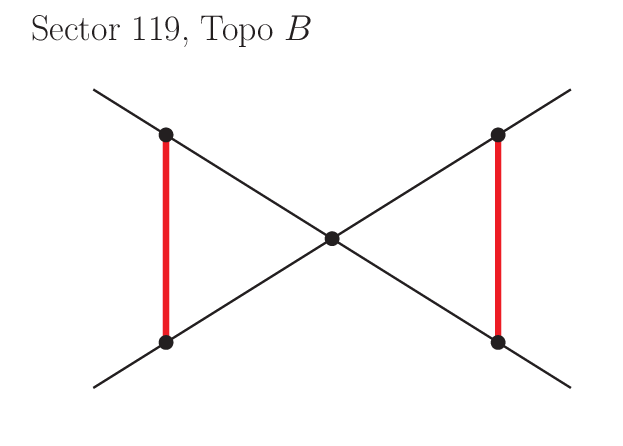}
\includegraphics[width=0.35\textwidth]{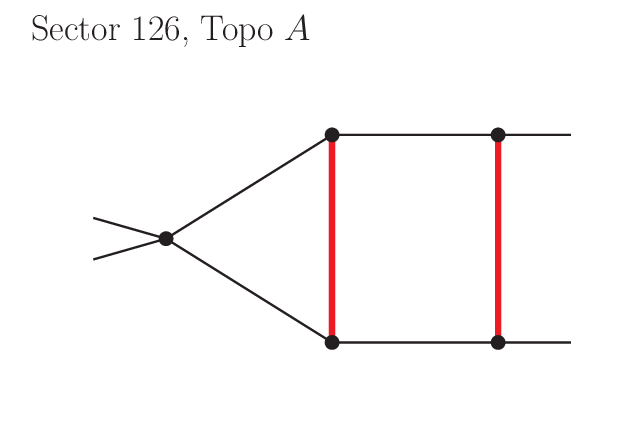}
\includegraphics[width=0.35\textwidth]{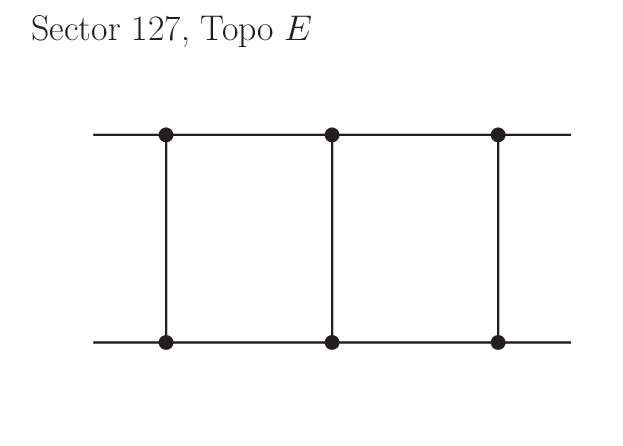}
\includegraphics[width=0.35\textwidth]{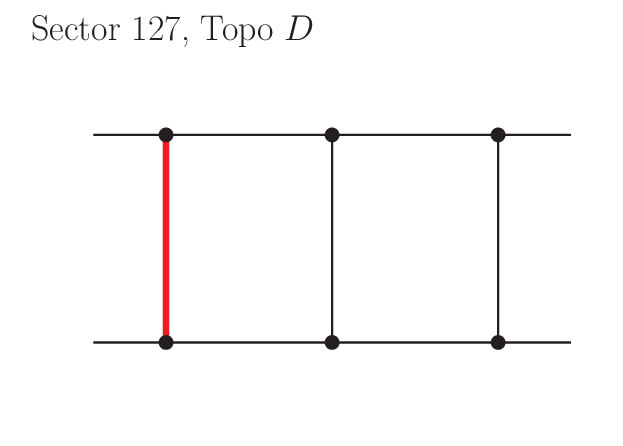}
\includegraphics[width=0.35\textwidth]{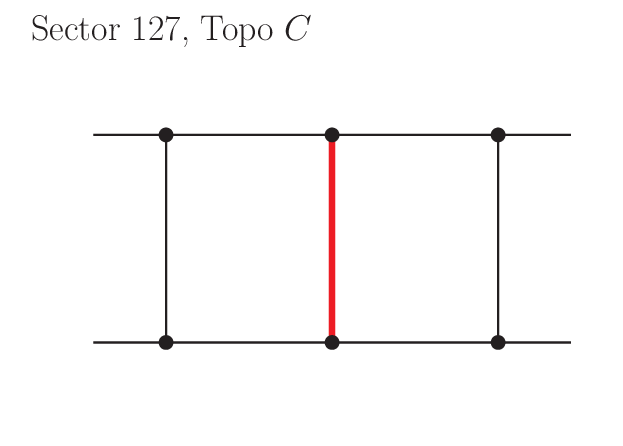}
\includegraphics[width=0.35\textwidth]{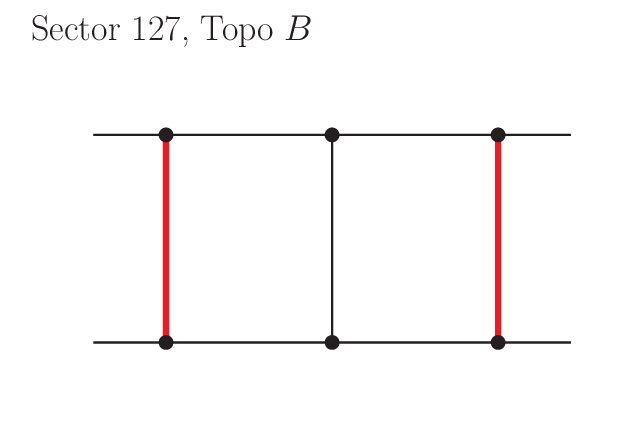}
\end{center}
\caption{\small
Master sectors for planar double-box integrals (part 5).
}
\end{figure}

\begin{figure}
\begin{center}
\includegraphics[width=0.35\textwidth]{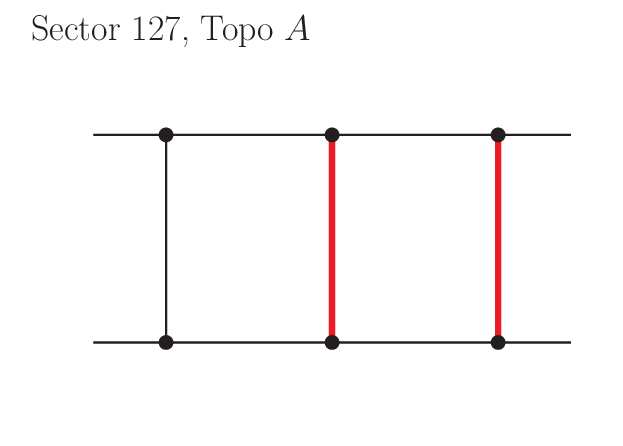}
\end{center}
\caption{\small
Master sectors for planar double-box integrals (part 6).
}
\end{figure}

\clearpage
\newpage
\subsection{Non-planar double-box integrals}
\begin{figure}[H]
\begin{center}
\includegraphics[width=0.35\textwidth]{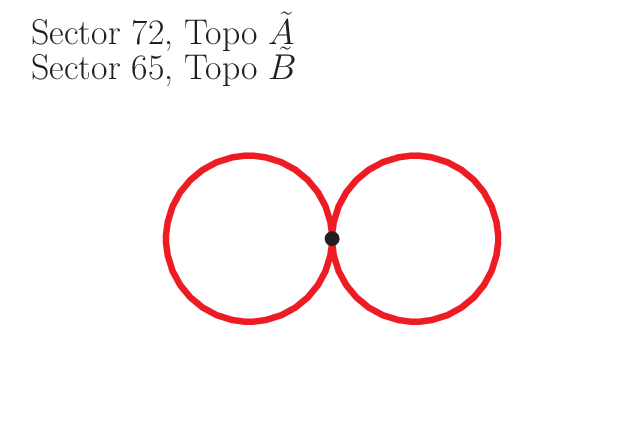}
\includegraphics[width=0.35\textwidth]{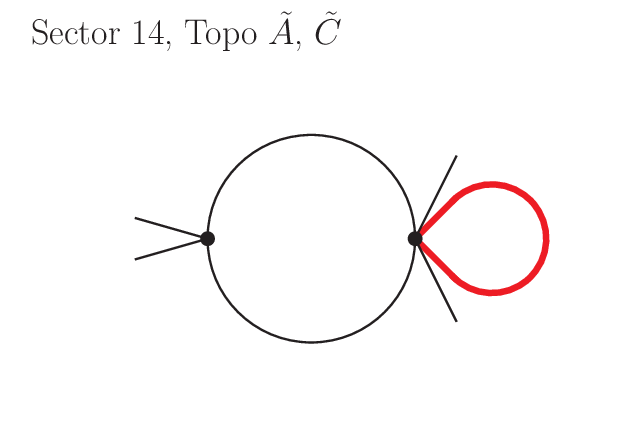}
\includegraphics[width=0.35\textwidth]{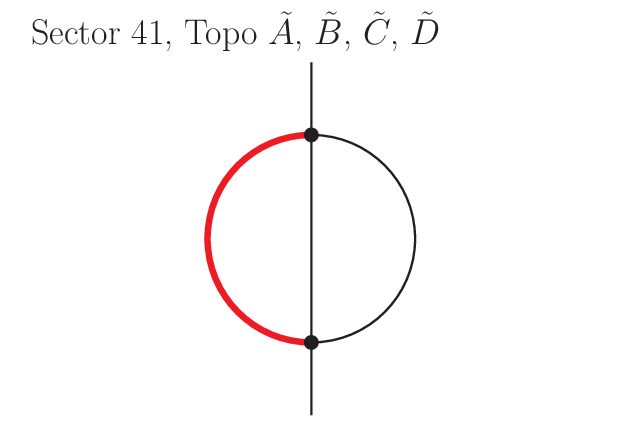}
\includegraphics[width=0.35\textwidth]{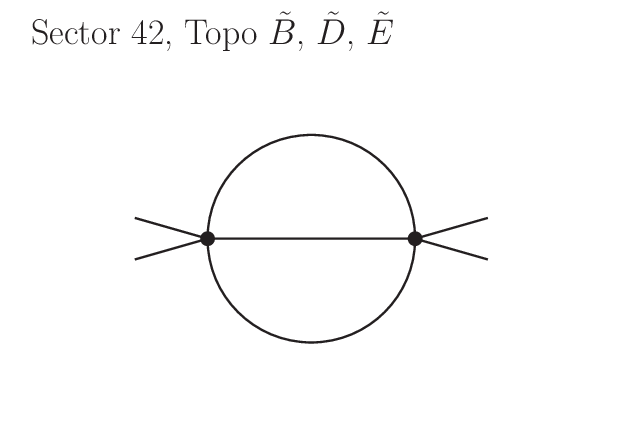}
\includegraphics[width=0.35\textwidth]{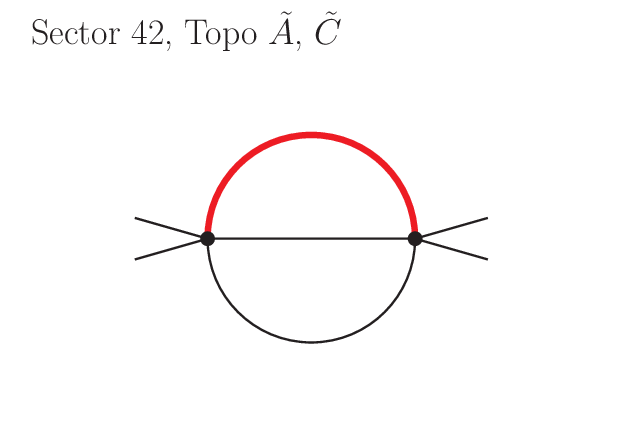}
\includegraphics[width=0.35\textwidth]{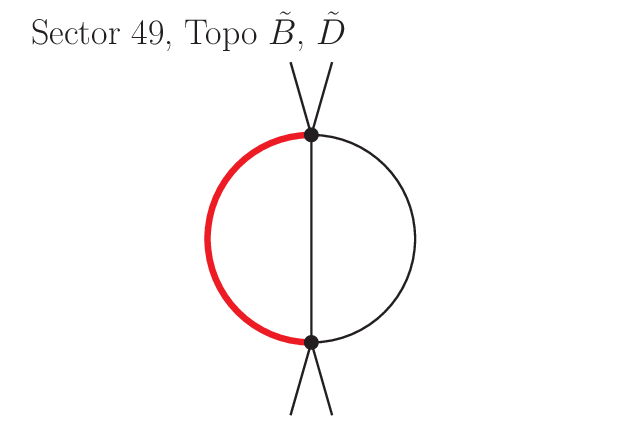}
\includegraphics[width=0.35\textwidth]{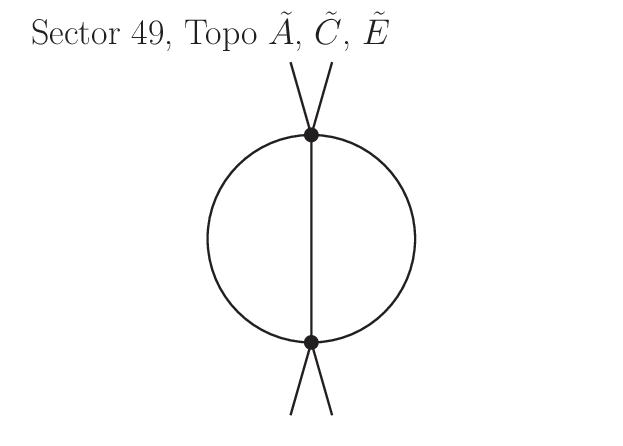}
\includegraphics[width=0.35\textwidth]{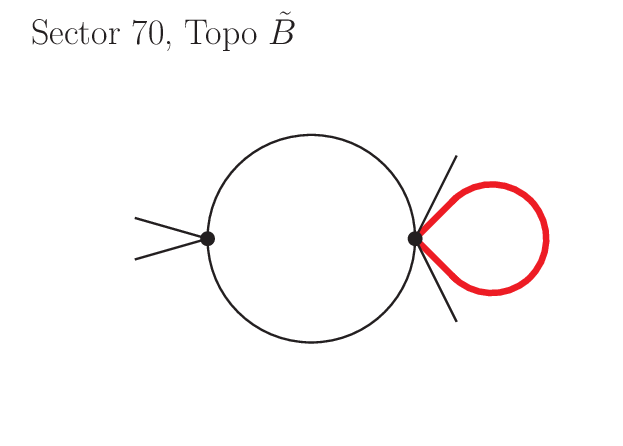}
\end{center}
\caption{\small
Master sectors for non-planar double-box integrals (part 1).
}
\end{figure}

\begin{figure}[H]
\begin{center}
\includegraphics[width=0.35\textwidth]{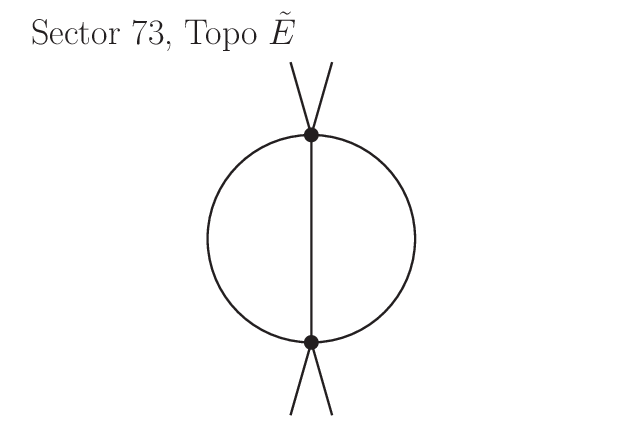}
\includegraphics[width=0.35\textwidth]{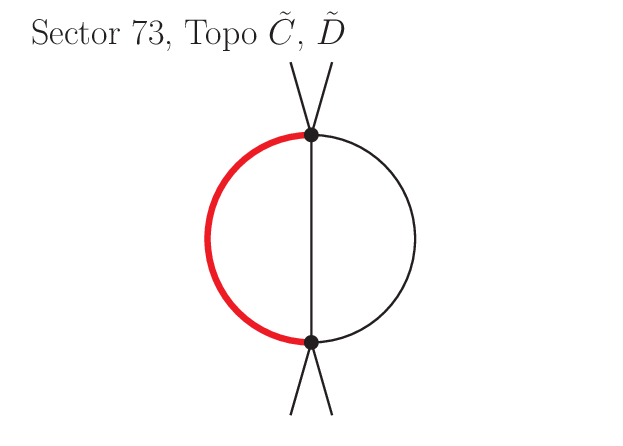}
\includegraphics[width=0.35\textwidth]{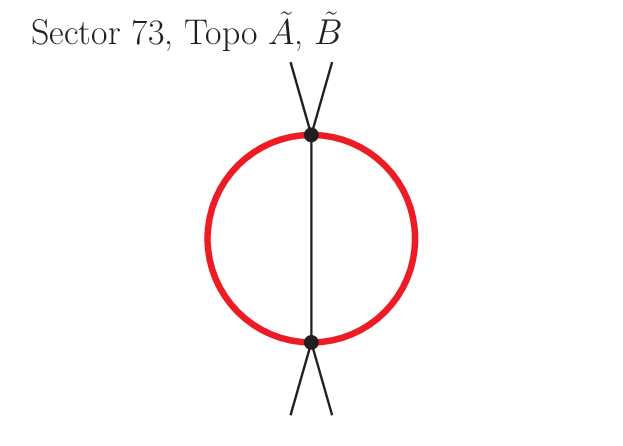}
\includegraphics[width=0.35\textwidth]{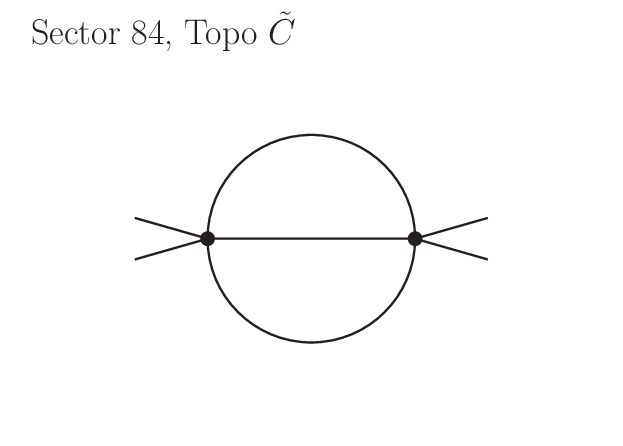}
\includegraphics[width=0.35\textwidth]{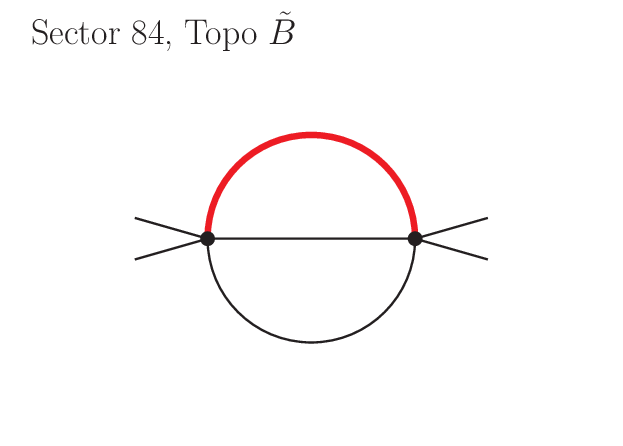}
\includegraphics[width=0.35\textwidth]{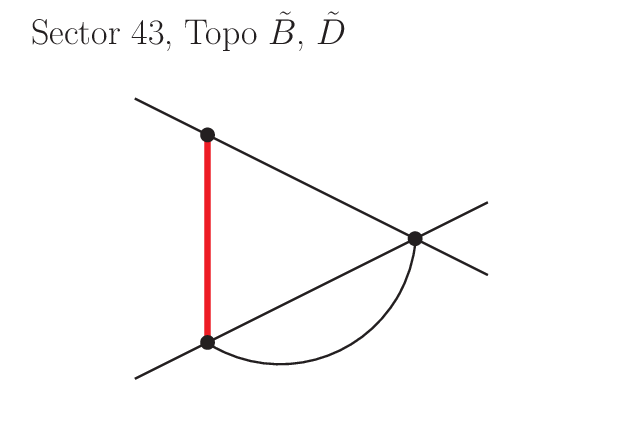}
\includegraphics[width=0.35\textwidth]{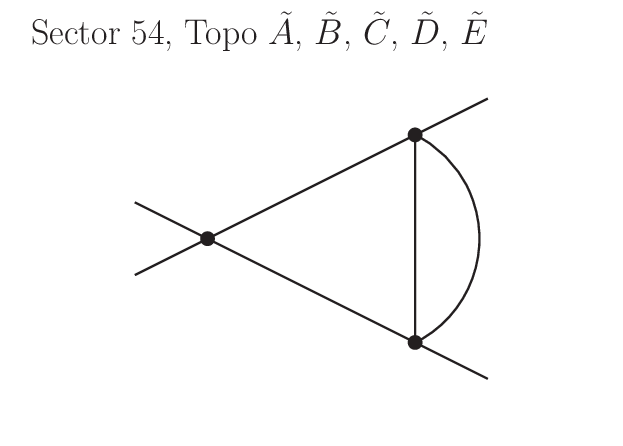}
\includegraphics[width=0.35\textwidth]{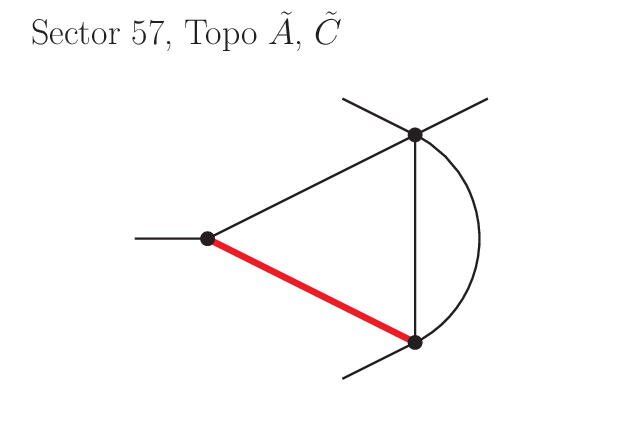}
\includegraphics[width=0.35\textwidth]{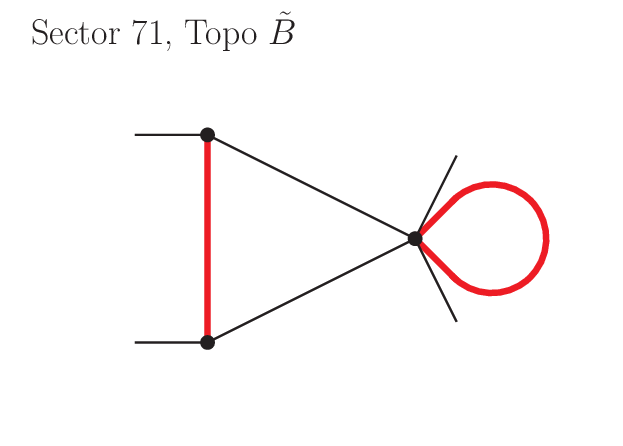}
\includegraphics[width=0.35\textwidth]{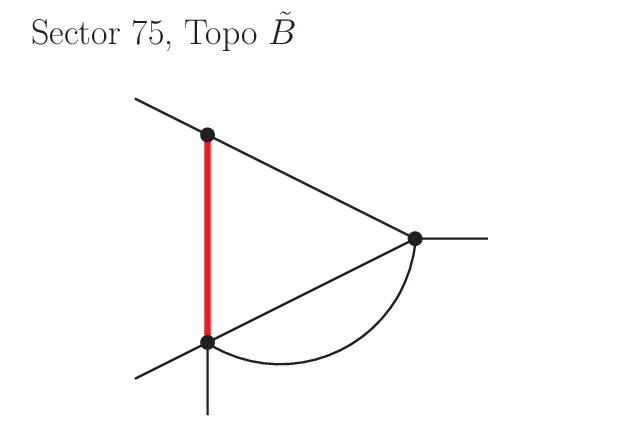}
\end{center}
\caption{\small
Master sectors for non-planar double-box integrals (part 2).
}
\end{figure}

\begin{figure}[H]
\begin{center}
\includegraphics[width=0.35\textwidth]{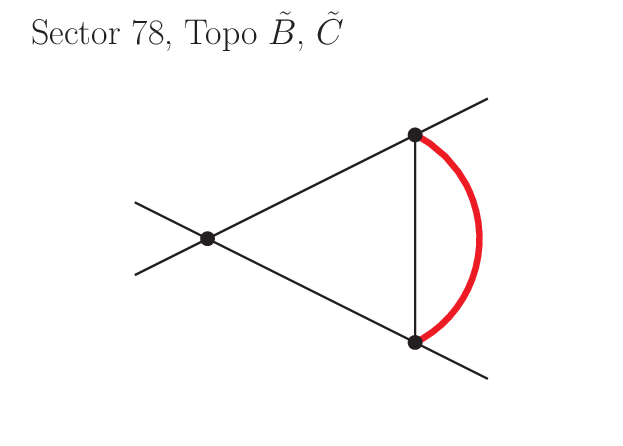}
\includegraphics[width=0.35\textwidth]{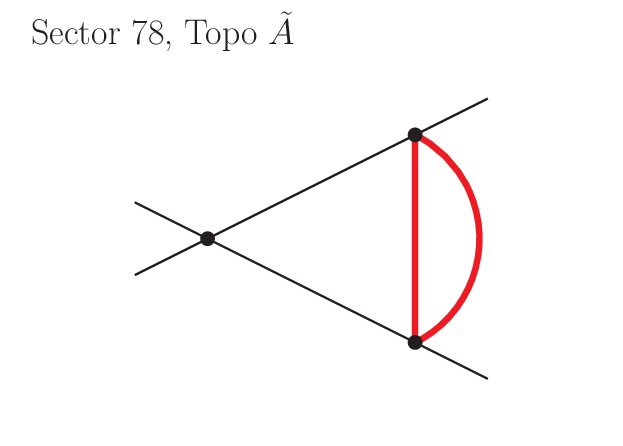}
\includegraphics[width=0.35\textwidth]{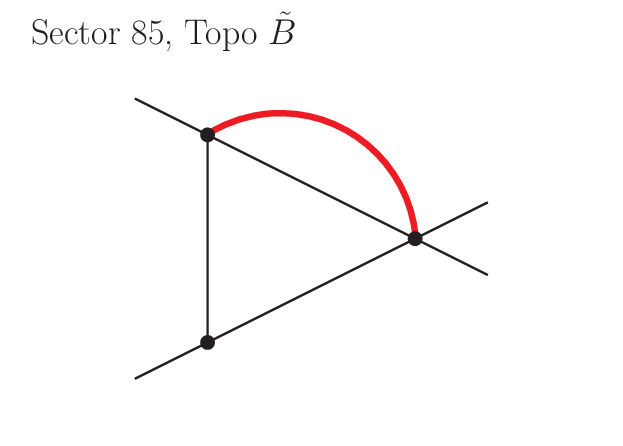}
\includegraphics[width=0.35\textwidth]{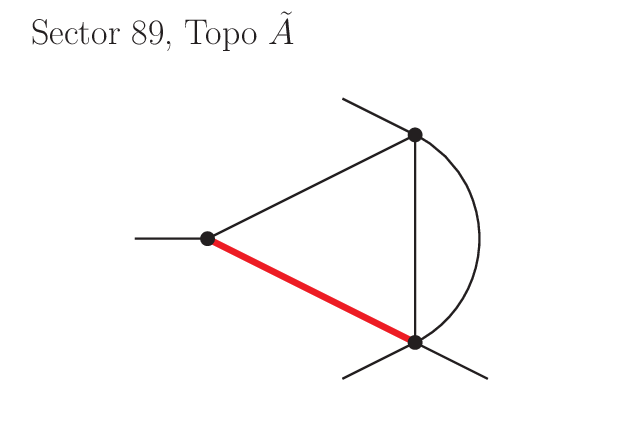}
\includegraphics[width=0.35\textwidth]{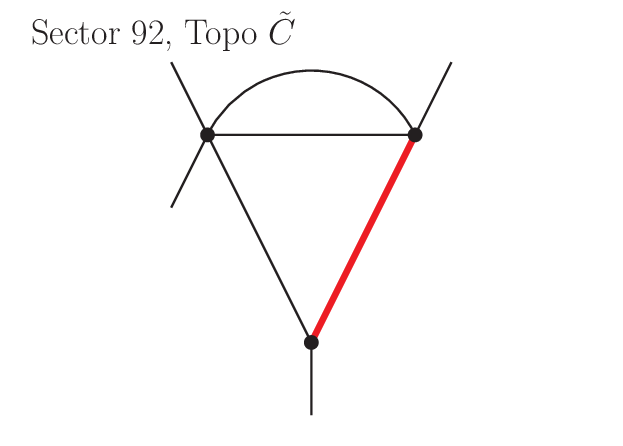}
\includegraphics[width=0.35\textwidth]{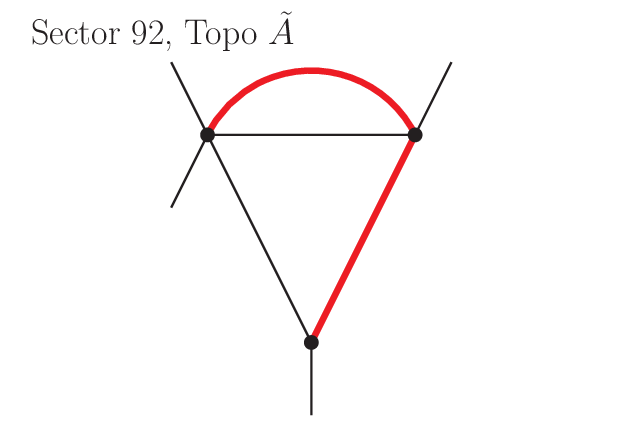}
\includegraphics[width=0.35\textwidth]{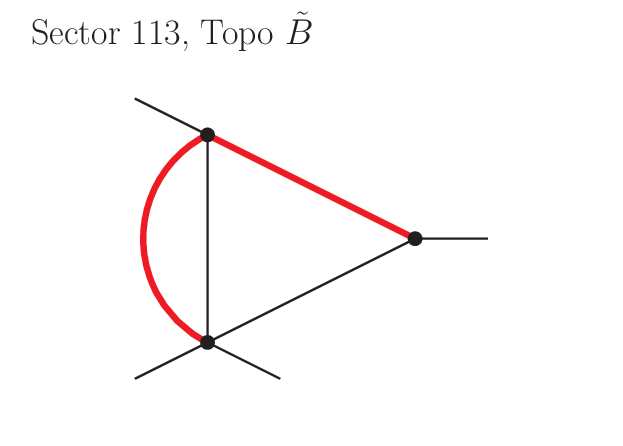}
\includegraphics[width=0.35\textwidth]{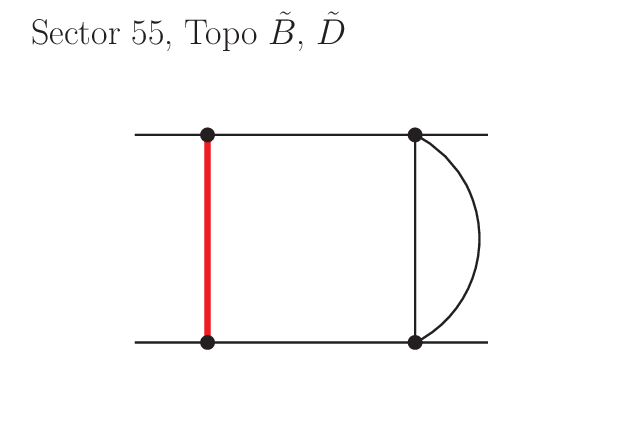}
\includegraphics[width=0.35\textwidth]{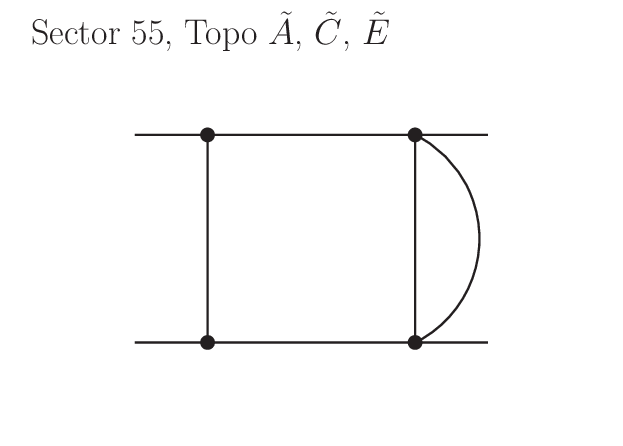}
\includegraphics[width=0.35\textwidth]{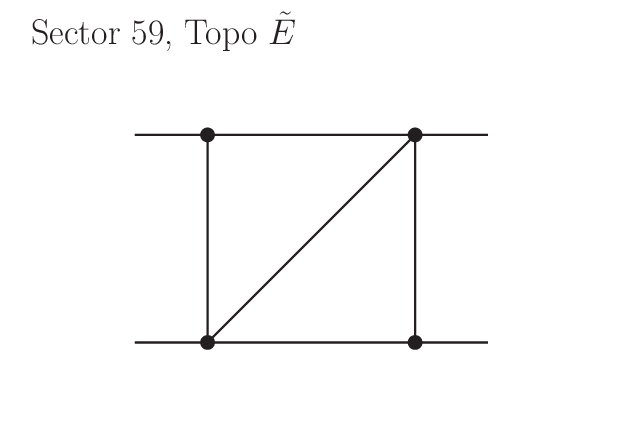}
\end{center}
\caption{\small
Master sectors for non-planar double-box integrals (part 3).
}
\end{figure}

\begin{figure}[H]
\begin{center}
\includegraphics[width=0.35\textwidth]{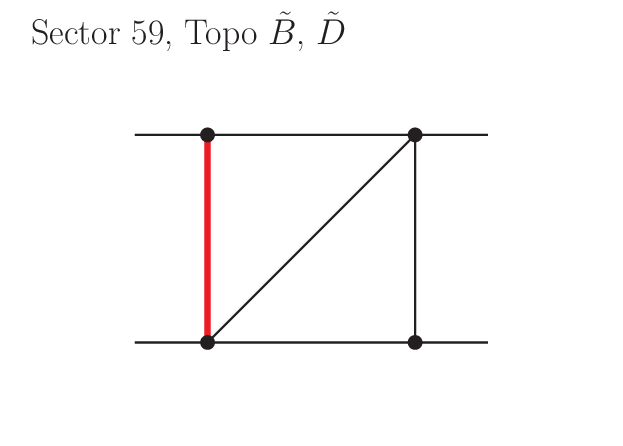}
\includegraphics[width=0.35\textwidth]{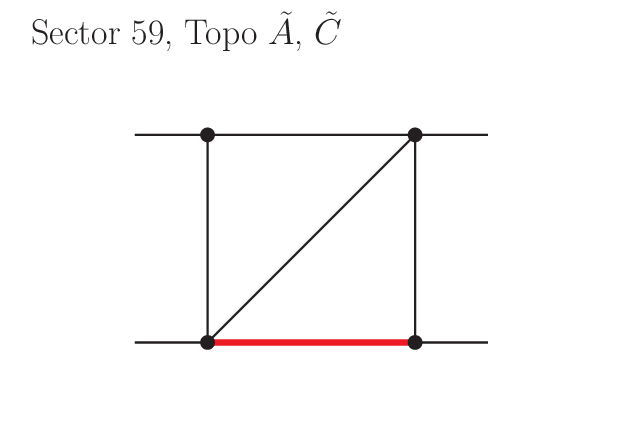}
\includegraphics[width=0.35\textwidth]{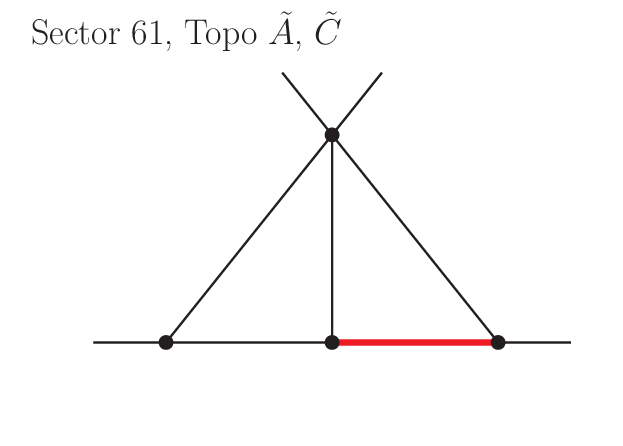}
\includegraphics[width=0.35\textwidth]{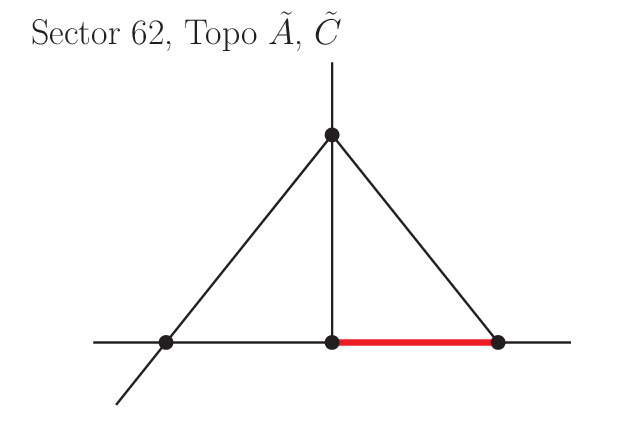}
\includegraphics[width=0.35\textwidth]{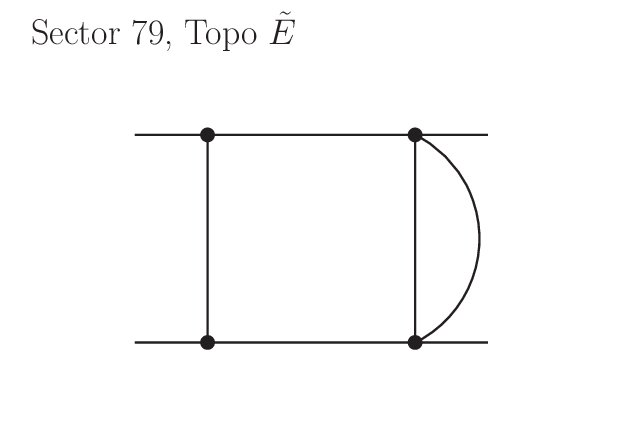}
\includegraphics[width=0.35\textwidth]{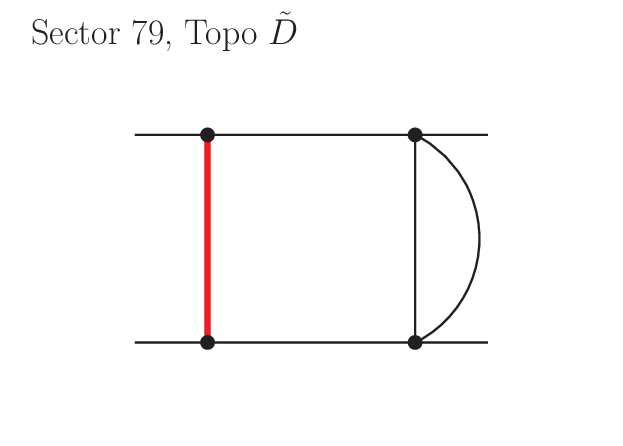}
\includegraphics[width=0.35\textwidth]{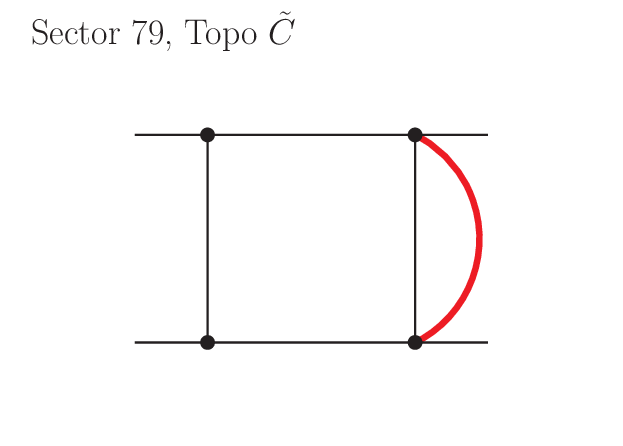}
\includegraphics[width=0.35\textwidth]{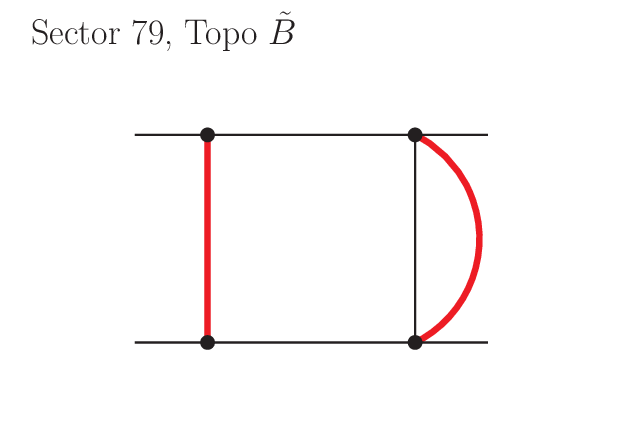}
\includegraphics[width=0.35\textwidth]{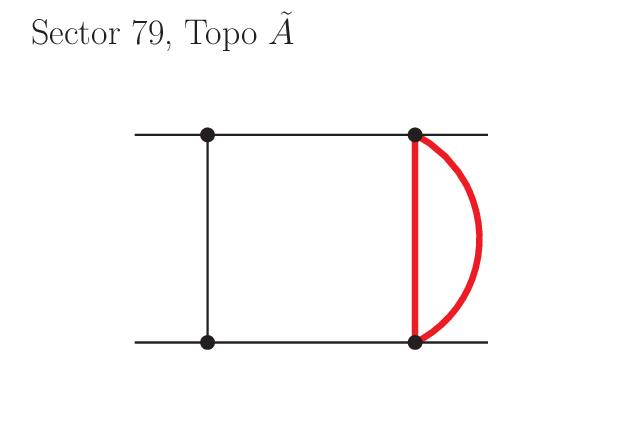}
\includegraphics[width=0.35\textwidth]{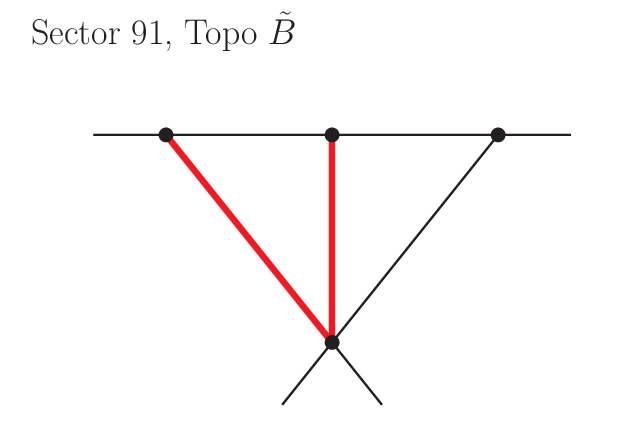}
\end{center}
\caption{\small
Master sectors for non-planar double-box integrals (part 4).
}
\end{figure}

\begin{figure}[H]
\begin{center}
\includegraphics[width=0.35\textwidth]{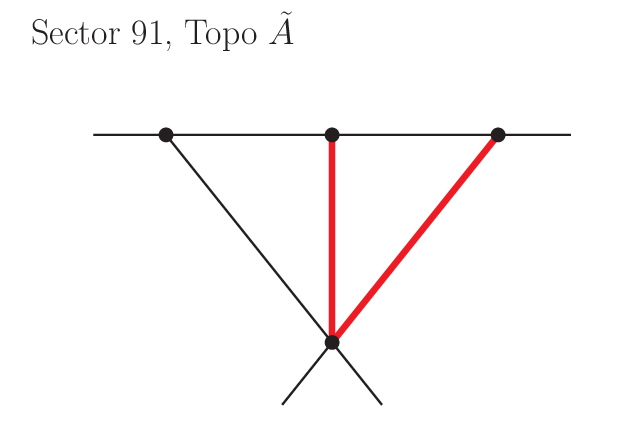}
\includegraphics[width=0.35\textwidth]{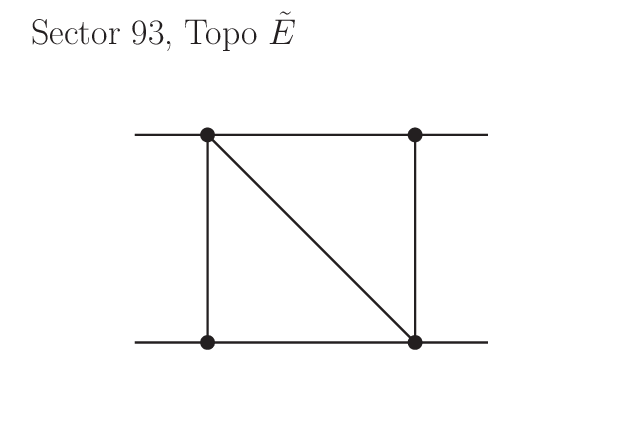}
\includegraphics[width=0.35\textwidth]{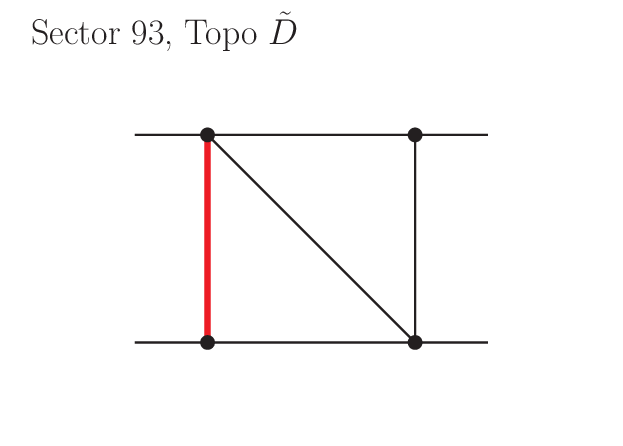}
\includegraphics[width=0.35\textwidth]{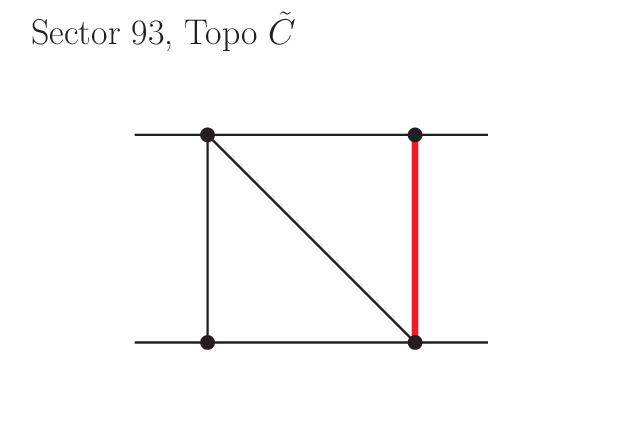}
\includegraphics[width=0.35\textwidth]{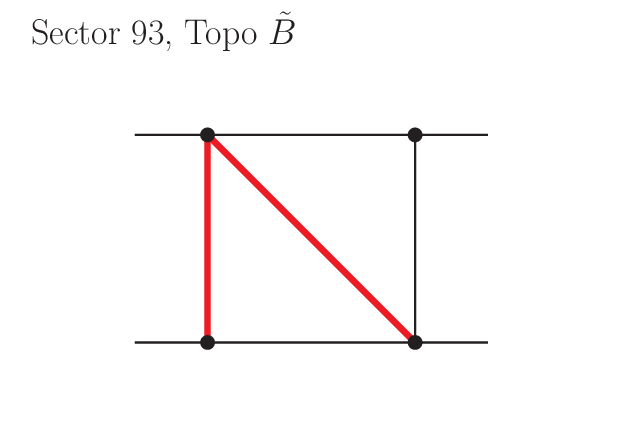}
\includegraphics[width=0.35\textwidth]{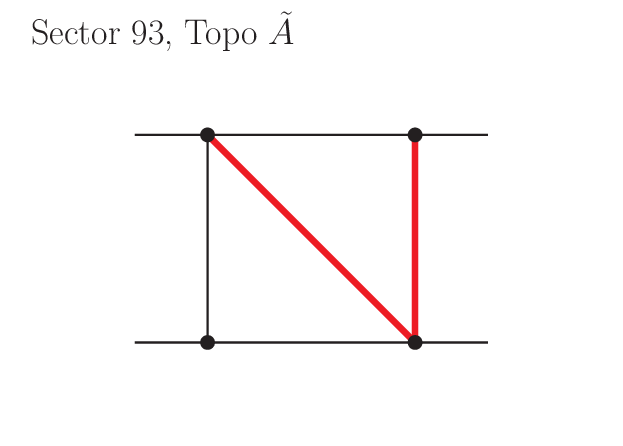}
\includegraphics[width=0.35\textwidth]{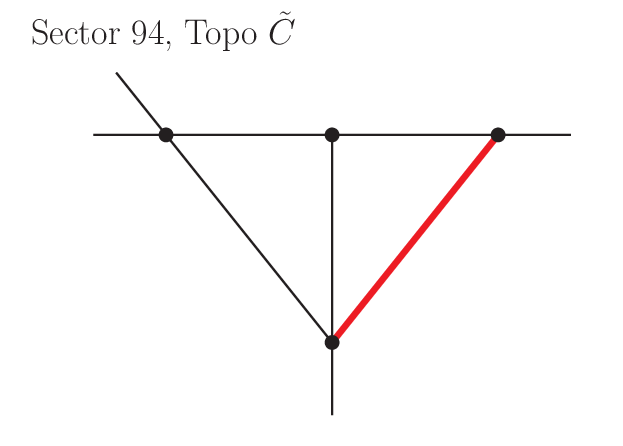}
\includegraphics[width=0.35\textwidth]{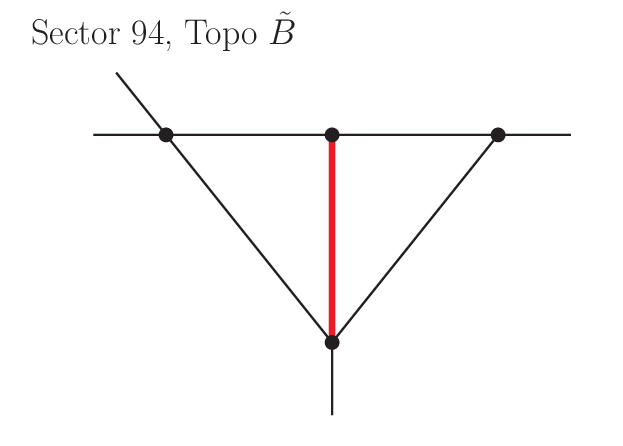}
\includegraphics[width=0.35\textwidth]{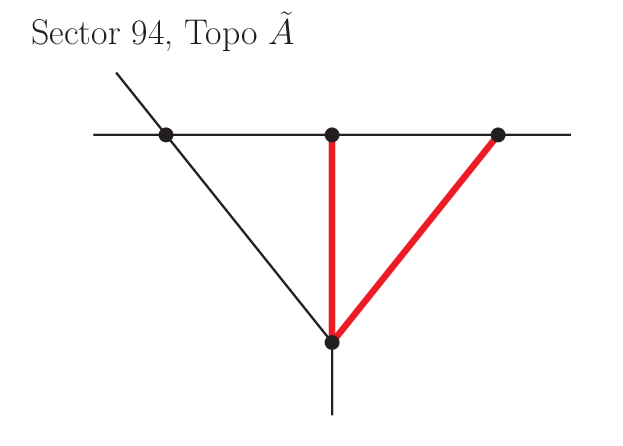}
\includegraphics[width=0.35\textwidth]{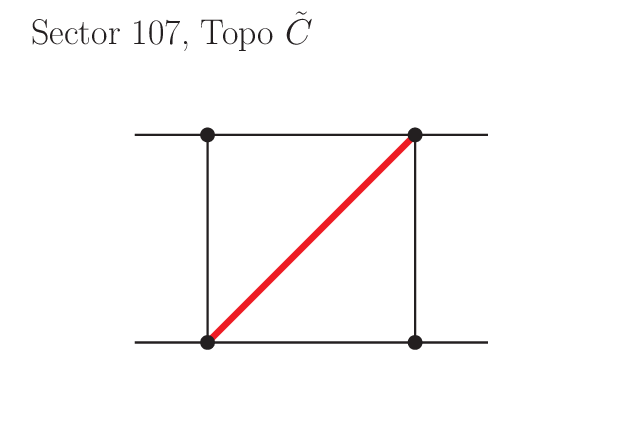}
\end{center}
\caption{\small
Master sectors for non-planar double-box integrals (part 5).
}
\end{figure}

\begin{figure}[H]
\begin{center}
\includegraphics[width=0.35\textwidth]{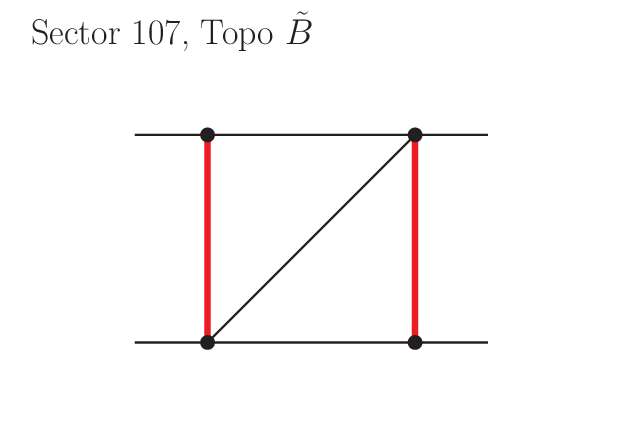}
\includegraphics[width=0.35\textwidth]{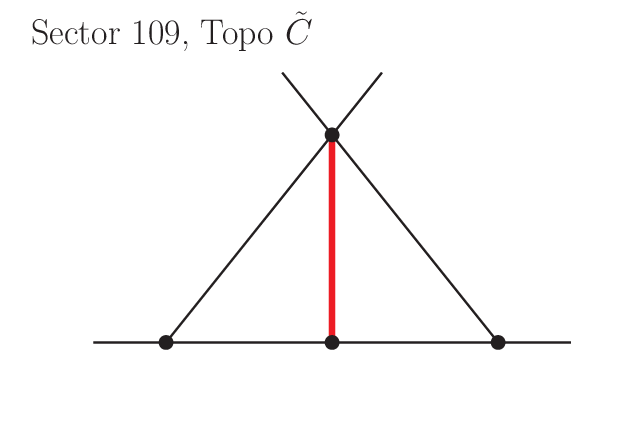}
\includegraphics[width=0.35\textwidth]{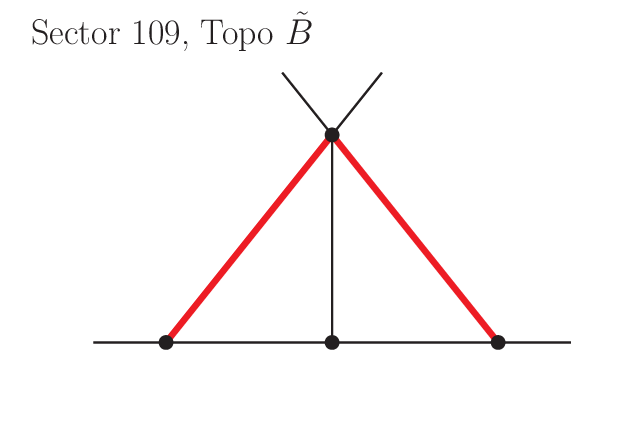}
\includegraphics[width=0.35\textwidth]{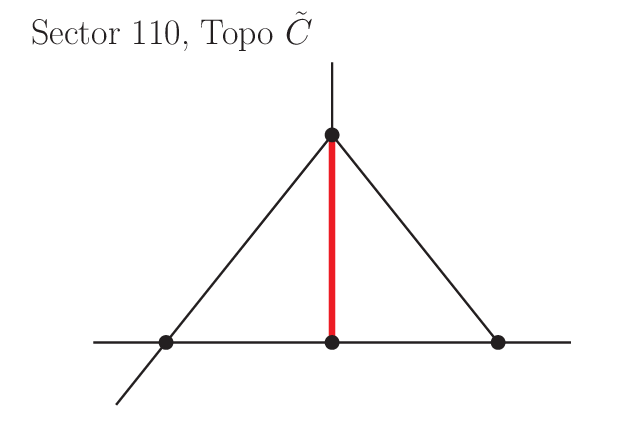}
\includegraphics[width=0.35\textwidth]{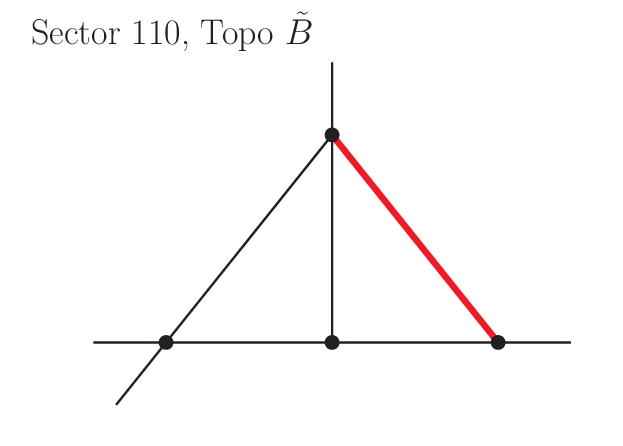}
\includegraphics[width=0.35\textwidth]{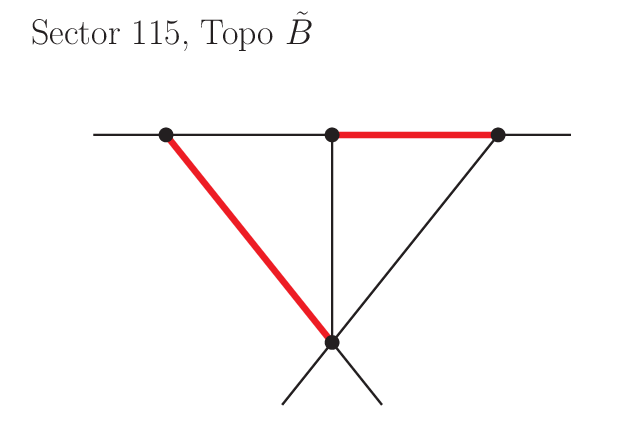}
\includegraphics[width=0.35\textwidth]{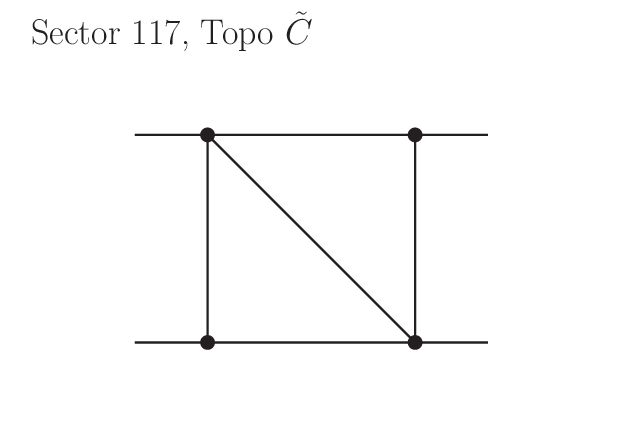}
\includegraphics[width=0.35\textwidth]{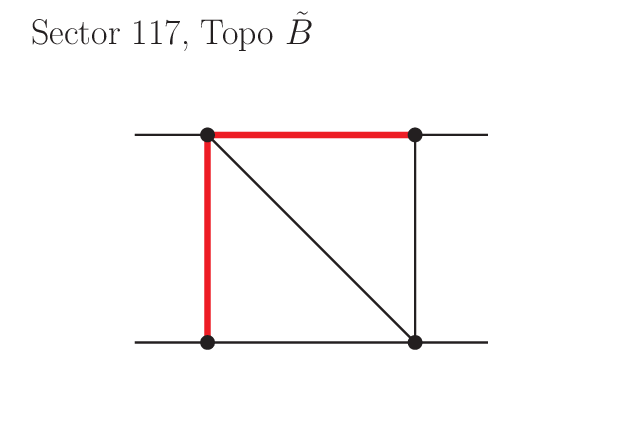}
\includegraphics[width=0.35\textwidth]{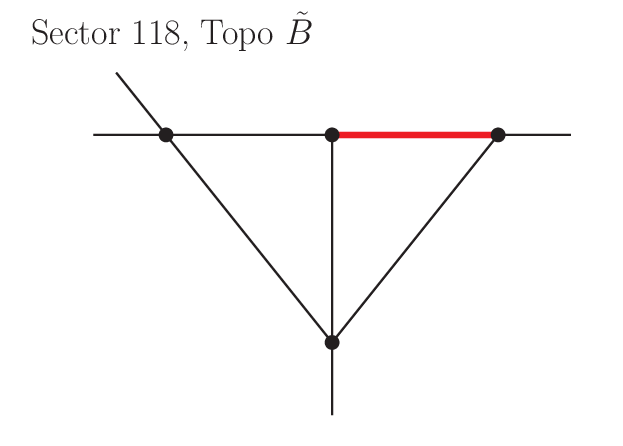}
\includegraphics[width=0.35\textwidth]{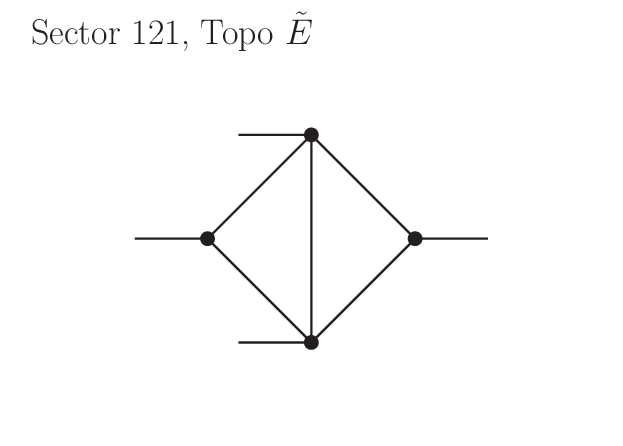}
\end{center}
\caption{\small
Master sectors for non-planar double-box integrals (part 6).
}
\end{figure}

\begin{figure}[H]
\begin{center}
\includegraphics[width=0.35\textwidth]{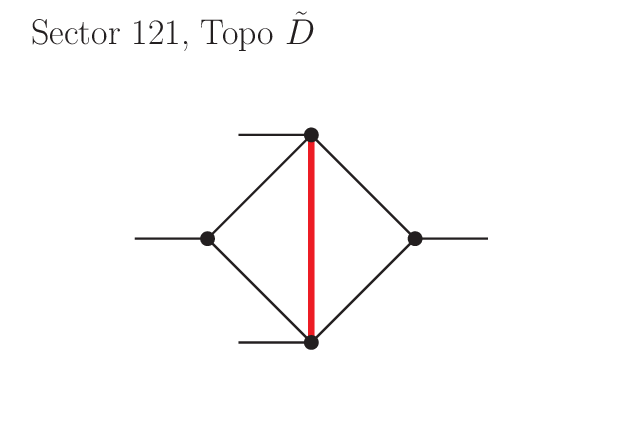}
\includegraphics[width=0.35\textwidth]{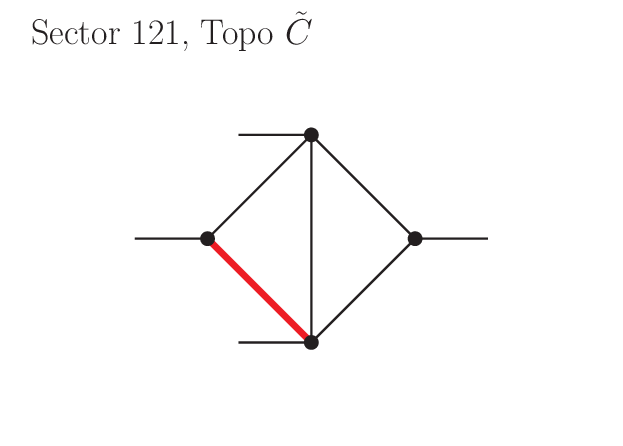}
\includegraphics[width=0.35\textwidth]{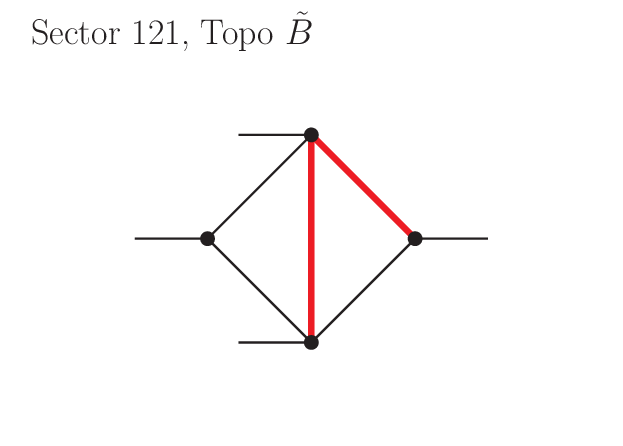}
\includegraphics[width=0.35\textwidth]{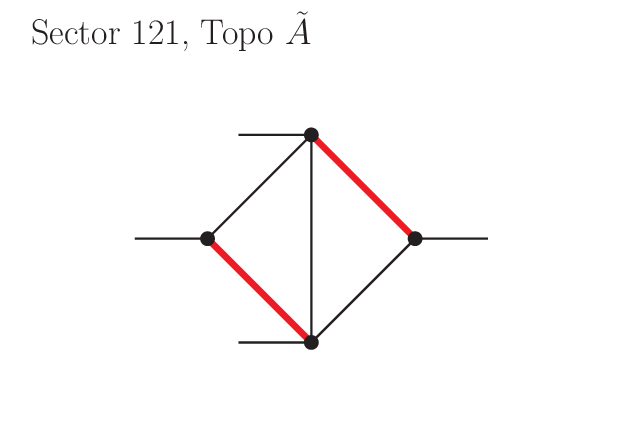}
\includegraphics[width=0.35\textwidth]{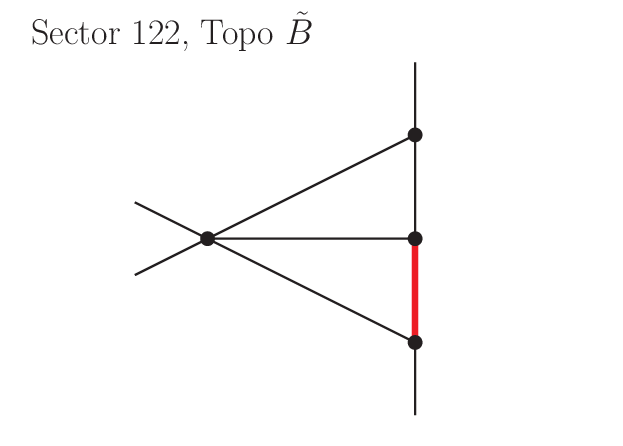}
\includegraphics[width=0.35\textwidth]{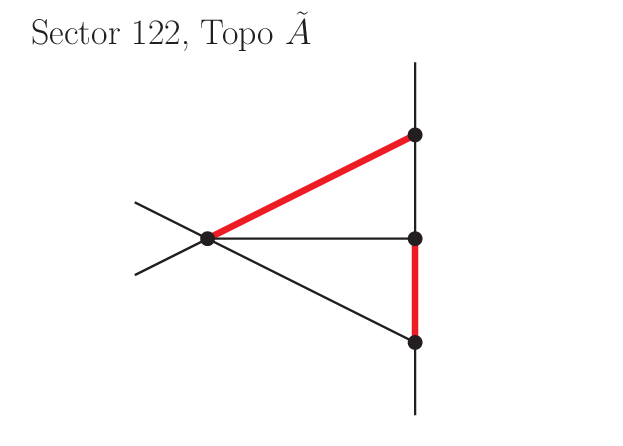}
\includegraphics[width=0.35\textwidth]{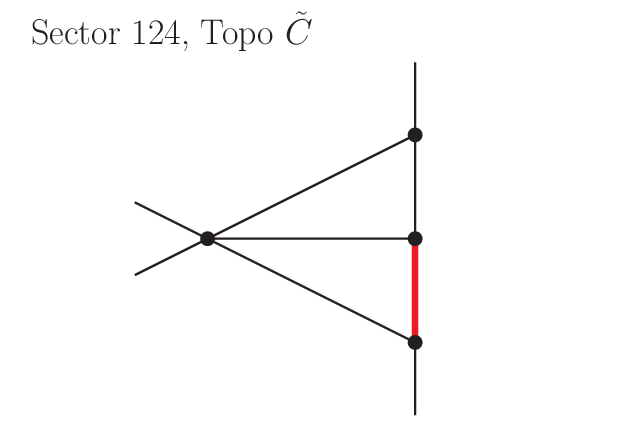}
\includegraphics[width=0.35\textwidth]{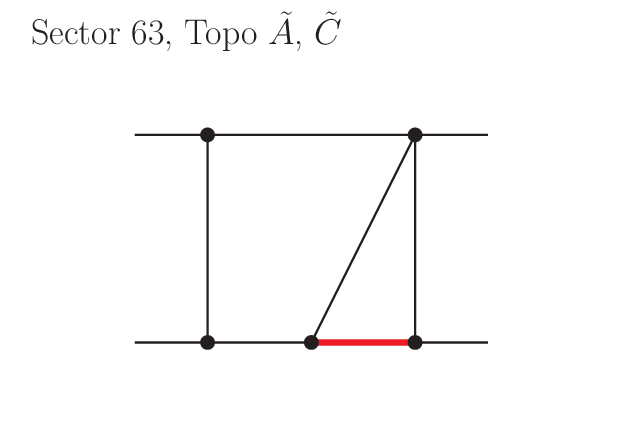}
\includegraphics[width=0.35\textwidth]{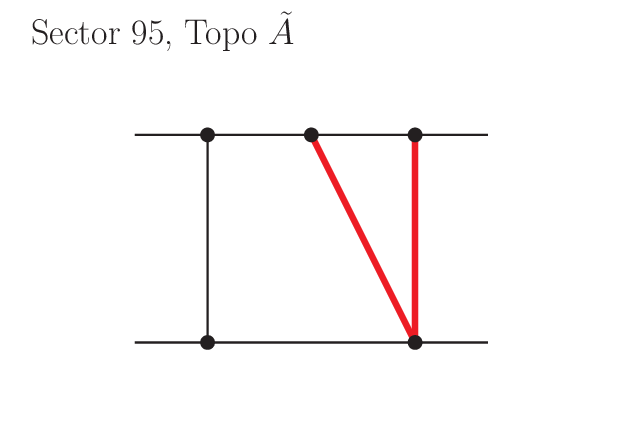}
\includegraphics[width=0.35\textwidth]{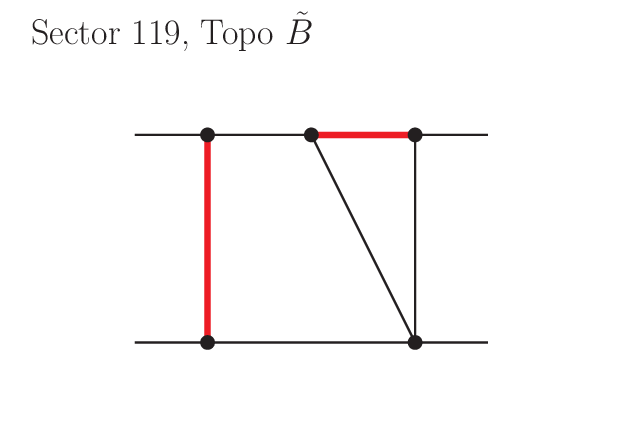}
\end{center}
\caption{\small
Master sectors for non-planar double-box integrals (part 7).
}
\end{figure}

\begin{figure}[H]
\begin{center}
\includegraphics[width=0.35\textwidth]{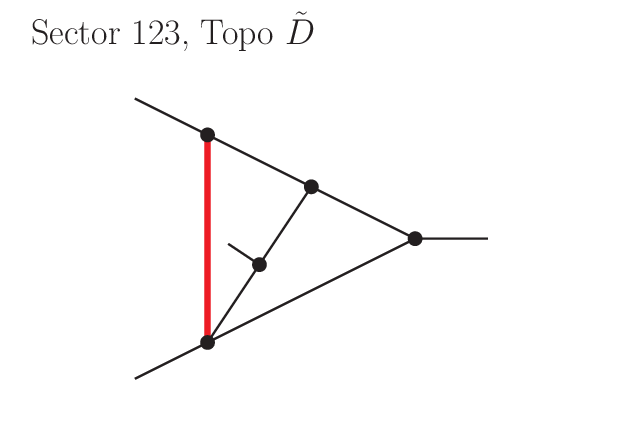}
\includegraphics[width=0.35\textwidth]{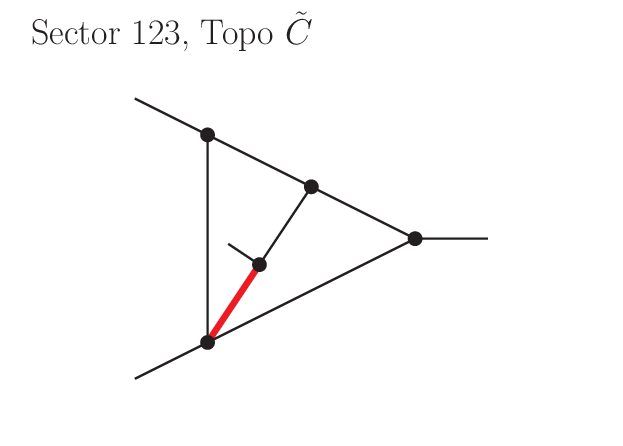}
\includegraphics[width=0.35\textwidth]{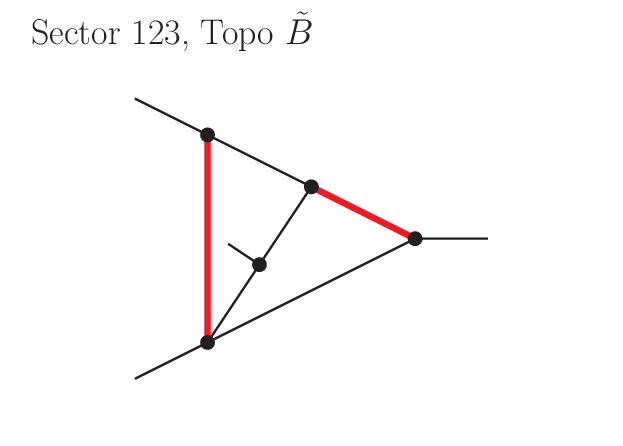}
\includegraphics[width=0.35\textwidth]{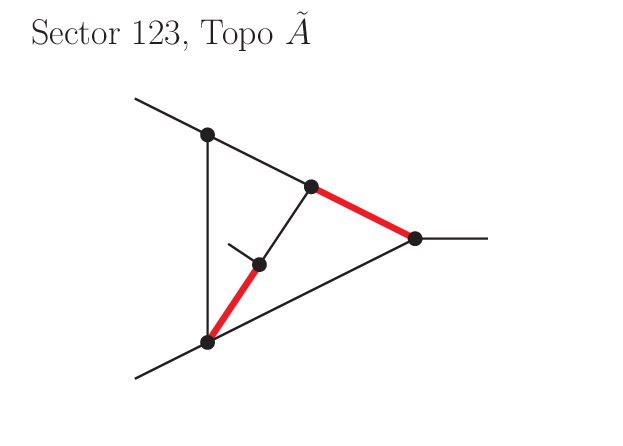}
\includegraphics[width=0.35\textwidth]{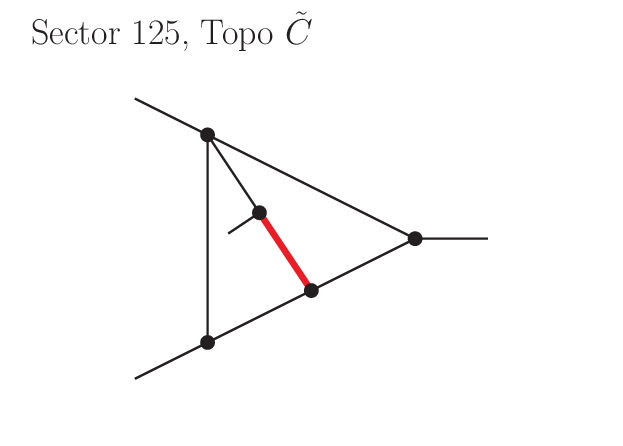}
\includegraphics[width=0.35\textwidth]{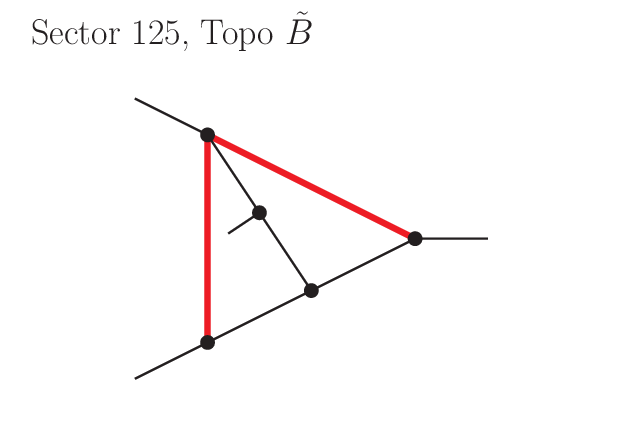}
\includegraphics[width=0.35\textwidth]{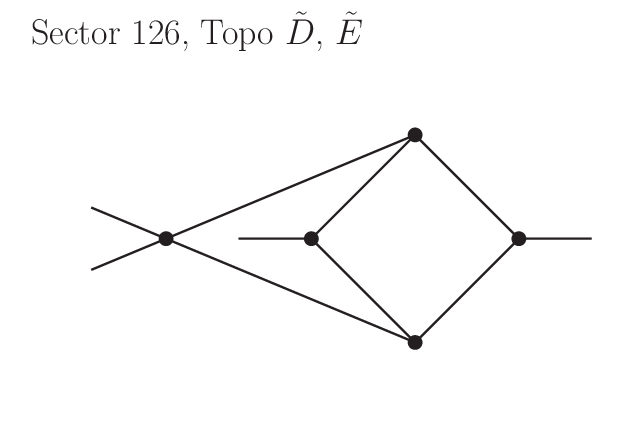}
\includegraphics[width=0.35\textwidth]{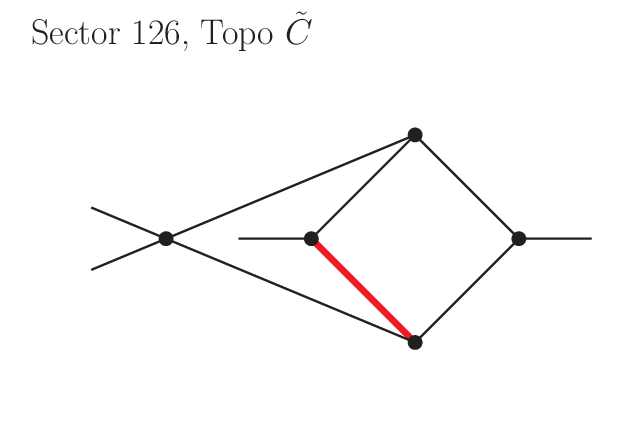}
\includegraphics[width=0.35\textwidth]{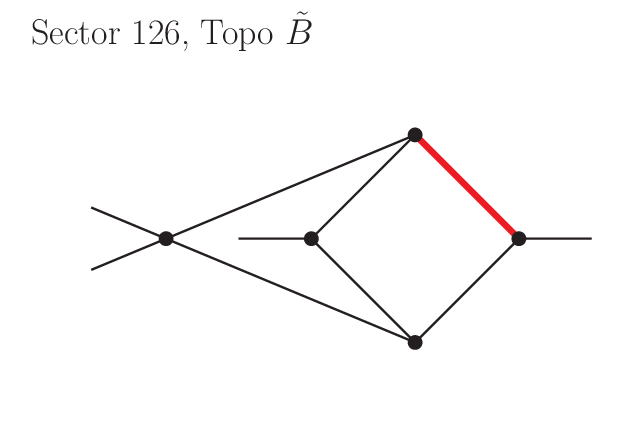}
\includegraphics[width=0.35\textwidth]{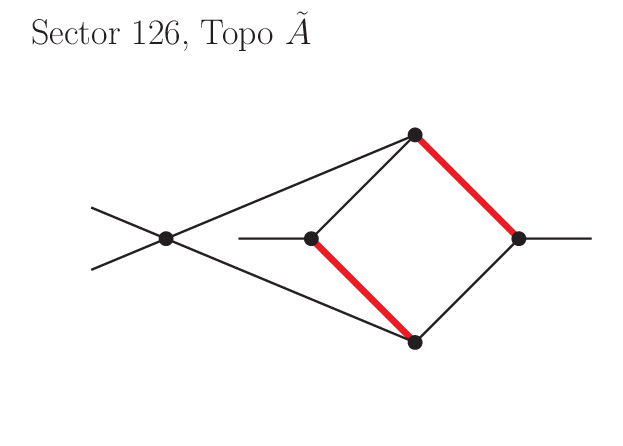}
\end{center}
\caption{\small
Master sectors for non-planar double-box integrals (part 8).
}
\end{figure}

\begin{figure}[H]
\begin{center}
\includegraphics[width=0.35\textwidth]{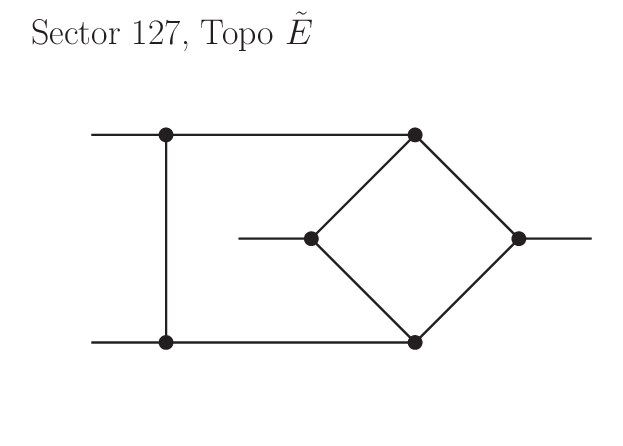}
\includegraphics[width=0.35\textwidth]{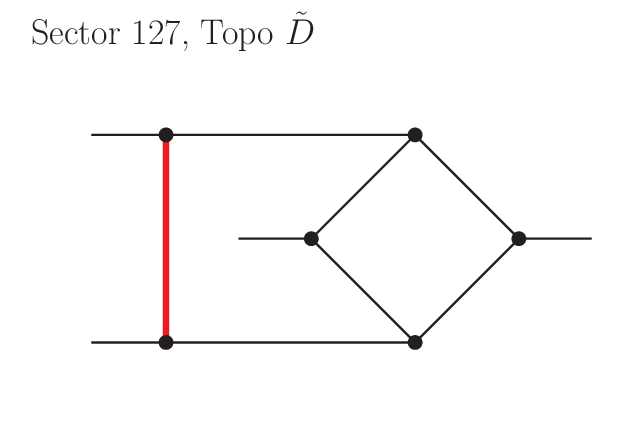}
\includegraphics[width=0.35\textwidth]{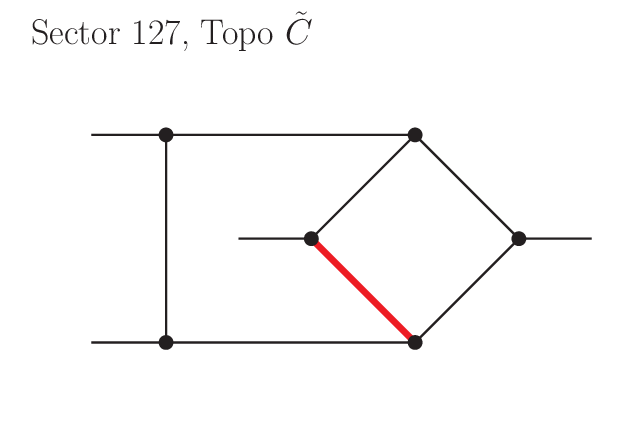}
\includegraphics[width=0.35\textwidth]{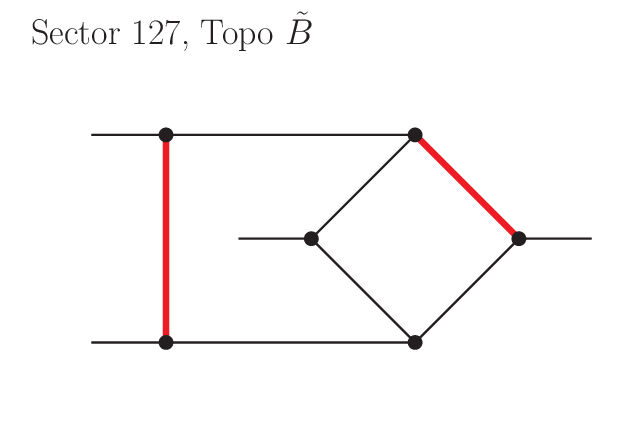}
\includegraphics[width=0.35\textwidth]{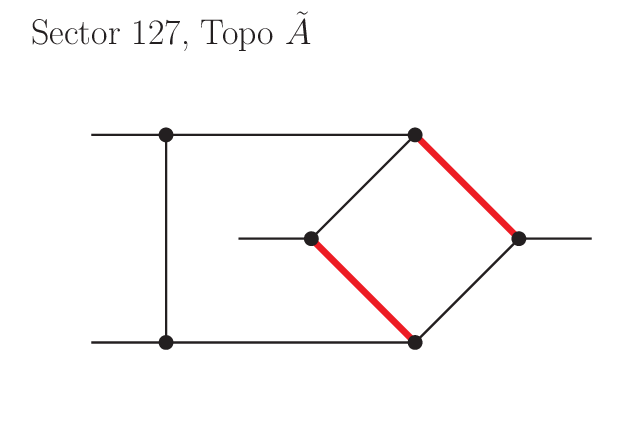}
\end{center}
\caption{\small
Master sectors for non-planar double-box integrals (part 9).
}
\end{figure}


\section{List of master integrals}
\label{sect:master_integrals_list}

In this appendix we present the master integrals, which lead to an $\eps$-factorised differential equation.
We list the master integrals for the planar topologies $A$, $B$, $C$, $D$ and $E$ in section~\ref{sect:planar}.
The master integrals for the non-planar topologies $\tilde{A}$, $\tilde{B}$, $\tilde{C}$, $\tilde{D}$ and $\tilde{E}$ are listed 
in section~\ref{sect:nonplanar}.

\subsection{The planar topologies}
\label{sect:planar}

\subsubsection{Topology $A$}

\begin{alignat*}{3}
&\text{Sector 72:} \hspace{0.5cm} &J^A_1 &= \epsilon^2 I^A_{000200200},\\
&\text{Sector 14:} &J^A_2 &= \epsilon^2 \left(\frac{-s}{\mu^2}\right) \mathbf{D^-} I^A_{011100000},\\
&\text{Sector 25:} &J^A_3 &= \epsilon^2 \left(\frac{m^2}{\mu^2}\right) \mathbf{D^-} I^A_{100110000},\\
&\text{Sector 28:} &J^A_4 &= \epsilon^2 \left(\frac{m^2-s}{\mu^2}\right) \mathbf{D^-} I^A_{001110000},\\
& &J^A_5 &= \epsilon^2 \left[\left(\frac{m^2}{\mu^2}\right) \mathbf{D^-} I^A_{001110000} + \mathbf{D^-} I^A_{00111(-1)000}\right],\\
&\text{Sector 73:} &J^A_6 &= \epsilon^2 \left[\left(\frac{r_1}{\mu^2}\right) \mathbf{D^-} I^A_{100100100} + \left(\frac{-t}{\mu^2}\right) I^A_{100200200}\right],\\
& &J^A_7 &= \epsilon^2 \left(\frac{-t}{\mu^2}\right) I^A_{100200200},\\
&\text{Sector 54:} &J^A_8 &= \epsilon^2 \left(\frac{s^2}{\mu^4}\right) \mathbf{D^-} I^A_{011011000},\\
&\text{Sector 57:} &J^A_9 &= \epsilon^3 \left(\frac{-s}{\mu^2}\right) I^A_{100211000},\\
& &J^A_{10} &= \epsilon^3 \left(\frac{-s}{\mu^2}\right) I^A_{200111000},\\
&\text{Sector 78:} &J^A_{11} &= \epsilon^2 \left(\frac{r_2}{\mu^2}\right) \left[\left(\frac{-s}{\mu^2}\right) I^A_{021100200} + \frac{\epsilon}{2(1+2\epsilon)} \left(\frac{\mu^2}{m^2}\right) I^A_{000200200}\right],\\
& &J^A_{12} &= \epsilon^3 \left(\frac{-s}{\mu^2}\right) I^A_{011100200},\\
&\text{Sector 89:} &J^A_{13} &= \epsilon^3 \left(\frac{-t}{\mu^2}\right) I^A_{100210100},\\
&\text{Sector 92:} &J^A_{14} &= \epsilon^2 \left(\frac{-s}{\mu^2}\right) \left[\left(\frac{m^2}{\mu^2}\right) \mathbf{D^-} I^A_{001110100} - I^A_{002120000} + \frac{1}{2} I^A_{002210000}\right],\\
& &J^A_{15} &= \epsilon^3 \left(\frac{-s}{\mu^2}\right) I^A_{001210100},\\
& \text{Sector 120:} &J^A_{16} &= \epsilon^3 \left(\frac{-s}{\mu^2}\right) I^A_{000211100},\\
&\text{Sector 59:} &J^A_{17} &= \epsilon^4 \left(\frac{-s}{\mu^2}\right) I^A_{110111000},\\
&\text{Sector 62:} \hspace{0.5cm} &J^A_{18} &= \epsilon^3 (-1+2\epsilon) \left(\frac{-s}{\mu^2}\right) I^A_{011111000},\\
&\text{Sector 79:} &J^A_{19} &= \epsilon^3 \left(\frac{-s}{\mu^2}\right) \left[I^A_{011100200} - I^A_{101100200} - I^A_{1111002(-1)0}\right],\\
& &J^A_{20} &= \epsilon^3 \left(\frac{-s}{\mu^2}\right) \left(\frac{r_1}{\mu^2}\right) I^A_{111100200},\\
&\text{Sector 91:} &J^A_{21} &= \epsilon^4 \left(\frac{-t}{\mu^2}\right) I^A_{110110100},\\
&\text{Sector 93:} &J^A_{22} &= \epsilon^4 \left(\frac{-s-t}{\mu^2}\right) I^A_{101110100},\\
& &J^A_{23} &= \epsilon^3 \left(\frac{r_3}{\mu^4}\right) I^A_{101210100},\\
& &J^A_{24} &= \epsilon^3 \left[\left(\frac{-s-t}{\mu^2}\right)\left(\frac{m^2}{\mu^2}\right) I^A_{101110200} - \left(\frac{s(2m^2-t)+t m^2}{2\mu^4}\right) I^A_{101210100}\right],\\
&\text{Sector 94:} &J^A_{25} &= \epsilon^4 \left(\frac{-s}{\mu^2}\right) I^A_{011110100},\\
&\text{Sector 118:} &J^A_{26} &= \epsilon^3 \left(\frac{s^2}{\mu^4}\right) I^A_{021011100},\\
&\text{Sector 121:} &J^A_{27} &= \epsilon^3 \left(\frac{-s}{\mu^2}\right) \left(\frac{m^2-t}{\mu^2}\right) I^A_{100211100},\\
& &J^A_{28} &= \epsilon^3 \left(\frac{r_4}{\mu^4}\right) \big[I^A_{200111100} + I^A_{100211100}\big],\\
& &J^A_{29} &= \epsilon^3 \left(\frac{-s}{\mu^2}\right) \Bigg[\Bigg(\frac{2m^2-t}{\mu^2}\Bigg)\big[I^A_{200111100} + I^A_{100211100}\big]\\
& & &\phantom{=} + \big[I^A_{20011110(-1)} + I^A_{10021110(-1)}\big]\Bigg] - \left(\frac{s}{t}\right) J^A_7,\\
&\text{Sector 95:} \hspace{0.5cm} &J^A_{30} &= \epsilon^4 \left(\frac{-s}{\mu^2}\right) \left(\frac{-t}{\mu^2}\right) I^A_{111110100},\\
&\text{Sector 126:} &J^A_{31} &= \epsilon^4 \left(\frac{s^2}{\mu^4}\right) I^A_{011111100},\\
&\text{Sector 127:} &J^A_{32} &= \epsilon^4 \left(\frac{-s}{\mu^2}\right) \left(\frac{r_4}{\mu^4}\right) I^A_{111111100},\\
& &J^A_{33} &= \epsilon^4 \left(\frac{s^2}{\mu^4}\right) \left[\left(\frac{2m^2-t}{\mu^2}\right) I^A_{111111100} + I^A_{11111110(-1)}\right] + \left(\frac{s}{t}\right) J^A_{30},\\
& &J^A_{34} &= \epsilon^4 \left(\frac{s^2}{\mu^4}\right) I^A_{1111111(-1)0} - \left(\frac{2m^2+s}{2(m^2-t)}\right) \left[J^A_3+\frac{3}{2}J^A_4-\frac{5}{2}J^A_5+3J^A_7\right.\\
& & &\phantom{=} \left. +2J^A_9 +4J^A_{13}-\frac{2}{3}J^A_{14} +\frac{2}{3}J^A_{15} + 2J^A_{17} + 2J^A_{21} - 6J^A_{22} + 4J^A_{24} -2 J^A_{27}\right],\\
& &J^A_{35} &= \epsilon^4 \left(\frac{-s}{\mu^2}\right) \left[I^A_{1111111(-1)(-1)} + \left(\frac{-t}{\mu^2}\right)\Bigg(\left(\frac{2m^2-t}{\mu^2}\right) I^A_{111111100}\right.\\
& & &\left. \phantom{=} + I^A_{11111110(-1)} \Bigg) + \left(\frac{m^2-t}{\mu^2}\right) I^A_{1111111(-1)0}\right] - \left(\frac{t+(4s+2t)\epsilon}{s(-1+2\epsilon)}\right) J^A_{18}\\
& & &\phantom{=} + \left(\frac{t}{s}\right) \left[J^A_{10}+2J^A_{25}+J^A_{26}\right] - \left(\frac{s}{t}\right) J^A_{21} - J^A_{30}\\
& & &\phantom{=} - \left(\frac{2tm^2 + s(2m^2+t)}{s^2}\right) J^A_{31}.
\end{alignat*}

\subsubsection{Topology $B$}

\begin{alignat*}{3}
&\text{Sector 65:} \hspace{0.5cm} &J^B_1 &= \epsilon^2 \, \mathbf{D^-} I^B_{100000100},\\
&\text{Sector 25:} &J^B_2 &= \epsilon^2 \left(\frac{m^2}{\mu^2}\right) \mathbf{D^-} I^B_{100110000},\\
&\text{Sector 28:} &J^B_3 &= \epsilon^2 \left(\frac{-s}{\mu^2}\right) \mathbf{D^-} I^B_{001110000},\\
&\text{Sector 49:} &J^B_4 &= \epsilon^2 \left(\frac{-s}{\mu^2}\right) \mathbf{D^-} I^B_{100011000},\\
&\text{Sector 73:} &J^B_5 &= \epsilon^2\left(\frac{r_1}{\mu^2}\right) \mathbf{D^-} I^B_{100100100},\\
& &J^B_6 &= \epsilon^2 \left(\frac{-t}{\mu^2}\right) I^B_{200100200},\\
&\text{Sector 29:} &J^B_7 &= \epsilon^3 \left(\frac{-s}{\mu^2}\right) I^B_{101210000},\\
&\text{Sector 54:} &J^B_8 &= \epsilon^2 \left(\frac{s^2}{\mu^4}\right) \mathbf{D^-} I^B_{011011000},\\
&\text{Sector 57:} &J^B_9 &= \epsilon^3 \left(\frac{-s}{\mu^2}\right) I^B_{100211000},\\
& &J^B_{10} &= \epsilon^3 \left(\frac{-s}{\mu^2}\right) I^B_{200111000},\\
&\text{Sector 71:} &J^B_{11} &= \epsilon^3 \left(\frac{-s}{\mu^2}\right) I^B_{111000200},\\
&\text{Sector 75:} &J^B_{12} &= \epsilon^3 \left(\frac{-t}{\mu^2}\right) I^B_{110100200},\\
&\text{Sector 55:} &J^B_{13} &= \epsilon^3 \left(\frac{s^2}{\mu^4}\right) I^B_{111021000},\\
&\text{Sector 59:} &J^B_{14} &= \epsilon^4 \left(\frac{-s}{\mu^2}\right) I^B_{110111000},\\
&\text{Sector 79:} &J^B_{15} &= \epsilon^3 \left(\frac{-s}{\mu^2}\right) \left(\frac{m^2-t}{\mu^2}\right) I^B_{111100200},\\
& &J^B_{16} &= \epsilon^3 \left(\frac{-s}{\mu^2}\right) \left[\left(\frac{m^2-t}{\mu^2}\right) I^B_{111100200} + I^B_{1111002(-1)0} + I^B_{110100200}\right],\\
& &J^B_{17} &= \epsilon^3 \left(\frac{r_4}{\mu^4}\right) \big[I^B_{111100200} + I^B_{111200100}\big]\\
&\text{Sector 91:} \hspace{0.5cm} &J^B_{18} &= \epsilon^4 \left(\frac{-t}{\mu^2}\right) I^B_{110110100},\\
&\text{Sector 93:} &J^B_{19} &= \epsilon^4 \left(\frac{-s-t}{\mu^2}\right) I^B_{101110100},\\
& &J^B_{20} &= \epsilon^3 \left(\frac{r_4}{\mu^4}\right) I^B_{101210100},\\
& &J^B_{21} &= \epsilon^3 \left(\frac{-s-t}{\mu^2}\right) I^B_{1(-1)1210100} + \epsilon^2 \left(\frac{-s}{\mu^2}\right) I^B_{200100200},\\
& &J^B_{22} &= \epsilon^3 \left(\frac{-s}{\mu^2}\right) \left[\left(\frac{2m^2-t}{\mu^2}\right) I^B_{101210100} + I^B_{1012101(-1)0}\right]\\
& & &\phantom{=} - \epsilon^2 \left(\frac{-s}{\mu^2}\right) I^B_{200100200},\\
&\text{Sector 119:} &J^B_{23} &= \epsilon^4 \left(\frac{s^2}{\mu^4}\right) I^B_{111011100},\\
&\text{Sector 127:} &J^B_{24} &= \epsilon^4 \left(\frac{-s}{\mu^2}\right) \left(\frac{r_4}{\mu^4}\right) I^B_{111111100},\\
& &J^B_{25} &= \epsilon^4 \left(\frac{s^2}{\mu^4}\right) \left[\left(\frac{2m^2-t}{\mu^2}\right) I^B_{111111100} + I^B_{11111110(-1)}\right]\\
& & &\phantom{=} - \left(\frac{2m^2+s}{m^2-t}\right) \left[2J^B_6 - 2J^B_7 + J^B_{10} + 2J^B_{12} + J^B_{14} - J^B_{15} + J^B_{18} - J^B_{19}\right.\\
& & &\phantom{=} \left. + 2J^B_{21} +J^B_{22}\right],\\
& &J^B_{26} &= \epsilon^4 \left(\frac{-s}{\mu^2}\right) \left[I^B_{1111111(-1)(-1)} + 2 \left(\frac{m^2-t}{\mu^2}\right) I^B_{11111110(-1)}\right.\\
& & &\phantom{=} \left. + \left(\frac{-t}{\mu^2}\right) \left(\frac{2m^2-t}{\mu^2}\right) I^B_{111111100}\right] + \left(\frac{2s+4t\epsilon}{3s(1-2\epsilon)}\right) J^B_3\\
& & &\phantom{=} - \left(\frac{2m^2(s+t)+st}{2s^2}\right) \left[8J^B_7 - 4J^B_9 - 4J^B_{13}\right]\\
& & &\phantom{=} - \left(\frac{(s+t)\epsilon}{s(1-2\epsilon)}\right) J^B_8 - \left(\frac{8m^2(s+t)}{s^2}\right) J^B_{14} - \left(\frac{s}{t}\right) J^B_{18} - \left(\frac{t}{s}\right) J^B_{23}.
\end{alignat*}

\subsubsection{Topology $C$}

\begin{alignat*}{3}
&\text{Sector 14:} \hspace{0.5cm} &J^C_1 &= \epsilon^2 \left(\frac{-s}{\mu^2}\right) \mathbf{D^-} I^C_{011100000}\\
&\text{Sector 25:} &J^C_2 &= \epsilon^2 \left(\frac{m^2}{\mu^2}\right) \mathbf{D^-} I^C_{100110000}\\
&\text{Sector 28:} &J^C_3 &= \epsilon^2 \left(\frac{m^2-s}{\mu^2}\right) \mathbf{D^-} I^C_{001110000}\\
& &J^C_4 &= \epsilon^2 \left[\left(\frac{m^2}{\mu^2}\right) \mathbf{D^-} I^C_{001110000} + \mathbf{D^-} I^C_{00111(-1)000}\right]\\
&\text{Sector 73:} &J^C_5 &= \epsilon^2 \left(\frac{m^2-t}{\mu^2}\right) \mathbf{D^-} I^C_{100100100}\\
& &J^C_6 &= \epsilon^2 \left(\frac{-t}{\mu^2}\right) I^C_{100200200}\\
&\text{Sector 54:} &J^C_7 &= \epsilon^2 \left(\frac{s^2}{\mu^4}\right) \mathbf{D^-} I^C_{011011000}\\
&\text{Sector 57:} &J^C_8 &= \epsilon^3 \left(\frac{-s}{\mu^2}\right) I^C_{200111000}\\
& &J^C_9 &= \epsilon^3 \left(\frac{-s}{\mu^2}\right) I^C_{100211000}\\
&\text{Sector 59:} &J^C_{10} &= \epsilon^4 \left(\frac{-s}{\mu^2}\right) I^C_{110111000}\\
&\text{Sector 62:} &J^C_{11} &= \epsilon^3 (-1+2\epsilon) \left(\frac{-s}{\mu^2}\right) I^C_{011111000}\\
&\text{Sector 79:} &J^C_{12} &= \epsilon^3 \left(\frac{-s}{\mu^2}\right) \left(\frac{-t}{\mu^2}\right) I^C_{111200100}\\
& &J^C_{13} &= \epsilon^3 \left(\frac{-s}{\mu^2}\right) I^C_{1112001(-1)0} - \frac{s}{2t} \left(J^C_2 - J^C_5 + 4J^C_6\right)\\
&\text{Sector 91:} &J^C_{14} &= \epsilon^4 \left(\frac{-t}{\mu^2}\right) I^C_{110110100}\\
&\text{Sector 93:} &J^C_{15} &= \epsilon^4 \left(\frac{-s-t}{\mu^2}\right) I^C_{101110100}\\
& &J^C_{16} &= \epsilon^3 \left(\frac{-m^2(s+t) + s t}{\mu^4}\right) I^C_{101210100}\\
&\text{Sector 127:} \hspace{0.5cm} &J^C_{17} &= \epsilon^4 \left(\frac{s^2}{\mu^4}\right) \left(\frac{m^2-t}{\mu^2}\right) I^C_{111111100}\\
& &J^C_{18} &= \epsilon^4 \left(\frac{s^2}{\mu^4}\right) I^C_{1111111(-1)0}\\
& & &\phantom{=} - \left(\frac{s}{2t}\right) \left[J^C_2 - J^C_3 + J^C_4 - J^C_5 + 2J^C_6 - 2J^C_9 - 2J^C_{10} + 2J^C_{12}\right.\\
& & &\phantom{=} \left. - 2J^C_{14} + 6J^C_{15} - 2J^C_{16}\right]\\
& &J^C_{19} &= \epsilon^4 \left(\frac{-s}{\mu^2}\right) \left[I^C_{1111111(-1)(-1)} + \left(\frac{-t}{\mu^2}\right) \left(\frac{m^2-t}{\mu^2}\right) I^C_{111111100}\right.\\
& & &\phantom{=} \left. + 2\left(\frac{-t}{\mu^2}\right) I^C_{1111111(-1)0}\right]\\
& & &\phantom{=} -\left(\frac{t}{4s}\right) \left(J^C_7 - 8J^C_8\right) - \frac{2\frac{t}{s} + 4\epsilon}{-1+2\epsilon} J^C_{11} -\left(\frac{s}{t}\right) J^C_{14}
\end{alignat*}

\subsubsection{Topology $D$}

\begin{alignat*}{3}
&\text{Sector 25:} \hspace{0.5cm} &J^D_1 &= \epsilon^2 \left(\frac{m^2}{\mu^2}\right) \mathbf{D^-} I^D_{100110000},\\
&\text{Sector 28:} &J^D_2 &= \epsilon^2 \left(\frac{-s}{\mu^2}\right) \mathbf{D^-} I^D_{001110000},\\
&\text{Sector 49:} &J^D_3 &= \epsilon^2 \left(\frac{-s}{\mu^2}\right) \mathbf{D^-} I^D_{100011000},\\
&\text{Sector 73:} &J^D_4 &= \epsilon^2 \left(\frac{m^2-t}{\mu^2}\right) \mathbf{D^-} I^D_{100100100},\\
& &J^D_5 &= \epsilon^2 \left(\frac{-t}{\mu^2}\right) I^D_{200100200},\\
&\text{Sector 29:} &J^D_6 &= \epsilon^3 \left(\frac{-s}{\mu^2}\right) I^D_{101210000},\\
&\text{Sector 54:} &J^D_7 &= \epsilon^2 \left(\frac{s^2}{\mu^4}\right) \mathbf{D^-} I^D_{011011000},\\
&\text{Sector 57:} &J^D_8 &= \epsilon^3 \left(\frac{-s}{\mu^2}\right) I^D_{100211000},\\
& &J^D_9 &= \epsilon^3 \left(\frac{-s}{\mu^2}\right) I^D_{200111000},\\
&\text{Sector 78:} &J^D_{10} &= \epsilon^3 \left(\frac{-s}{\mu^2}\right) I^D_{011200100},\\
&\text{Sector 55:} &J^D_{11} &= \epsilon^3 \left(\frac{s^2}{\mu^4}\right) I^D_{111021000},\\
&\text{Sector 59:} &J^D_{12} &= \epsilon^4 \left(\frac{-s}{\mu^2}\right) I^D_{110111000},\\
&\text{Sector 79:} &J^D_{13} &= \epsilon^3 \left(\frac{-s}{\mu^2}\right) \left(\frac{m^2-t}{\mu^2}\right) I^D_{111200100} - J^D_5,\\
& &J^D_{14} &= \epsilon^3 \left(\frac{-s}{\mu^2}\right) \left[\left(\frac{m^2-t}{\mu^2}\right) I^D_{111200100} - I^D_{1112001(-1)0}\right] + \frac{s}{2t} J^D_5,\\
&\text{Sector 93:} &J^D_{15} &= \epsilon^4 \left(\frac{-s-t}{\mu^2}\right) I^D_{101110100},\\
& &J^D_{16} &= \epsilon^3 \left(\frac{-s}{\mu^2}\right) \left(\frac{m^2-t}{\mu^2}\right) I^D_{101210100},\\
&\text{Sector 121:} \hspace{0.5cm} &J^D_{17} &= \epsilon^3 \left(\frac{-s}{\mu^2}\right) \left[\left(\frac{-t}{\mu^2}\right) I^D_{200111100} + I^D_{20011110(-1)}\right]\\
& & &\phantom{=} -\left(\frac{s}{2t}\right) \left[J^D_1 - J^D_4 + 4J^D_5\right],\\
& &J^D_{18} &= \epsilon^3 \left(\frac{-s}{\mu^2}\right) I^D_{20011110(-1)} - \left(\frac{s}{2t}\right) \left[J^D_1 - J^D_4 + 4J^D_5\right],\\
&\text{Sector 127:} &J^D_{19} &= \epsilon^4 \left(\frac{s^2}{\mu^4}\right) \left[\left(\frac{m^2-t}{\mu^2}\right) I^D_{111111100} + I^D_{11111110(-1)}\right]\\
& & &\phantom{=} + \frac{1}{2}\left(\frac{2m^2+s}{m^2-t}\right) \left[2J^D_5 + 2J^D_{13} - J^D_{16}\right],\\
& &J^D_{20} &= \epsilon^4 \left(\frac{s^2}{\mu^4}\right) I^D_{11111110(-1)} + \frac{1}{2} \left(\frac{2m^2+s}{m^2-t}\right) \left[2J^D_5 + 2J^D_{13} - J^D_{16}\right],\\
& &J^D_{21} &= \epsilon^4 \left(\frac{s^2}{\mu^4}\right) \left[\left(\frac{m^2-t}{\mu^2}\right) I^D_{111111100} + I^D_{1111111(-1)0}\right]\\
& & &\phantom{=} - \left(\frac{s}{24t}\right) \left[9J^D_1 - J^D_2 - 9J^D_4 + 36J^D_5 - 12J^D_6 - 24J^D_9 - 24J^D_{12}\right.\\
& & &\phantom{=} \left. - 36J^D_{15} - 6J^D_{16} + 24J^D_{17} - 24J^D_{18}\right].
\end{alignat*}

\subsubsection{Topology $E$}

\begin{alignat*}{3}
&\text{Sector 28:} \hspace{0.5cm} &J^E_1 &= \epsilon^2 \left(\frac{-s}{\mu^2}\right) \mathbf{D^-} I^E_{001110000},\\
&\text{Sector 73:} &J^E_2 &= \epsilon^2 \left(\frac{-t}{\mu^2}\right) \mathbf{D^-} I^E_{100100100},\\
&\text{Sector 54:} &J^E_3 &= \epsilon^2 \left(\frac{s^2}{\mu^4}\right) \mathbf{D^-} I^E_{011011000},\\
&\text{Sector 57:} &J^E_4 &= \epsilon^3 \left(\frac{-s}{\mu^2}\right) I^E_{200111000},\\
&\text{Sector 79:} &J^E_5 &= \epsilon^3 \left(\frac{-s}{\mu^2}\right) \left(\frac{-t}{\mu^2}\right) I^E_{111100200},\\
&\text{Sector 93:} &J^E_6 &= \epsilon^4 \left(\frac{-s-t}{\mu^2}\right) I^E_{101110100},\\
&\text{Sector 127:} &J^E_7 &= \epsilon^4 \left(\frac{s^2}{\mu^4}\right) \left(\frac{-t}{\mu^2}\right) I^E_{111111100},\\
& &J^E_8 &= \epsilon^4 \left(\frac{s^2}{\mu^4}\right) I^E_{1111111(-1)0} - \frac{s}{4t}\left[J^E_1+J^E_2+4J^E_5-12J^E_6\right].
\end{alignat*}


\subsection{The non-planar topologies}
\label{sect:nonplanar}

\subsubsection{Topology $\tilde{A}$}

\begin{alignat*}{3}
&\text{Sector 72:} \hspace{0.5cm} &J^{\tilde{A}}_1 &= \eps^2 \, \mathbf{D^-} I^{\tilde{A}}_{000100100},\\
&\text{Sector 14:} &J^{\tilde{A}}_2 &= \eps^2 \left(\frac{-s}{\mu^2}\right) \mathbf{D^-} I^{\tilde{A}}_{011100000},\\
&\text{Sector 41:} &J^{\tilde{A}}_3 &= \eps^2 \left(\frac{m^2}{\mu^2}\right) \mathbf{D^-} I^{\tilde{A}}_{100101000},\\
&\text{Sector 42:} &J^{\tilde{A}}_4 &= \eps^2 \left(\frac{-s}{\mu^2}\right) I^{\tilde{A}}_{020201000},\\
& &J^{\tilde{A}}_5 &= \eps^2 \left(\frac{m^2-s}{\mu^2}\right) \mathbf{D^-} I^{\tilde{A}}_{010101000},\\
&\text{Sector 49:} &J^{\tilde{A}}_6 &= \eps^2 \left(\frac{-s-t}{\mu^2}\right) \mathbf{D^-} I^{\tilde{A}}_{100011000},\\
&\text{Sector 73:} &J^{\tilde{A}}_7 &= \eps^2 \left(\frac{r_1}{\mu^2}\right) \mathbf{D^-} I^{\tilde{A}}_{100100100},\\
& &J^{\tilde{A}}_8 &= \eps^2 \left(\frac{-t}{\mu^2}\right) I^{\tilde{A}}_{100200200},\\
&\text{Sector 54:} &J^{\tilde{A}}_9 &= \eps^3 \left(\frac{-s}{\mu^2}\right) I^{\tilde{A}}_{011021000},\\
&\text{Sector 57:} &J^{\tilde{A}}_{10} &= \eps^3 \left(\frac{-s-t}{\mu^2}\right) I^{\tilde{A}}_{100112000},\\
&\text{Sector 78:} &J^{\tilde{A}}_{11} &= \eps^2 \left(\frac{r_2}{\mu^2}\right) \left[\left(\frac{-s}{\mu^2}\right) I^{\tilde{A}}_{021100200} + \left(\frac{\mu^2}{m^2}\right) \frac{\eps}{2(1+2\eps)} I^{\tilde{A}}_{000200200}\right],\\
& &J^{\tilde{A}}_{12} &= \eps^3 \left(\frac{-s}{\mu^2}\right) I^{\tilde{A}}_{011100200},\\
&\text{Sector 89:} &J^{\tilde{A}}_{13} &= \eps^3 \left(\frac{-t}{\mu^2}\right) I^{\tilde{A}}_{100110200},\\
&\text{Sector 92:} &J^{\tilde{A}}_{14} &= \eps^3 \left(\frac{-s}{\mu^2}\right) I^{\tilde{A}}_{001110200},\\
& &J^{\tilde{A}}_{15} &= \eps^3 \left(\frac{-s}{\mu^2}\right) I^{\tilde{A}}_{002110100},\\
&\text{Sector 55:} &J^{\tilde{A}}_{16} &= \eps^3 \left(\frac{-s}{\mu^2}\right) \left(\frac{-s-t}{\mu^2}\right) I^{\tilde{A}}_{111021000},\\
&\text{Sector 59:} &J^{\tilde{A}}_{17} &= \eps^4 \left(\frac{-t}{\mu^2}\right) I^{\tilde{A}}_{110111000},\\
& &J^{\tilde{A}}_{18} &= \eps^3 \left(\frac{-t}{\mu^2}\right) \left(\frac{m^2}{\mu^2}\right) I^{\tilde{A}}_{110211000},\\
&\text{Sector 61:} \hspace{0.5cm} &J^{\tilde{A}}_{19} &= \eps^4 \left(\frac{-s-t}{\mu^2}\right) I^{\tilde{A}}_{101111000},\\
&\text{Sector 62:} &J^{\tilde{A}}_{20} &= \eps^4 \left(\frac{-s}{\mu^2}\right) I^{\tilde{A}}_{011111000},\\
& &J^{\tilde{A}}_{21} &= \eps^3 \left(\frac{-s}{\mu^2}\right) \left(\frac{m^2}{\mu^2}\right) I^{\tilde{A}}_{011211000},\\ 
&\text{Sector 79:} &J^{\tilde{A}}_{22} &= \eps^3 \left(\frac{-s}{\mu^2}\right) \left(\frac{r_1}{\mu^2}\right) I^{\tilde{A}}_{111100200},\\
& &J^{\tilde{A}}_{23} &= \eps^3 \left(\frac{-s}{\mu^2}\right) I^{\tilde{A}}_{1111002(-1)0} - \left(\frac{s}{2t}\right) J^{\tilde{A}}_8,\\
&\text{Sector 91:} &J^{\tilde{A}}_{24} &= \eps^4 \left(\frac{-t}{\mu^2}\right) I^{\tilde{A}}_{110110100},\\
&\text{Sector 93:} &J^{\tilde{A}}_{25} &= \eps^4 \left(\frac{-s-t}{\mu^2}\right) I^{\tilde{A}}_{101110100},\\
& &J^{\tilde{A}}_{26} &= \eps^3 \left(\frac{r_3}{\mu^2}\right) I^{\tilde{A}}_{101110200},\\
& &J^{\tilde{A}}_{27} &= \eps^3 \left[\left(\frac{-s-t}{\mu^2}\right) \left(\frac{m^2}{\mu^2}\right) I^{\tilde{A}}_{101210100} + \frac{1}{2}\left(\frac{st - m^2(2s+t)}{\mu^2} I^{\tilde{A}}_{101110200}\right)\right],\\
&\text{Sector 94:} &J^{\tilde{A}}_{28} &= \eps^4 \left(\frac{-s}{\mu^2}\right) I^{\tilde{A}}_{011110100},\\
&\text{Sector 121:} &J^{\tilde{A}}_{29} &= \eps^4 \left(\frac{-s}{\mu^2}\right) I^{\tilde{A}}_{100111100},\\
& &J^{\tilde{A}}_{30} &= \eps^3 \left(\frac{r_6}{\mu^4}\right) I^{\tilde{A}}_{200111100},\\
& &J^{\tilde{A}}_{31} &= \eps^3 \left(\frac{-s}{\mu^2}\right) \left(\frac{m^2}{\mu^2}\right) I^{\tilde{A}}_{100211100},\\
& &J^{\tilde{A}}_{32} &= \eps^3 \left(\frac{m^2}{\mu^2}\right) \left[\left(\frac{-s-t}{\mu^2}\right) I^{\tilde{A}}_{100121100} + \left(\frac{-t}{\mu^2}\right) I^{\tilde{A}}_{100211100}\right],\\
&\text{Sector 122:} &J^{\tilde{A}}_{33} &= \eps^4 \left(\frac{-s}{\mu^2}\right) I^{\tilde{A}}_{010111100},\\
& &J^{\tilde{A}}_{34} &= \eps^3 \left(\frac{-s}{\mu^2}\right) \left(\frac{m^2}{\mu^2}\right) I^{\tilde{A}}_{020111100},\\
&\text{Sector 63:} &J^{\tilde{A}}_{35} &= \eps^4 \left(\frac{-s}{\mu^2}\right) \left(\frac{-s-t}{\mu^2}\right) I^{\tilde{A}}_{111111000},\\
& &J^{\tilde{A}}_{36} &= \eps^4 \left(\frac{-s}{\mu^2}\right) I^{\tilde{A}}_{1111110(-1)0},\\
&\text{Sector 95:} \hspace{0.5cm} &J^{\tilde{A}}_{37} &= \eps^4 \left(\frac{-s}{\mu^2}\right) \left(\frac{-t}{\mu^2}\right) I^{\tilde{A}}_{111110100},\\
&\text{Sector 123:} &J^{\tilde{A}}_{38} &= \eps^4 \frac{\pi}{\psi^{(b)}_1} \left(\frac{-s}{\mu^2}\right) I^{\tilde{A}}_{110111100},\\
& &J^{\tilde{A}}_{39} &= \eps^4 \left(\frac{-s-t}{\mu^2}\right) I^{\tilde{A}}_{1101111(-1)0} + F^{\tilde{A}}_{39,38} J^{\tilde{A}}_{38} - \left(\frac{t}{s}\right) J^{\tilde{A}}_{33},\\
 & &J^{\tilde{A}}_{40} &= \eps^4 \Bigg[\bigg(\frac{-t}{\mu^2}\bigg) I^{\tilde{A}}_{11(-1)111100} + \bigg(\frac{-s}{\mu^2}\bigg) I^{\tilde{A}}_{1101111(-1)0} - \bigg(\frac{-t}{\mu^2}\bigg) I^{\tilde{A}}_{100111100}\Bigg],\\
& &J^{\tilde{A}}_{41} &= \frac{\big(\psi^{(b)}_1\big)^2}{2 \pi i W^{(b)}_{m^2} \eps} \frac{\partial}{\partial m^2} J^{\tilde{A}}_{38} + F^{\tilde{A}}_{41,38} J^{\tilde{A}}_{38}\\
  & & &\phantom{=} + \frac{1}{4} F^{\tilde{A}}_{41,40} \Big[-12 J^{\tilde{A}}_3 - 53J^{\tilde{A}}_4 + 12J^{\tilde{A}}_5 - 35J^{\tilde{A}}_8 - 59J^{\tilde{A}}_{10} + 7J^{\tilde{A}}_{13} - 26J^{\tilde{A}}_{14}\\
  & & &\phantom{=} + 20J^{\tilde{A}}_{15} + 18J^{\tilde{A}}_{17} + 10J^{\tilde{A}}_{18} + 26J^{\tilde{A}}_{19} - 26J^{\tilde{A}}_{24} + 18J^{\tilde{A}}_{25} - 46J^{\tilde{A}}_{27} - 120J^{\tilde{A}}_{29}\\
  & & &\phantom{=} + 104J^{\tilde{A}}_{31} + 51J^{\tilde{A}}_{32} + 38J^{\tilde{A}}_{33} - 10J^{\tilde{A}}_{34} - 84J^{\tilde{A}}_{39} + 60J^{\tilde{A}}_{40}\Big]\\
  & & &\phantom{=} + F^{\tilde{A}}_{41,26} J^{\tilde{A}}_{26} + F^{\tilde{A}}_{41,30} J^{\tilde{A}}_{30},\\
&\text{Sector 126:} \hspace{0.5cm} &J^{\tilde{A}}_{42} &= \eps^4 \frac{\pi}{\psi^{(c)}_1} \bigg(\frac{-s}{\mu^2}\bigg) I^{\tilde{A}}_{011111100},\\
& &J^{\tilde{A}}_{43} &= \eps^4 \bigg(\frac{s^2}{\mu^4}\bigg) \bigg(\frac{\mu^2}{-s-2t}\bigg) I^{\tilde{A}}_{(-1)11111100} + \frac{1}{3}\Bigg(2\frac{m^2}{\mu^2} - \frac{s(2s+t)}{s+2t}\Bigg) \frac{\psi^{(c)}_1}{\pi} J^{\tilde{A}}_{42}\\
  & & &\phantom{=} - \left(\frac{s}{s+2t}\right) J^{\tilde{A}}_{33},\\
& &J^{\tilde{A}}_{44} &= \frac{\big(\psi^{(c)}_1\big)^2}{2\pi i W^{(c)}_{m^2} \eps} \frac{\partial}{\partial m^2} J^{\tilde{A}}_{42} + \frac{\big(\psi^{(c)}_1\big)^2}{\pi} \frac{76 m^4 - 44m^2 s - 5s^2 }{6 i m^2 \big(8m^4 - 7m^2 s - s^2\big) W^{(c)}_{m^2}} J^{\tilde{A}}_{42},\\
&\text{Sector 127} \hspace{0.5cm} &J^{\tilde{A}}_{45} &= \eps^4 \frac{\pi}{\psi^{(c)}_1}  
  \left[  \left(\frac{-s}{\mu^2}\right) I^{\tilde{A}}_{1111111(-1)0} - \frac{\left(s+2t\right)}{\mu^2} I^{\tilde{A}}_{110111100} \right],\\
& &J^{\tilde{A}}_{46} &= \eps^4 \left(\frac{-s}{\mu^2}\right) \left(\frac{r_6}{\mu^2}\right) I^{\tilde{A}}_{111111100} + F^{\tilde{A}}_{46,45} \left[J^{\tilde{A}}_{45} - J^{\tilde{A}}_{42}\right] + F^{\tilde{A}}_{46,38} J^{\tilde{A}}_{38},\\
& &J^{\tilde{A}}_{47} &= \eps^4 \left(\frac{-s}{\mu^2}\right) I_{1111111(-2)0} + \frac{\psi^{(c)}_1}{\pi} \frac{2m^2 - 2s - 3t}{3 \mu^2} J^{\tilde{A}}_{45}
 - 2\left(\frac{t}{s}\right) J^{\tilde{A}}_{43} \\
& & &\phantom{=} + \frac{\psi^{(c)}_1}{\pi} \frac{4(m^2-s)(s+t)}{3s} J^{\tilde{A}}_{42} - \left(\frac{s}{t}\right) J^{\tilde{A}}_{40} + \left(\frac{s}{t} + \frac{t}{s+t}\right) J^{\tilde{A}}_{39}\\
& & &\phantom{=} + F^{\tilde{A}}_{47,38} J^{\tilde{A}}_{38} + \frac{s}{s+t} J^{\tilde{A}}_{33},\\
& &J^{\tilde{A}}_{48} &= \frac{\left(\psi^{(c)}_1\right)^2}{2\pi i W^{(c)}_{m^2} \eps} \frac{\partial}{\partial m^2} J^{\tilde{A}}_{45} 
 + F^{\tilde{A}}_{48,46} J^{\tilde{A}}_{46}
 + F^{\tilde{A}}_{48,45} J^{\tilde{A}}_{45} 
 + F^{\tilde{A}}_{48,42} J^{\tilde{A}}_{42}
 + F^{\tilde{A}}_{48,38} J^{\tilde{A}}_{38}
 + F^{\tilde{A}}_{48,30} J^{\tilde{A}}_{30}.
\end{alignat*}
The functions
$F^{\tilde{A}}_{39,38}$,
$F^{\tilde{A}}_{41,30}$,
$F^{\tilde{A}}_{41,26}$,
$F^{\tilde{A}}_{46,45}$,
$F^{\tilde{A}}_{48,42}$,
$F^{\tilde{A}}_{46,38}$,
$F^{\tilde{A}}_{47,38}$,
$F^{\tilde{A}}_{48,38}$ and
$F^{\tilde{A}}_{48,30}$
are determined by a triangular system of first-order differential equations.
The functions
$F^{\tilde{A}}_{41,40}$,
$F^{\tilde{A}}_{41,38}$,
$F^{\tilde{A}}_{48,46}$ and
$F^{\tilde{A}}_{48,45}$
are related algebraically to the ones above.
As the explicit expressions are rather long, these functions are given in the supplementary electronic file
attached to the arxiv version of this article, see appendix~\ref{sect:supplement}.

\subsubsection{Topology $\tilde{B}$}

\begin{alignat*}{3}
&\text{Sector 65:} \hspace{0.5cm} &J^{\tilde{B}}_1 &= \eps^2 I^{\tilde{B}}_{200000200},\\
&\text{Sector 41:} &J^{\tilde{B}}_2 &= \eps^2 \left(\frac{m^2}{\mu^2}\right) \mathbf{D^-}I^{\tilde{B}}_{100101000},\\
&\text{Sector 42:} &J^{\tilde{B}}_3 &= \eps^2 \left(\frac{-s}{\mu^2}\right) \mathbf{D^-}I^{\tilde{B}}_{010101000},\\
&\text{Sector 49:} &J^{\tilde{B}}_4 &= \eps^2 \left(\frac{m^2+s+t}{\mu^2}\right) \mathbf{D^-}I^{\tilde{B}}_{100011000} ,\\
& &J^{\tilde{B}}_5 &= \eps^2 \left[\left(\frac{m^2}{\mu^2}\right)  \mathbf{D^-}I^{\tilde{B}}_{100011000} + \mathbf{D^-}I^{\tilde{B}}_{10001100(-1)}\right],\\
&\text{Sector 70:} &J^{\tilde{B}}_6 &= \eps^2 \left(\frac{-s}{\mu^2}\right) \mathbf{D^-}I^{\tilde{B}}_{011000100},\\
&\text{Sector 73:} &J^{\tilde{B}}_7 &= \eps^2 \left(\frac{r_1}{\mu^2}\right) \mathbf{D^-}I^{\tilde{B}}_{100100100} ,\\
& &J^{\tilde{B}}_8 &= \eps^2 \left(\frac{-t}{\mu^2}\right) I^{\tilde{B}}_{200100200},\\
&\text{Sector 84:} &J^{\tilde{B}}_9 &= \eps^2 \left(\frac{m^2-s}{\mu^2}\right) \mathbf{D^-}I^{\tilde{B}}_{001010100},\\
& &J^{\tilde{B}}_{10} &= \eps^2 \left(\frac{-s}{\mu^2}\right) I^{\tilde{B}}_{001020200},\\
&\text{Sector 43:} &J^{\tilde{B}}_{11} &= \eps^3 \left(\frac{-s}{\mu^2}\right) I^{\tilde{B}}_{110201000},\\
&\text{Sector 54:} &J^{\tilde{B}}_{12} &= \eps^3 \left(\frac{-s}{\mu^2}\right) I^{\tilde{B}}_{011021000},\\
&\text{Sector 71:} &J^{\tilde{B}}_{13} &= \eps^3 \left(\frac{-s}{\mu^2}\right) I^{\tilde{B}}_{111000200},\\
&\text{Sector 75:} &J^{\tilde{B}}_{14} &= \eps^3 \left(\frac{-t}{\mu^2}\right) I^{\tilde{B}}_{110100200},\\
&\text{Sector 78:} &J^{\tilde{B}}_{15} &= \eps^3 \left(\frac{-s}{\mu^2}\right) I^{\tilde{B}}_{011200100},\\
& &J^{\tilde{B}}_{16} &= \eps^3 \left(\frac{-s}{\mu^2}\right) I^{\tilde{B}}_{011100200},\\
&\text{Sector 85:} \hspace{0.5cm} &J^{\tilde{B}}_{17} &= \eps^3 \left(\frac{-s}{\mu^2}\right) I^{\tilde{B}}_{101020100},\\
& &J^{\tilde{B}}_{18} &= \eps^3 \left(\frac{-s}{\mu^2}\right) I^{\tilde{B}}_{101010200},\\
&\text{Sector 113:} &J^{\tilde{B}}_{19} &= \eps^3 \left(\frac{-s-t}{\mu^2}\right) I^{\tilde{B}}_{200011100},\\
& &J^{\tilde{B}}_{20} &= \eps^3 \left(\frac{-s-t}{\mu^2}\right) I^{\tilde{B}}_{100021100},\\
&\text{Sector 55:} &J^{\tilde{B}}_{21} &= \eps^3 (1 - 2 \eps) \left(\frac{-s}{\mu^2}\right) I^{\tilde{B}}_{111011000},\\
& &J^{\tilde{B}}_{22} &= \eps^3 \left(\frac{-s}{\mu^2}\right) \left(\frac{m^2+s+t}{\mu^2}\right) I^{\tilde{B}}_{111021000},\\
&\text{Sector 59:} &J^{\tilde{B}}_{23} &= \eps^4 \left(\frac{-t}{\mu^2}\right) I^{\tilde{B}}_{110111000},\\
& &J^{\tilde{B}}_{24} &= \eps^3 \left(\frac{-t}{\mu^2}\right) \left(\frac{m^2}{\mu^2}\right) I^{\tilde{B}}_{210111000},\\
&\text{Sector 79:} &J^{\tilde{B}}_{25} &= \eps^3 \left(\frac{-s}{\mu^2}\right) \left(\frac{m^2-t}{\mu^2}\right) I^{\tilde{B}}_{111100200},\\
& &J^{\tilde{B}}_{26} &= \eps^3 \left(\frac{-s}{\mu^2}\right) \left[\left(\frac{m^2-t}{\mu^2}\right) I^{\tilde{B}}_{111100200} + I^{\tilde{B}}_{1111002(-1)0} + I^{\tilde{B}}_{110100200}\right],\\
& &J^{\tilde{B}}_{27} &= \eps^3 \left(\frac{r_4}{\mu^2}\right) \left[I^{\tilde{B}}_{111100200}+I^{\tilde{B}}_{111200100}\right],\\
&\text{Sector 91:} &J^{\tilde{B}}_{28} &= \eps^4 \left(\frac{-t}{\mu^2}\right) I^{\tilde{B}}_{110110100},\\
&\text{Sector 93:} &J^{\tilde{B}}_{29} &= \eps^4 \left(\frac{-s-t}{\mu^2}\right) I^{\tilde{B}}_{101110100},\\
& &J^{\tilde{B}}_{30} &= \eps^3 \left(\frac{r_3}{\mu^4}\right) I^{\tilde{B}}_{101110200},\\
& &J^{\tilde{B}}_{31} &= \eps^3 \left(\frac{m^2}{\mu^2}\right) \left(\frac{-s-t}{\mu^2}\right) I^{\tilde{B}}_{201110100} - \left(\frac{-s(-2m^2 + t) + m^2 t}{2 r_3}\right) J^{\tilde{B}}_{30},\\
&\text{Sector 94:} &J^{\tilde{B}}_{32} &= \eps^4 \left(\frac{-s}{\mu^2}\right) I^{\tilde{B}}_{011110100},\\
&\text{Sector 107:} \hspace{0.5cm} &J^{\tilde{B}}_{33} &= \eps^4 \left(\frac{-s-t}{\mu^2}\right) I^{\tilde{B}}_{110101100},\\
& &J^{\tilde{B}}_{34} &= \eps^3 \left(\frac{r_4}{\mu^4}\right) I^{\tilde{B}}_{110201100},\\
& &J^{\tilde{B}}_{35} &= \eps^3 \left(\frac{m^2}{\mu^2}\right) \left(\frac{-s-t}{\mu^2}\right) I^{\tilde{B}}_{110101200},\\
& &J^{\tilde{B}}_{36} &= \eps^3 \left(\frac{m^2}{\mu^2}\right) \left[\left(\frac{-s}{\mu^2}\right) I^{\tilde{B}}_{110102100} - \left(\frac{-t}{\mu^2}\right) I^{\tilde{B}}_{110101200}\right],\\
&\text{Sector 109:} &J^{\tilde{B}}_{37} &= \eps^4 \left(\frac{-t}{\mu^2}\right) I^{\tilde{B}}_{101101100},\\
&\text{Sector 110:} &J^{\tilde{B}}_{38} &= \eps^4 \left(\frac{-s}{\mu^2}\right) I^{\tilde{B}}_{011101100},\\
&\text{Sector 115:} &J^{\tilde{B}}_{39} &= \eps^4 \left(\frac{-s-t}{\mu^2}\right) I^{\tilde{B}}_{110011100} ,\\
& &J^{\tilde{B}}_{40} &= \eps ^3 \left(\frac{m^2}{\mu^2}\right) \left(\frac{-s-t}{\mu^2}\right) I^{\tilde{B}}_{110021100},\\
&\text{Sector 117:} &J^{\tilde{B}}_{41} &= \eps^4 \left(\frac{-t}{\mu^2}\right) I^{\tilde{B}}_{101011100},\\
& &J^{\tilde{B}}_{42} &= \eps^3 \left(\frac{-t}{\mu^2}\right) \left(\frac{m^2}{\mu^2}\right) I^{\tilde{B}}_{201011100},\\
& &J^{\tilde{B}}_{43} &= \eps^3 \left(\frac{-m^2 t+s^2+s t}{\mu^2}\right) I^{\tilde{B}}_{101021100} ,\\
& &J^{\tilde{B}}_{44} &= \eps^3 \left(\frac{-t}{\mu^2}\right) \left(\frac{m^2}{\mu^2}\right) I^{\tilde{B}}_{101011200},\\
&\text{Sector 118:} &J^{\tilde{B}}_{45} &= \eps^4 \left(\frac{-s}{\mu^2}\right) I^{\tilde{B}}_{011011100},\\
& &J^{\tilde{B}}_{46} &= \eps^3 \left(\frac{-s}{\mu^2}\right) \left(\frac{m^2}{\mu^2}\right) I^{\tilde{B}}_{011021100},\\
&\text{Sector 121:} &J^{\tilde{B}}_{47} &= \eps^4 \left(\frac{-s}{\mu^2}\right) I^{\tilde{B}}_{100111100},\\
& &J^{\tilde{B}}_{48} &= \eps^3 \left(\frac{r_7}{\mu^4}\right) I^{\tilde{B}}_{200111100},\\
& &J^{\tilde{B}}_{49} &= \eps^3 \left[\left(\frac{m^2}{\mu^2}\right) \left(\frac{-s}{\mu^2}\right) I^{\tilde{B}}_{100111200}\right.\\
& & &\phantom{=} \left. - \frac{1}{2} \left(\frac{2 m^2 s+m^2 t-s t-t^2}{\mu^4}\right) I^{\tilde{B}}_{200111100}\right],\\
&\text{Sector 122:} &J^{\tilde{B}}_{50} &= \eps^4 \left(\frac{-s}{\mu^2}\right) I^{\tilde{B}}_{010111100},\\
&\text{Sector 119:} \hspace{0.5cm} &J^{\tilde{B}}_{51} &= \eps^4 \left(\frac{-s}{\mu^2}\right) \left(\frac{m^2+s+t}{\mu^2}\right) I^{\tilde{B}}_{111011100},\\
& &J^{\tilde{B}}_{52} &= \eps^3 \left(\frac{r_8}{\mu^4}\right) \left(\frac{m^2}{\mu^2}\right) \left[2 I^{\tilde{B}}_{111021100} + I^{\tilde{B}}_{111011200}\right]\\
& & &\phantom{=} + \left(\frac{r_8}{10s (m^2+s+t)}\right) \left[-5J^{\tilde{B}}_2 + 6J^{\tilde{B}}_4 - 6J^{\tilde{B}}_5 + 5J^{\tilde{B}}_9 - 18J^{\tilde{B}}_{10}\right.\\
& & &\phantom{=} \left. + 4J^{\tilde{B}}_{17} + 8J^{\tilde{B}}_{18} - 28J^{\tilde{B}}_{19} - 24J^{\tilde{B}}_{20} + 20J^{\tilde{B}}_{22} - 20J^{\tilde{B}}_{39} - 20J^{\tilde{B}}_{40} - 36J^{\tilde{B}}_{41}\right.\\
& & &\phantom{=} \left. + 24J^{\tilde{B}}_{42} + \left(\frac{4(4s(s+t)+m^2(5s+t))}{s(s+t)-m^2 t}\right) J^{\tilde{B}}_{43} + 4J^{\tilde{B}}_{44}\right],\\
& &J^{\tilde{B}}_{53} &= \eps^4 \left(\frac{-s}{\mu^2}\right) \left[2\left(\frac{-s-t}{\mu^2}\right) I^{\tilde{B}}_{111011100} - I^{\tilde{B}}_{011011100} + I^{\tilde{B}}_{1110111(-1)0}\right.\\
& & &\phantom{=} \left.  + I^{\tilde{B}}_{101011100}\right] + \eps^4 \left(\frac{-t}{\mu^2}\right) I^{\tilde{B}}_{110011100},\\
&\text{Sector 123:} &J^{\tilde{B}}_{54} &= \eps^4 \frac{\pi}{\psi^{(a)}_1} \left(\frac{-s}{\mu^2}\right) I^{\tilde{B}}_{110111100},\\
& &J^{\tilde{B}}_{55} &= \eps^4 \left[\left(\frac{-s-t}{\mu^2}\right) I^{\tilde{B}}_{1101111(-1)0} - \left(\frac{-t}{\mu^2}\right) I^{\tilde{B}}_{010111100}\right] + F^{\tilde{B}}_{55,54} J^{\tilde{B}}_{54},\\
& &J^{\tilde{B}}_{56} &= \eps^4 \left(\frac{-t}{\mu^2}\right) \left[I^{\tilde{B}}_{11(-1)111100} - I^{\tilde{B}}_{1101111(-1)0} + I^{\tilde{B}}_{010111100}\right] + F^{\tilde{B}}_{56,54} J^{\tilde{B}}_{54}\\
& & &\phantom{=} - \left(\frac{t}{s}\right) J^{\tilde{B}}_{47},\\
& &J^{\tilde{B}}_{57} &= \frac{\left(\psi^{(a)}_1\right)^2}{2\pi i W^{(a)}_{m^2} \eps} \frac{\partial}{\partial m^2} J^{\tilde{B}}_{54} + \frac{1}{8} F^{\tilde{B}}_{57,55} \left[5J^{\tilde{B}}_4 - 5J^{\tilde{B}}_5 - 2J^{\tilde{B}}_8 - 22J^{\tilde{B}}_{11} - 14J^{\tilde{B}}_{14}\right.\\
& & &\phantom{=} \left. - 4J^{\tilde{B}}_{19} - 8J^{\tilde{B}}_{20} + 36J^{\tilde{B}}_{23} - 20J^{\tilde{B}}_{24} - 20J^{\tilde{B}}_{28} + 16J^{\tilde{B}}_{35} + 6J^{\tilde{B}}_{36} - 20J^{\tilde{B}}_{39}\right.\\
& & &\phantom{=} \left. - 20J^{\tilde{B}}_{40} - 12J^{\tilde{B}}_{47} + 4J^{\tilde{B}}_{49} - 44J^{\tilde{B}}_{50} + 48J^{\tilde{B}}_{55} + 24J^{\tilde{B}}_{56}\right]\\
& & &\phantom{=} - F^{\tilde{B}}_{57,34} J^{\tilde{B}}_{34} - F^{\tilde{B}}_{57,48} J^{\tilde{B}}_{48} + F^{\tilde{B}}_{57,54} J^{\tilde{B}}_{54},\\
&\text{Sector 125:} &J^{\tilde{B}}_{58} &= \eps^4 \left(\frac{-t}{\mu^2}\right) \left(\frac{m^2}{\mu^2}\right) I^{\tilde{B}}_{101111100},\\
& &J^{\tilde{B}}_{59} &= \eps^4 \left(\frac{-t}{\mu^2}\right) \left[\left(\frac{-s-t}{\mu^2}\right) I^{\tilde{B}}_{101111100} + I^{\tilde{B}}_{1011111(-1)0}\right]
 -\frac{t}{4s} \Big[J^{\tilde{B}}_{2}-J^{\tilde{B}}_{9}+4J^{\tilde{B}}_{10}\Big],\\
&\text{Sector 126:} &J^{\tilde{B}}_{60} &= \eps^4 \left(\frac{-s}{\mu^2}\right) \left(\frac{m^2+s}{\mu^2}\right) I^{\tilde{B}}_{011111100},\\
&\text{Sector 127:} \hspace{0.5cm} &J^{\tilde{B}}_{61} &= \eps^4 \left(\frac{r_4}{\mu^4}\right) \left[\left(\frac{m^2-s-t}{\mu^2}\right) I^{\tilde{B}}_{111111100} + I^{\tilde{B}}_{1111111(-1)0} - I^{\tilde{B}}_{011111100}\right]\\
& & &\phantom{=} + F^{\tilde{B}}_{61,54} J^{\tilde{B}}_{54},\\
& &J^{\tilde{B}}_{62} &= \eps^4 \left(\frac{-s}{\mu^2}\right)\Bigg\{\left(\frac{m^2+s+t}{\mu^2}\right)\Bigg[\left(\frac{2m^2-t}{\mu^2}\right) I^{\tilde{B}}_{111111100}\\
& & &\phantom{=} + I^{\tilde{B}}_{1111111(-1)0} + I^{\tilde{B}}_{101111100}\Bigg] + \left(\frac{-t}{\mu^2}\right) I^{\tilde{B}}_{011111100}\Bigg\} - F^{\tilde{B}}_{62,54} J^{\tilde{B}}_{54},\\
& &J^{\tilde{B}}_{63} &= \eps^4 \left(\frac{-s}{\mu^2}\right) \Bigg[\left(\frac{-t}{\mu^2}\right) I^{\tilde{B}}_{1111111(-1)0} + \left(\frac{-s}{\mu^2}\right) I^{\tilde{B}}_{11111110(-1)}\\
& & &\phantom{=} - \left(\frac{2m^2-t}{\mu^2}\right) \left(\frac{s+t}{\mu^2}\right) I^{\tilde{B}}_{111111100} - \left(\frac{m^2-t}{\mu^2}\right)I^{\tilde{B}}_{011111100}\Bigg]\\
& & &\phantom{=} - \left(\frac{s(s+t)}{m^2 t}\right) J^{\tilde{B}}_{58} + F^{\tilde{B}}_{63,54} J^{\tilde{B}}_{54}\\
& & &\phantom{=} + \frac{1}{12}\left(\frac{m^2}{m^2+s+t}\right) \Big[12J^{\tilde{B}}_{51} - 24J^{\tilde{B}}_{24} + 36J^{\tilde{B}}_{23} + 12J^{\tilde{B}}_{22} + 3J^{\tilde{B}}_4 + J^{\tilde{B}}_3\\
& & &\phantom{=} - 3J^{\tilde{B}}_2\Big] + \left(\frac{s+t}{m^2-t}\right) \Big[J^{\tilde{B}}_{38} + J^{\tilde{B}}_{37} - J^{\tilde{B}}_{36} - 3J^{\tilde{B}}_{35} + 3J^{\tilde{B}}_{33}\Big]\\
& & &\phantom{=} + \frac{1}{2}\left(\frac{t}{m^2-t}\right) \Big[-2J^{\tilde{B}}_{32} - 4J^{\tilde{B}}_{31} + 6J^{\tilde{B}}_{29} - 2J^{\tilde{B}}_{28} + 2J^{\tilde{B}}_{17} - 4J^{\tilde{B}}_{10}\\
& & &\phantom{=} + J^{\tilde{B}}_9 - 3J^{\tilde{B}}_8 - J^{\tilde{B}}_2\Big]\\
& & &\phantom{=} + \left(\frac{s}{m^2-t}\right) \Big[-J^{\tilde{B}}_{25} + J^{\tilde{B}}_{16} + 2J^{\tilde{B}}_{14}\Big] + \left(\frac{(2m^2+s)(s+t)}{(m^2-t)(m^2+s+t)}\right) J^{\tilde{B}}_{11},\\
& &J^{\tilde{B}}_{64} &= \eps^4 \left(\frac{-s}{\mu^2}\right) \Bigg[I^{\tilde{B}}_{1111111(-2)0} + \left(\frac{4m^2-t}{\mu^2}\right) I^{\tilde{B}}_{1111111(-1)0}\\
& & &\phantom{=} + \left(\frac{2m^2}{\mu^2}\right) \left(\frac{2m^2-t}{\mu^2}\right) I^{\tilde{B}}_{111111100} + \left(\frac{m^2-s}{\mu^2}\right) \left(\frac{t}{s}\right) I^{\tilde{B}}_{011111100}\Bigg]\\
& & &\phantom{=} + \left(\frac{s}{t}\right) J^{\tilde{B}}_{59} + \left(\frac{s(2m^2+s+t)}{m^2 t}\right) J^{\tilde{B}}_{58} - \frac{1}{2}\left(\frac{s-t}{s+t}\right) \Big[J^{\tilde{B}}_{55} - J^{\tilde{B}}_{50}\Big]\\
& & &\phantom{=} - F^{\tilde{B}}_{64,54} J^{\tilde{B}}_{54} + \frac{1}{12}\left(\frac{t}{s}\right) \Big[12J^{\tilde{B}}_{38} + 12J^{\tilde{B}}_{32} + 12J^{\tilde{B}}_{16} + 12J^{\tilde{B}}_{12} + 12J^{\tilde{B}}_{10}\\
& & &\phantom{=} - 3J^{\tilde{B}}_9 + J^{\tilde{B}}_3 + 3J^{\tilde{B}}_2\Big].
\end{alignat*}
The functions
$F^{\tilde{B}}_{55,54}$,
$F^{\tilde{B}}_{56,54}$,
$F^{\tilde{B}}_{57,48}$,
$F^{\tilde{B}}_{57,34}$,
$F^{\tilde{B}}_{61,54}$,
$F^{\tilde{B}}_{62,54}$ and
$F^{\tilde{B}}_{64,54}$
are determined by a triangular system of first-order differential equations.
The functions
$F^{\tilde{B}}_{57,55}$,
$F^{\tilde{B}}_{57,54}$ and
$F^{\tilde{B}}_{63,54}$
are related algebraically to the ones above.
As the explicit expressions are rather long, these functions are given in the supplementary electronic file
attached to the arxiv version of this article, see appendix~\ref{sect:supplement}.

\subsubsection{Topology $\tilde{C}$}

\begin{alignat*}{3}
&\text{Sector 14:} \hspace{0.5cm} &J^{\tilde{C}}_1 &= \eps^2 \left(\frac{-s}{\mu^2}\right) \mathbf{D^-} I^{\tilde{C}}_{011100000},\\
&\text{Sector 41:} &J^{\tilde{C}}_2 &= \eps^2 \left(\frac{m^2}{\mu^2}\right) \mathbf{D^-} I^{\tilde{C}}_{100101000},\\
&\text{Sector 42:} &J^{\tilde{C}}_3 &= \eps^2 \left(\frac{-s}{\mu^2}\right) I^{\tilde{C}}_{020201000},\\
& &J^{\tilde{C}}_4 &= \eps^2 \left(\frac{m^2-s}{\mu^2}\right) \mathbf{D^-} I^{\tilde{C}}_{010101000},\\
&\text{Sector 49:} &J^{\tilde{C}}_5 &= \eps^2 \left(\frac{-s-t}{\mu^2}\right) \mathbf{D^-} I^{\tilde{C}}_{100011000},\\
&\text{Sector 73:} &J^{\tilde{C}}_6 &= \eps^2 \left(\frac{m^2-t}{\mu^2}\right) \mathbf{D^-} I^{\tilde{C}}_{100100100},\\
& &J^{\tilde{C}}_7 &= \eps^2 \left(\frac{-t}{\mu^2}\right) I^{\tilde{C}}_{100200200},\\
&\text{Sector 84:} &J^{\tilde{C}}_8 &= \eps^2 \left(\frac{-s}{\mu^2}\right) \mathbf{D^-} I^{\tilde{C}}_{001010100},\\
&\text{Sector 54:} &J^{\tilde{C}}_9 &= \eps^3 \left(\frac{-s}{\mu^2}\right) I^{\tilde{C}}_{011021000},\\
&\text{Sector 57:} &J^{\tilde{C}}_{10} &= \eps^3 \left(\frac{-s-t}{\mu^2}\right) I^{\tilde{C}}_{100112000},\\
&\text{Sector 78:} &J^{\tilde{C}}_{11} &= \eps^3 \left(\frac{-s}{\mu^2}\right) I^{\tilde{C}}_{011200100},\\
& &J^{\tilde{C}}_{12} &= \eps^3 \left(\frac{-s}{\mu^2}\right) I^{\tilde{C}}_{011100200},\\
&\text{Sector 92:} &J^{\tilde{C}}_{13} &= \eps^3 \left(\frac{-s}{\mu^2}\right) I^{\tilde{C}}_{001110200},\\
&\text{Sector 55:} &J^{\tilde{C}}_{14} &= \eps^3 \left(\frac{-s}{\mu^2}\right) \left(\frac{-s-t}{\mu^2}\right) I^{\tilde{C}}_{111021000},\\
&\text{Sector 59:} &J^{\tilde{C}}_{15} &= \eps^4 \left(\frac{-t}{\mu^2}\right) I^{\tilde{C}}_{110111000},\\
& &J^{\tilde{C}}_{16} &= \eps^3 \left(\frac{-t}{\mu^2}\right) \left(\frac{m^2}{\mu^2}\right) I^{\tilde{C}}_{110211000},\\
&\text{Sector 61:} &J^{\tilde{C}}_{17} &= \eps^4 \left(\frac{-s-t}{\mu^2}\right) I^{\tilde{C}}_{101111000},\\
&\text{Sector 62:} \hspace{0.5cm} &J^{\tilde{C}}_{18} &= \eps^4 \left(\frac{-s}{\mu^2}\right) I^{\tilde{C}}_{011111000},\\
& &J^{\tilde{C}}_{19} &= \eps^3 \left(\frac{-s}{\mu^2}\right) \left(\frac{m^2}{\mu^2}\right) I^{\tilde{C}}_{011211000},\\
&\text{Sector 79:} &J^{\tilde{C}}_{20} &= \eps^3 \left(\frac{-s}{\mu^2}\right) \left(\frac{-t}{\mu^2}\right) I^{\tilde{C}}_{111200100},\\
& &J^{\tilde{C}}_{21} &= \eps^3 \left(\frac{-s}{\mu^2}\right) I^{\tilde{C}}_{1112001(-1)0} - \left(\frac{s}{2t}\right) \left[J^{\tilde{C}}_2 - J^{\tilde{C}}_6 + 4J^{\tilde{C}}_7\right],\\
&\text{Sector 93:} &J^{\tilde{C}}_{22} &= \eps^4 \left(\frac{-s-t}{\mu^2}\right) I^{\tilde{C}}_{101110100},\\
& &J^{\tilde{C}}_{23} &= \eps^3 \left(\frac{-s-t}{\mu^2}\right) \left(\frac{m^2}{\mu^2}\right) I^{\tilde{C}}_{101210100},\\
&\text{Sector 94:} &J^{\tilde{C}}_{24} &= \eps^4 \left(\frac{-s}{\mu^2}\right) I^{\tilde{C}}_{011110100},\\
&\text{Sector 107:} &J^{\tilde{C}}_{25} &= \eps^4 \left(\frac{-s-t}{\mu^2}\right) I^{\tilde{C}}_{110101100},\\
& &J^{\tilde{C}}_{26} &= \eps^3 \left(\frac{(-s-t)m^2+s t}{\mu^4}\right) I^{\tilde{C}}_{110201100},\\
&\text{Sector 109:} &J^{\tilde{C}}_{27} &= \eps^4 \left(\frac{-t}{\mu^2}\right) I^{\tilde{C}}_{101101100},\\
&\text{Sector 110:} &J^{\tilde{C}}_{28} &= \eps^4 \left(\frac{-s}{\mu^2}\right) I^{\tilde{C}}_{011101100},\\
&\text{Sector 117:} &J^{\tilde{C}}_{29} &= \eps^4 \left(\frac{-t}{\mu^2}\right) I^{\tilde{C}}_{101011100},\\
&\text{Sector 121:} &J^{\tilde{C}}_{30} &= \eps^4 \left(\frac{-s}{\mu^2}\right) I^{\tilde{C}}_{100111100},\\
& &J^{\tilde{C}}_{31} &= \eps^3 \left(\frac{-s}{\mu^2}\right) \left(\frac{m^2}{\mu^2}\right) I^{\tilde{C}}_{100211100},\\
&\text{Sector 124:} &J^{\tilde{C}}_{32} &= \eps^4 \left(\frac{-s}{\mu^2}\right) I^{\tilde{C}}_{001111100},\\
&\text{Sector 63:} &J^{\tilde{C}}_{33} &= \eps^4 \left(\frac{-s}{\mu^2}\right) \left(\frac{-s-t}{\mu^2}\right) I^{\tilde{C}}_{111111000},\\
& &J^{\tilde{C}}_{34} &= \eps^4 \left(\frac{-s}{\mu^2}\right) I^{\tilde{C}}_{1111110(-1)0},\\
&\text{Sector 123:} &J^{\tilde{C}}_{35} &= \eps^4 \left(\frac{-s-t}{\mu^2}\right) \left(\frac{m^2}{\mu^2}\right) I^{\tilde{C}}_{110111100},\\
&\text{Sector 125:} \hspace{0.5cm} &J^{\tilde{C}}_{36} &= \eps^4 \left(\frac{r_5}{\mu^4}\right) I^{\tilde{C}}_{101111100},\\
& &J^{\tilde{C}}_{37} &= \eps^4 \left[\left(\frac{-s-t}{\mu^2}\right) I^{\tilde{C}}_{1(-1)1111100} - \frac{1}{2} \left(\frac{-s}{\mu^2}\right) \left(\frac{m^2-2(s+t)}{\mu^2}\right) I^{\tilde{C}}_{101111100}\right]\\
& & &\phantom{=} - \left(\frac{s}{2t}\right) \left[J^{\tilde{C}}_{10}+J^{\tilde{C}}_{13}-J^{\tilde{C}}_{17}+3J^{\tilde{C}}_{22}-J^{\tilde{C}}_{23}+J^{\tilde{C}}_{27}-J^{\tilde{C}}_{29}+3J^{\tilde{C}}_{30}-J^{\tilde{C}}_{31}\right.\\
& & &\phantom{=} \left. -J^{\tilde{C}}_{32}\right] - \left(\frac{t}{s}\right) J^{\tilde{C}}_{30},\\
&\text{Sector 126:} &J^{\tilde{C}}_{38} &= \eps^4 \left(\frac{-s}{\mu^2}\right) \left(\frac{m^2+s}{\mu^2}\right) I^{\tilde{C}}_{011111100},\\
&\text{Sector 127:} &J^{\tilde{C}}_{39} &= \eps^4 \left(\frac{s^2}{\mu^4}\right) I^{\tilde{C}}_{11111110(-1)}\\
& & &\phantom{=} - \left(\frac{s}{2t}\right) \left[-J^{\tilde{C}}_2 + 2J^{\tilde{C}}_3 + J^{\tilde{C}}_6 - 2J^{\tilde{C}}_7 + 2J^{\tilde{C}}_{11} - 2J^{\tilde{C}}_{20} - 6J^{\tilde{C}}_{25} + 2J^{\tilde{C}}_{26}\right.\\
& & &\phantom{=} \left. + 2J^{\tilde{C}}_{27} + 2J^{\tilde{C}}_{28}\right] + \left(\frac{2s^2 + t m^2 + 2s t}{2r_5}\right) J^{\tilde{C}}_{36},\\
& &J^{\tilde{C}}_{40} &= \eps^4 \left(\frac{-s}{\mu^2}\right) \left(\frac{m^2-t}{\mu^2}\right) \left[\left(\frac{-s-t}{\mu^2}\right) I^{\tilde{C}}_{111111100} + I^{\tilde{C}}_{1111111(-1)0}\right]\\
& & &\phantom{=} - \left(\frac{2s t - m^2 (s-t)}{2m^2(-s-t)}\right) J^{\tilde{C}}_{35} - \left(\frac{(2s+t) m^2}{2 r_5}\right) J^{\tilde{C}}_{36} - \left(\frac{m^2-t}{m^2+s}\right) J^{\tilde{C}}_{38},\\
& &J^{\tilde{C}}_{41} &= \eps^4 \left(\frac{-s}{\mu^2}\right) \left[I^{\tilde{C}}_{1111111(-2)0} - \left(\frac{-t}{\mu^2}\right) I^{\tilde{C}}_{11111110(-1)}\right.\\
& & &\phantom{=} \left. + \left(\frac{m^2-t}{\mu^2}\right) I^{\tilde{C}}_{1111111(-1)0} - I^{\tilde{C}}_{1111111(-1)(-1)}\right]\\
& & &\phantom{=} +\frac{1}{8}\left(\frac{t}{s}\right) \left(\frac{\mu^2}{-s-t}\right) \left[\left(\frac{-14s-8t}{\mu^2}\right)J^{\tilde{C}}_3 + \left(\frac{3s+2t}{\mu^2}\right)J^{\tilde{C}}_4\right]\\
& & &\phantom{=} + \left(\frac{s}{t}\right) \left[-J^{\tilde{C}}_{27} + J^{\tilde{C}}_{29}\right]\\
& & &\phantom{=} + \left(\frac{t}{8s}\right) \left[2J^{\tilde{C}}_2 + J^{\tilde{C}}_8 + 12J^{\tilde{C}}_9 + 4J^{\tilde{C}}_{11} + 8J^{\tilde{C}}_{18} - 4J^{\tilde{C}}_{24} + 12J^{\tilde{C}}_{28} - 4J^{\tilde{C}}_{32}\right]\\
& & &\phantom{=} + \left(\frac{t}{24(s+t)}\right) \left[J^{\tilde{C}}_5 + 3J^{\tilde{C}}_6 - 18J^{\tilde{C}}_7 + 36J^{\tilde{C}}_{15} - 12J^{\tilde{C}}_{16} - 12J^{\tilde{C}}_{25} - 36J^{\tilde{C}}_{30}\right.\\
& & &\phantom{=} \left. + 12J^{\tilde{C}}_{31} + 12J^{\tilde{C}}_{35}\right] + \left(\frac{m^2 t}{2r_5}\right) J^{\tilde{C}}_{36} + \left(\frac{2(m^2-t)}{m^2+s} + \frac{3t}{2s}\right) J^{\tilde{C}}_{38}.
\end{alignat*}

\subsubsection{Topology $\tilde{D}$}

\begin{alignat*}{3}
&\text{Sector 41:} \hspace{0.5cm}&J^{\tilde{D}}_1 &= \eps^2 \left(\frac{m^2}{\mu^2}\right) \mathbf{D^-} I^{\tilde{D}}_{100101000},\\
&\text{Sector 42:} &J^{\tilde{D}}_2 &= \eps^2 \left(\frac{-s}{\mu^2}\right) \mathbf{D^-} I^{\tilde{D}}_{010101000},\\
&\text{Sector 49:} &J^{\tilde{D}}_3 &= \eps^2 \left(\frac{m^2+s+t}{\mu^2}\right) \mathbf{D^-} I^{\tilde{D}}_{100011000},\\
& &J^{\tilde{D}}_4 &= \eps^2 \left[\left(\frac{m^2}{\mu^2}\right) \mathbf{D^-} I^{\tilde{D}}_{100011000} + \mathbf{D^-} I^{\tilde{D}}_{10001100(-1)}\right],\\
&\text{Sector 73:} &J^{\tilde{D}}_5 &= \eps^2 \left(\frac{m^2-t}{\mu^2}\right) \mathbf{D^-} I^{\tilde{D}}_{100100100}\\
& &J^{\tilde{D}}_6 &= \eps^2 \left(\frac{-t}{\mu^2}\right) I^{\tilde{D}}_{200100200},\\
&\text{Sector 43:} &J^{\tilde{D}}_7 &= \eps^3 \left(\frac{-s}{\mu^2}\right) I^{\tilde{D}}_{110102000},\\
&\text{Sector 54:} &J^{\tilde{D}}_8 &= \eps^3 \left(\frac{-s}{\mu^2}\right) I^{\tilde{D}}_{011021000},\\
&\text{Sector 55:} &J^{\tilde{D}}_9 &= \eps^3 (-1+2\eps) \left(\frac{-s}{\mu^2}\right) I^{\tilde{D}}_{111011000},\\
& &J^{\tilde{D}}_{10} &= \eps^3 \left(\frac{-s}{\mu^2}\right) \left(\frac{m^2+s+t}{\mu^2}\right) I^{\tilde{D}}_{111021000},\\
&\text{Sector 59:} &J^{\tilde{D}}_{11} &= \eps^4 \left(\frac{-t}{\mu^2}\right) I^{\tilde{D}}_{110111000},\\
& &J^{\tilde{D}}_{12} &= \eps^3 \left(\frac{-t}{\mu^2}\right) \left(\frac{m^2}{\mu^2}\right) I^{\tilde{D}}_{210111000},\\
&\text{Sector 79:} &J^{\tilde{D}}_{13} &= \eps^3 \left(\frac{-s}{\mu^2}\right) \left(\frac{m^2-t}{\mu^2}\right) I^{\tilde{D}}_{111200100} - J^{\tilde{D}}_6,\\
& &J^{\tilde{D}}_{14} &= \eps^3 \left(\frac{-s}{\mu^2}\right) \left[\left(\frac{m^2-t}{\mu^2}\right) I^{\tilde{D}}_{111200100} - I^{\tilde{D}}_{1112001(-1)0}\right] + \left(\frac{s}{2t}\right) J^{\tilde{D}}_6,\\
&\text{Sector 93:} &J^{\tilde{D}}_{15} &= \eps^4 \left(\frac{-s-t}{\mu^2}\right) I^{\tilde{D}}_{101110100},\\
& &J^{\tilde{D}}_{16} &= \eps^3 \left(\frac{-s-t}{\mu^2}\right) \left(\frac{m^2}{\mu^2}\right) I^{\tilde{D}}_{201110100},\\
&\text{Sector 121:} \hspace{0.5cm} &J^{\tilde{D}}_{17} &= \eps^4 \left(\frac{-s}{\mu^2}\right) I^{\tilde{D}}_{100111100},\\
& &J^{\tilde{D}}_{18} &= \eps^3 \left(\frac{s(m^2-t)-t^2}{\mu^4}\right) I^{\tilde{D}}_{200111100},\\
&\text{Sector 123:} &J^{\tilde{D}}_{19} &= \eps^4 \left(\frac{-s}{\mu^2}\right) \left(\frac{m^2}{\mu^2}\right) I^{\tilde{D}}_{110111100},\\
&\text{Sector 126:} &J^{\tilde{D}}_{20} &= \eps^4 \left(\frac{s^2}{\mu^4}\right) I^{\tilde{D}}_{011111100},\\
&\text{Sector 127:} &J^{\tilde{D}}_{21} &= \eps^4 \left(\frac{-s}{\mu^2}\right) \left(\frac{m^2+s+t}{\mu^2}\right) \left[\left(\frac{m^2-t}{\mu^2}\right) I^{\tilde{D}}_{111111100} + I^{\tilde{D}}_{1111111(-1)0}\right]\\
& & &\phantom{=} - \left(\frac{-s-t}{m^2}\right) J^{\tilde{D}}_{19} - \left(\frac{m^2+t}{-s}\right) J^{\tilde{D}}_{20},\\
& &J^{\tilde{D}}_{22} &= \eps^4 \left(\frac{-s}{\mu^2}\right) \left(\frac{m^2-t}{\mu^2}\right) \left[\left(\frac{m^2-s-t}{\mu^2}\right) I^{\tilde{D}}_{111111100} + I^{\tilde{D}}_{1111111(-1)0} \right]\\
& & &\phantom{=} - \left(\frac{-t}{m^2}\right) J^{\tilde{D}}_{19} - \left(\frac{m^2-t}{-s}\right) J^{\tilde{D}}_{20},\\
& &J^{\tilde{D}}_{23} &= \eps^4 \left(\frac{-s}{\mu^2}\right) \left[I^{\tilde{D}}_{1111111(-2)0} + \left(\frac{3m^2-t}{\mu^2}\right) I^{\tilde{D}}_{1111111(-1)0}\right.\\
& & &\phantom{=} \left. + 2\left(\frac{m^2}{\mu^2}\right) \left(\frac{m^2-t}{\mu^2}\right) I^{\tilde{D}}_{111111100} \right]\\
& & &\phantom{=} -\left(\frac{\mu^2}{-t}\right) \left(\frac{\mu^2}{-s-t}\right) \left[\left(\frac{s^2+st+t^2}{72\mu^4}\right)\left(36J^{\tilde{D}}_1 - J^{\tilde{D}}_2 + 48J^{\tilde{D}}_7 + 12J^{\tilde{D}}_{19}\right)\right.\\
& & &\phantom{=} \left. -\left(\frac{-s}{8 \mu^2}\right) \left(\frac{2t-s}{\mu^2}\right) (J^{\tilde{D}}_4 + 4J^{\tilde{D}}_{11})\right.\\
& & &\phantom{=} \left. + \left(\frac{-s}{8\mu^2}\right) \left(\frac{3s+2t}{\mu^2}\right) (J^{\tilde{D}}_5 - 2J^{\tilde{D}}_6 - 4J^{\tilde{D}}_{15})\right.\\
& & &\phantom{=} \left. - \left(\frac{s^2+st-5t^2}{6\mu^4}\right) J^{\tilde{D}}_{12} + \left(\frac{-s}{6\mu^2}\right) \left(\frac{5s+6t}{\mu^2}\right) J^{\tilde{D}}_{16}\right.\\
& & &\phantom{=} \left. - \left(\frac{s^2}{6\mu^4}\right) (3J^{\tilde{D}}_{17} + 2J^{\tilde{D}}_{18})\right] - \left(\frac{-s-t}{m^2}\right) J^{\tilde{D}}_{19} - \left(\frac{2m^2+t}{-s}\right) J^{\tilde{D}}_{20}.
\end{alignat*}

\subsubsection{Topology $\tilde{E}$}

\begin{alignat*}{3}
&\text{Sector 42:} \hspace{0.5cm} &J_1^{\tilde{E}} &= \eps^2 \left(\frac{-s}{\mu^2}\right) \mathbf{D^-} I^{\tilde{E}}_{010101000},\\
&\text{Sector 49:} &J^{\tilde{E}}_2 &= \eps^2 \left(\frac{-s-t}{\mu^2}\right) \mathbf{D^-} I^{\tilde{E}}_{100011000},\\
&\text{Sector 73:} &J^{\tilde{E}}_3 &= \eps^2 \left(\frac{-t}{\mu^2}\right) \mathbf{D^-} I^{\tilde{E}}_{100100100},\\
&\text{Sector 54:} &J^{\tilde{E}}_4 &= \eps^3 \left(\frac{-s}{\mu^2}\right) I^{\tilde{E}}_{011021000},\\
&\text{Sector 55:} &J^{\tilde{E}}_5 &= \eps^3 \left(\frac{-s}{\mu^2}\right) \left(\frac{-s-t}{\mu^2}\right) I^{\tilde{E}}_{111021000},\\
&\text{Sector 59:} &J^{\tilde{E}}_6 &= \eps^4 \left(\frac{-t}{\mu^2}\right) I^{\tilde{E}}_{110111000},\\
&\text{Sector 79:} &J^{\tilde{E}}_7 &= \eps^3 \left(\frac{-s}{\mu^2}\right) \left(\frac{-t}{\mu^2}\right) I^{\tilde{E}}_{111200100},\\
&\text{Sector 93:} &J^{\tilde{E}}_8 &= \eps^4 \left(\frac{-s-t}{\mu^2}\right) I^{\tilde{E}}_{101110100},\\
&\text{Sector 121:} &J^{\tilde{E}}_9 &= \eps^4 \left(\frac{-s}{\mu^2}\right) I^{\tilde{E}}_{100111100},\\
&\text{Sector 126:} &J^{\tilde{E}}_{10} &= \eps^4 \left(\frac{s^2}{\mu^4}\right) I^{\tilde{E}}_{011111100},\\
&\text{Sector 127:} &J^{\tilde{E}}_{11} &= \eps^4 \left(\frac{s^2}{\mu^2}\right) I^{\tilde{E}}_{11111110(-1)}\\
& & &\phantom{=} + \frac{1}{4(1+4\eps)}  \left(\frac{\mu^2}{t}\right)\left[2\left(\frac{-t}{\mu^2}\right) \big((1+3\eps) J^{\tilde{E}}_2 - \eps J^{\tilde{E}}_3 + 12 \eps J^{\tilde{E}}_6 + 12 \eps J^{\tilde{E}}_8\big)\right.\\
	& & &\phantom{=} \left. + \left(\frac{s}{\mu^2}\right) \big((1+6\eps) J^{\tilde{E}}_1 + J^{\tilde{E}}_3 + 4J^{\tilde{E}}_7 - 12 J^{\tilde{E}}_8\right.\\
	& & &\phantom{=} \left. + 2\eps \big[J^{\tilde{E}}_2 + 2J^{\tilde{E}}_3 + 8J^{\tilde{E}}_7 - 36 J^{\tilde{E}}_8 - 12 J^{\tilde{E}}_9\big]\big)\right],\\
& &J^{\tilde{E}}_{12} &= \eps^4 \left(\frac{-s}{\mu^2}\right) \left(\frac{-s-t}{\mu^2}\right) \left[\left(\frac{-t}{\mu^2}\right) I^{\tilde{E}}_{111111100} + I^{\tilde{E}}_{1111111(-1)0}\right] - \frac{t}{s} J^{\tilde{E}}_{10}\\
	& & &\phantom{=} + \frac{1}{1+4\eps} \left(\frac{\mu^2}{t}\right) \left[\frac{\eps}{2} \left(\frac{s}{\mu^2}\right)\big(J^{\tilde{E}}_1 + J^{\tilde{E}}_2\big)\right.\\
	& & &\phantom{=} \left. + \left(\frac{t}{\mu^2}\right) \left(-\frac{4+15\eps}{2} J^{\tilde{E}}_2 + \frac{1+6\eps}{4} J^{\tilde{E}}_3 - 6\eps J^{\tilde{E}}_6 - 6\eps J^{\tilde{E}}_8\right) \right.\\
	& & &\phantom{=} \left.- 6\eps \left(\frac{s}{\mu^2}\right) \big(J^{\tilde{E}}_8 + J^{\tilde{E}}_9\big) \right].
\end{alignat*}


\section{Supplementary material}
\label{sect:supplement}

Attached to the arxiv version of this article are for each topology
\bq
 X & \in & \left\{ A,B,C,D,E,\tilde{A},\tilde{B},\tilde{C},\tilde{D},\tilde{E}, \right\}
\eq
the electronic files
\begin{center}
 \verb|topo_X_symbolic.mpl|, \; \verb|topo_X_numeric.cc|, \; \verb|topo_X_numeric.out|
\end{center}
and a common file
\begin{center}
 \verb|input_common.mpl|
\end{center}
The files \verb|topo_X_symbolic.mpl| are in {\tt Maple} syntax and 
define for each topology $X$
the transformation matrix $U^X$ appearing in eq.~(\ref{def_fibre_transformation}),
its inverse $(U^X)^{-1}$,
the pre-canonical basis $I^X$ appearing in the same equation 
(as pre-canonical basis $I^X$ we choose the {\tt Kira} basis with integral ordering $5$),
the matrix $A^X$ appearing in the differential equation~(\ref{eps_factorised_deq}),
the boundary values $J_{\mathrm{boundary}}^X$ at the boundary point defined by eq.~(\ref{def_boundary_point}),
the alphabets for the various topologies (denoted by \verb|alphabet_lst_X|),
the explicit expressions for the differential one forms (given in \verb|alphabet_subs_lst_X|)
and the matrices $\tilde{M}_1^X$ determining the large logarithms (see section~\ref{sect:logarithms}).
These quantities are denoted as 
\begin{flushleft}
 \hspace*{10mm}
 \verb|U_X|, \; \verb|Uinv_X|, \; \verb|I_X|, \; \verb|A_X|, \; \verb|J_boundary_X|, \\
 \hspace*{10mm}
 \verb|alphabet_lst_X|, \; \verb|alphabet_subs_lst_X|, \; \verb|Mtilde_1_X|.
\end{flushleft}
The pre-canonical master integrals in the vector \verb|I_X| are for example denoted as
\begin{center}
 \verb|I_1111111m10|,
\end{center}
which stands for $I_{1111111(-1)0}$. 
The zeta value $\zeta_3$ is denoted by \verb|zeta_3|.
The square roots $r_i$, the periods $\psi^{(x)}_1$ and the functions $F^{X}_{ij}$ are denoted 
for example as
\begin{center}
 \verb|R1(s,t,m2)|, \; \verb|Psi1_a(s,t,m2)|, \; \verb|F_Atilde_39_38(s,t,m2)|,
\end{center}
and similar for the other ones.
The file Maple \verb|input_common.mpl|
collects the relevant information on these functions.
Without loss of generality we have set $\mu=1$ in the Maple files.

The file \verb|topo_X_numeric.cc| is a {\tt C++}-program and provides numerical evaluation routines
for all master integrals of a given topology.
This {\tt C++}-program requires the \verb|GiNaC|-library \cite{Bauer:2000cp}.
The file \verb|topo_X_numeric.out| gives the output of the corresponding {\tt C++}-program with the default parameters.
This allows to verify that the compiled programs execute correctly.

\end{appendix}

{\footnotesize
\bibliography{/home/stefanw/notes/biblio}
\bibliographystyle{/home/stefanw/latex-style/h-physrev5}
}

\end{document}